# Between Fear and Trust:
# Factors Influencing Older Adults' Evaluation of Socially Assistive Robots

Thesis submitted in partial fulfillment
of the requirements for the degree of
"DOCTOR OF PHILOSOPHY"

by

# Oded Zafrani

Submitted to the Senate of
Ben-Gurion University of the Negev

06.03.2022

Beer-Sheva

# Between Fear and Trust:
# Factors Influencing Older Adults' Evaluation of Socially Assistive Robots

Thesis submitted in partial fulfillment

of the requirements for the degree of

"DOCTOR OF PHILOSOPHY"

by

# Oded Zafrani

Submitted to the Senate of

Ben-Gurion University of the Negev

Approved by the advisor: Prof. Galit Nimrod 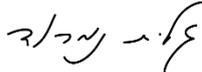

Approved by the advisor: Prof. Yael Edan 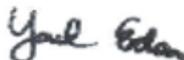

Approved by the Dean of the Kreitman School of Advanced Graduate Studies

06.03.2022

Beer-Sheva

i

This work was carried out under the supervision of

Prof. Galit Nimrod

Prof. Yael Edan

In the Department of Industrial Engineering and Management

Faculty of Engineering Sciences

Ben-Gurion University of the Negev

# **Research-Student's Affidavit when Submitting the Doctoral Thesis for Judgment**

I, Oded Zafrani, whose signature appears below, hereby declare that

✓ I have written this Thesis by myself, except for the help and guidance offered by my Thesis Advisors.

✓ The scientific materials included in this Thesis are products of my own research, culled <u>from the period during which I was a research student</u>.

Date: O6.03.2022     Student's name: Oded Zafrani     Signature: 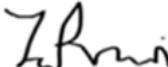

**Acknowledgements**

First and foremost, I am grateful and thankful to Almighty God for the infinite blessings in my life and for giving me the health, power, diligence, perseverance, and determination in each step of my life to continue achieving my goals.

I would like to express special appreciation to my supervisors, Prof. Galit Nimrod and Prof. Yael Edan, your mentorship has been tremendous. I am grateful for your close professional guidance, dedication to the work, promptness in reviews, commitment to excellence, vast knowledge and helpful insights, and devotedness to my all-round development that contributed to the success of this research and kept me inspired all through this PhD journey. Aside from being educated, brilliant and groundbreaking researchers, I have been blessed with peaceful, pleasant, enabling, supportive, and considerate supervisors. I had the privilege and pleasure of being your student. Thank you for always being there.

A big thank you to my PhD guidance committee, Prof. Noam Tractinsky and Prof. Norm O'Rourke for their dedication to ensure high quality research at every stage of the work. I cherish you for providing me with detailed feedback and valuable inputs at every stage of the research.

I want to also express huge appreciation to all my lab members, and to the administrative and technical staff at the department of Industrial Engineering and Management for all their support, collaboration, friendship, and assistance.

I would also like to thank the 19 charming and kind older adults who participated in this research. They opened their hearts and homes to participate in a new and challenging digital experience. I appreciate each and every one of them, and wish them a healthy, happy and long life.

Last but certainly not least, my endless appreciation and love goes to my parents, sisters and brother for their unconditional love, patience, continuous support and constant encouragement. You have always believed in me and my capabilities and pushed me to achieve more when I did not think it was possible.

This research was partially supported by the EU funded Innovative Training Network (ITN) in the Marie Skłodowska-Curie People Programme (Horizon2020): SOCRATES (Social Cognitive Robotics in a European Society training research network), by a grant from the Israeli Ministry for Science and Technology [grant number: 3-15713, and by Ben-Gurion University of the Negev through the Agricultural, Biological, and Cognitive Robotics Initiative, the Marcus Endowment Fund, the Rabbi W. Plaut Chair for Manufacturing Engineering.

**Dedication**

I would like to thank my family for their unconditional love, encouragement, and empowerment to realize myself and achieve my goals. I dedicate this work to my loving parents Mercedes and Raphael, for instilling in me the virtues of discipline, hard work and achievement. You are the best parents anyone can ever have. Likewise, I dedicate this work to my sisters Avital and Reut, and to my brother Itay, for their constant encouragement, support and by standing with me during the difficult times. I love you all and may Almighty God continue showering you with His blessings.

**Table of Contents**

# List of Publications

**Thesis journal publications**

**Zafrani, O**., & Nimrod, G. (2019). Towards a holistic approach to studying Human Robot interaction in later life. *The Gerontologist, 59*(1), e26–e36.
<https://doi.org/10.1093/geront/gny077>

**Journal publications related to the thesis in which I had contributions**

Krakovsky, M., Kumar, S., Givati, S., Bardea, M., **Zafrani, O**., Nimrod, G., Bar-Haim, S., & Edan, Y. (2021). "Gymmy": Designing and testing a robot for physical and cognitive training of older adults. *Applied Sciences, 11*(14), 6431. https://doi.org/10.3390/app11146431

**Manuscripts in preparation**

**Zafrani, O**., G. Nimrod, Y. Edan. Between fear and trust: Older adults' evaluation of socially assistive robots.

**Zafrani, O**., Edan. Y., Krakovsky, M., Kumar, S., Givati, S., Bardea, M., & Nimrod, G. Fans vs. Skeptics: Older adults' assimilation of socially assistive robots.

# List of Figures

# List of Tables

# List of Acronyms

HRI: Human-Robot Interaction

ICT: Information and Communication technology

QE: Quality Evaluation

SARs: Socially Assistive Robots

SWB: Subjective Well-Being

UEQ: User Experience Questionnaire

UX: User experience

# Abstract

The world's population is aging rapidly, and the number of older adults is projected to increase dramatically in the years to come. Socially Assistive Robots (SARs) are expected to help humanity face the challenges posed by this trend by supporting autonomy, aging in place, and wellbeing in later life. For successful acceptance and assimilation of SARs, it is necessary to understand the factors affecting older adults' Quality Evaluations (QEs) of them. Previous studies examining Human-Robot Interaction (HRI) in later life indicated that trust in robots significantly enhances QE, while aspects of technophobia considerably decrease it. Yet, previous research examined separately the impacts of trust and of technophobia on SARs QE among older adults, ignoring the possibility that these key factors can coexist and neutralize each other's influence. Moreover, contrary to trust, technophobia has hardly been investigated in the context of HRI. In addition, the existing literature suggests that older adults' overall QE of SARs is shaped by three aspects: their uses, constraints to beneficial use, and use outcomes. However, these studies were usually limited in duration, focused on acceptance aspects rather than on assimilation, and typically focused on only one aspect of the interaction between robots and older adults, i.e., examining uses, constraints, and outcomes separately. Furthermore, most HRI studies to date relied on either quantitative or qualitative analyses and did not apply a mixed-methods approach.

This dissertation aimed to bridge the gaps in the existing literature based on two complementary studies. First, an acceptance study simultaneously examined the effect of trust and technophobia on older adults' QE of SARs through an online survey of 384 individuals. Then, an assimilation study with nineteen community-dwelling older adults explored how the QE is shaped following actual interaction with SARs by a simultaneous exploration of uses, constraints, and outcomes in real-life conditions over a long period. This study relied on in-depth interviews, weekly surveys and use reports produced by the SAR. In both parts of the research, we used "Gymmy", a robotic system for the physical and cognitive training of older adults, developed in our lab. Both parts of the research were conducted along the COVID-19 pandemic.

The results indicated that the relative impact of technophobia on older adults' QE of SARs is significantly more substantial than that of trust, and that robot-related technophobia constituted a most influential antecedent constraint to SARs use. In addition, two patterns were found in the assimilation study: (A) The 'Fans' - participants who enjoyed using Gymmy, trusted it, attributed added value to it, and experienced a successful assimilation process, and (B) The 'Skeptics' - participants who did not like Gymmy, negatively evaluated its use, and experienced a disappointing assimilation process. This group expressed technophobia before participation in the study, whereas the 'Fans' did not report any fear or reservation concerning robots.

The combination of acceptance and assimilation research suggested that an assimilation pattern can be predicted according to the level of acceptance. The findings highlight the importance of investigating technophobia in HRI studies and propose that implementing robotics technology in later life strongly depends on reducing older adults' sense of technophobia. The conceptual framework of the dissertation made it possible to understand in depth the interrelationships and the mutual influence of facilitators and inhibitors factors, as well as their relative impact on older adults' evaluations of SARs. Furthermore, both practically and theoretically, this dissertation demonstrates the usefulness of a holistic approach in the research of older technology users and sheds light on the value of simultaneous explorations of uses, constraints, and outcomes, longitudinal methods in real-life conditions in assimilation studies, and the use of a mixed-methods approach in HRI studies.

**KEYWORDS:** Aging, human-robot interaction, quality evaluation, socially assistive robots, technophobia, trust, acceptance, assimilation, older adults.

# 1. Introduction

## 1.1. Problem description

Population aging is expected to be the most significant demographic transformation of the twenty-first century, with implications for nearly all sectors of society (Morina & Grima, 2021). Recent forecasts suggest that by 2050, one in six people worldwide will be aged 65 years or above (Backman et al., 2021), and the ratio is projected to increase even more in the future (Broniatowska, 2019; Taieb-Maimon & Vaisman-Fairstein, 2022). This trend yields numerous social and economic challenges related to health and quality of life in old age in general, and a healthy life expectancy in particular (Zhu & Walker, 2021). To address the challenges associated with aging, smart technology solutions can prove beneficial (Ghorayeb et al., 2021), wherein Socially Assistive Robots (SARs) are expected to play a central role (Beer et al., 2012; Cortellessa et al., 2021). Indeed, in the past two decades the use of SARs has been gradually evolving, and a variety of robotic technologies have either been created particularly for older adults or adapted to their needs (Allaban et al., 2020; Vandemeulebroucke et al., 2021). SARs are designed to develop effective and close interactions with humans for the purpose of providing assistance in convalescence, learning, rehabilitation, or therapy (Feil-Seifer & Mataric, 2005), and have great potential to support autonomy, aging in place, and wellbeing in later life (Fields et al., 2021). However, to facilitate acceptance and assimilation of SARs, realize their potential benefits, and reduce their potential risks, it is necessary to understand factors affecting older adults' Quality Evaluation (QE) of SARs (Andriella et al., 2021), including pragmatic and hedonic evaluations and overall attractiveness (Khalid, 2006; Mlekus et al., 2020; Santoso et al., 2016).

Previous studies that examined Human-Robot Interaction (HRI) in later life (Naneva et al., 2020; Rogers et al., 2021; Tussyadiah, et al., 2020) indicated that trust in robots significantly enhances QE (Martelaro et al., 2016; Naneva et al., 2020; Salem et al., 2015), while aspects of technophobia considerably decrease it (Syrdal et al., 2009; Tussyadiah, et al., 2020). In addition, they suggest that older adults' overall QE of SARs is shaped by three aspects: their uses, constraints to beneficial use, and use outcomes (Zafrani & Nimrod, 2019). However, previous research has four significant weaknesses: (1) although trust and technophobia may affect each other, studies examined *separately* the impacts of *trust* and *aspects of technophobia on SARs QE among older adults,* ignoring the possibility that these key factors can coexist and neutralize each other's influence. In this context, it is important to note that contrary to trust, *technophobia has hardly been investigated in the context of HRI*. (2) studies typically focused on *only one aspect of the interaction* between robots and older adults, (i.e., examining uses, constraints, and outcomes separately). (3) most studies were *limited in duration*, and thus mainly focused on acceptance aspects rather than assimilation; and (4) very few studies have investigated HRI using a mixed- methods approach (quantitative and qualitative), which are necessary for a more comprehensive understanding of the reciprocity between humans and robots (Seibt et al., 2021). Accordingly, no study thus far has explored how the QE of SARs among older adults is shaped, both with regard to anticipated interaction and to actual interaction.

### 1.2. Research objectives

This research aimed to bridge the gaps in the existing literature that explores what affects SARs QE among older adults. Accordingly, it was carried out in two parts: (a) an *acceptance study* that applied an online survey to explore the coexistence and possible relative *effects of trust and technophobia* on SAR's QE as shaped by anticipated interaction; and (b) an *assimilation study* examining how the QE is shaped following actual interaction with the SAR by a *simultaneous exploration of uses, constraints, and outcomes* in real-life conditions over a long period.

The acceptance study performed using quantitative analyses aimed at exploring the following questions:

1. Is trust in robots and robot-related technophobia associated with QE of SARs among older adults, and if so, how?
2. To what extent and how is the combination of trust in robots and robot-related technophobia associated with QE of SARs among older adults?

The assimilation study performed using qualitative methods was designed to examine the following questions:

1. What are the simultaneous exploration of uses, constraints and outcomes that older adults experience while assimilating a SAR into their lives?
2. How do older adults' experiences with a SAR over a long period and in real life conditions affect their QE of that SAR in particular, and of SARs in general?

In both parts of the study, we used Gymmy—a personal trainer robot developed in our lab (Krakovsky et al., 2021; Krakovsky, 2022, Figure 1). With Gymmy, older adults can exercise independently at home according to their schedule. Gymmy serves as a robotic physical training coach that demonstrates the physical exercises. In addition, during the training sessions, cognitive training activities, such as memory and math exercises, are randomly presented to the users. Furthermore, the system offers users relaxation exercises to release stress and relieve pressure, according to Jacobson's relaxation technique (Jacobson, 1938). This robot was specifically developed and used as a test case, so that the participants' reference and evaluation will be towards a concrete robot, and not to a robot as a conceptual idea.

### 1.3. Research contribution and innovation

This research significantly contributed to bridging the gaps in the existing literature by performing for the first time:

(a) an *investigation of* how the *QE of SARs* among older adults is shaped *both* with regard to *anticipated interaction* (acceptance study) and to *actual interaction* over time (assimilation study) .

(b) *Simultaneous examination* of the effect of *trust* in robots and *robot-related technophobia* on older adults' QE of SARs.

(c) An assimilation study examining what characterizes the process of older adults' QE of SARs by a *simultaneous exploration of* the SAR's *uses, constraints, and outcomes in real-life conditions over a long period*.

In addition, this study applied *a mixed-methods approach*, utilizing *both quantitative* and *qualitative methods*. A mixed-methods approach can more accurately, deeply and reasonably reflect the dynamic complexity and subtlety of human experience in HRI. The present study demonstrated the usefulness of this approach.

### 1.4. Dissertation structure

The thesis is structured as follows: Chapter 2 presents a thorough literature review that provides the relevant background for the research reported in this dissertation. Chapter 3 presents the research overview, the system description, and the goals and questions for each part of the study. Chapters 4 and 5 are devoted to the acceptance study and assimilation study respectively. They present their methods and results and offer discussions of the findings. Finally, Chapter 6 presents an integrative general discussion of the two parts of the study, including main conclusions, their limitations, and suggestions for future research.

## 2. Literature Review

### 2.1. The challenges associated with later life

In most cases, older adulthood is a period of life associated with inevitable physical, cognitive, and social deterioration, accompanied by new challenges (Munukka et al., 2021; Nowlan et al., 2015; Shaw et al., 2016; Segel-Karpas et al., 2021). Approximately 85% of older adults aged 65 years and above have at least one chronic health condition, and this percentage is expected to increase as the population ages (Curtis, 2021; McKay et al., 2020; Mitzner et al., 2011). Age brings with it a decline in the performance of cognitive abilities such as memory and processing speed, problem solving, focused and divided attention, flexibility, eye-hand coordination, and decision making (Sebri et al., 2019; Yow et al., 2021; Zhang et al., 2017), as well as impairment of physical abilities such as walking, vision, hearing, balance, aerobic capacity, flexibility and strength (Alici & Donmez, 2020; de Carvalho Fonseca et al., 2018; Hutchinson et al., 2020; Schmidt, 2015). Additionally, this population is often faced with stressful life events such as retirement, death of loved ones, reduced financial resources, declining health and loss of independence (Hannaford et al., 2018) that affect their quality of life and Subjective Well-Being (SWB, Gupta & Sharma, 2018; Jennison, 1992; Patel et al., 2021).

The older adult population represent one of the most rapidly growing segments of the world's population (Aiello et al., 2021; Damluji et al., 2017; Stone et al., 2018). Recent forecasts suggest that the number of individuals in this age group is expected to more than double by 2050 and more than triple by 2100, rising from 962 million globally in 2017 to 2.1 billion in 2050 and 3.1 billion in 2100 (Trombetti et al., 2016; United Nations, 2019).

"Older adult" is a broad definition, uniting under the same label a heterogeneous group of individuals with different needs (Gonçalves et al., 2009; Rogers, 2019). Most gerontology and geriatrics literature set minimum age limits of 60 or 65 years as the threshold for older adulthood (e.g., Ediev et al., 2016; Pathak et al., 2017). This threshold is consistent with official definitions of old age (e.g., United Nations, 2019; World Health Organization, 2002). The term "older adults" contains within it two main age groups: "young-old" (65-74 years), and "old-old" (75 years and older; Boot et al., 2020; Kubota et al., 2012). Yet, there are meaningful differences between the two groups in various aspects such as functional status, social involvement, leisure activities and everyday life patterns (Krakovsky et al., 2021; Menec & Chipperfield, 1997; Nimrod & Shrira, 2014; Zarit et al., 1995).

The literature suggests that old-old adults are typically more physically and mentally challenged than young-old adults (Droz et al., 2008; Gottlieb & Gillespie, 2008; Hammer et al., 2010; Kaushik, 2021; Suji & Sivakami, 2004). Specifically, more than 50% of those over age 75 are unable to perform at least one of the Activities of Daily Living (ADL, e.g., getting in and out of bed, dressing, bathing, walking) without help (Brummel-Smith, 2013; Teggi, 2020). Similarly, old-old adults sometimes have a lower level of cognitive capability, compared to the young-old adults (Xu et al., 2016). Furthermore, old-old adults are influenced by a number of biological, cultural, social, psychological and lifestyle factors that affect their ability to deal with life transitions and adjusting to new circumstances by adopting new technologies (Nimrod et al., 2008; Thalacker-Mercer, 2016; Tu et al., 2021). In contrast, young-old adults are more willing to learn about new technologies and to find ways to integrate them into their repertoire of life skills (Boulton-Lewis et al., 2016; Chen et al., 2020). The old-old prefer to live independently at home, in their familiar surroundings rather than in nursing homes (Ettema et al., 2016; Koru, 2021). However, studies have consistently found that older adults who live alone are more likely to have lower levels of SWB (e.g., Kharicha et al., 2007; Nauck & Ren, 2021; Stahl et al., 2017; Xiu-Ying et al., 2012).

To address the challenges associated with aging, smart technology solutions can prove beneficial, wherein Socially Assistive Robots (SARs) are expected to play a central role (Beer et al., 2012; Cortellessa et al., 2021). In the past two decades, various robotic technologies have been created particularly for older people or adapted to their needs (Abou Allaban et al., 2020; Vandemeulebroucke et al., 2021). These robotic systems have helped improve older people's quality of life and physical and cognitive function and compensated for the existing shortage of caregivers (Fasola & Matarić, 2013; Padir et al., 2015; Pu et al., 2019; Tsardoulias et al., 2017). Assisting users through social as well as physical interaction, SARs are generally designed to provide the appropriate emotional, cognitive, and social cues to encourage individuals' development, learning, or therapy (Feil-Seifer & Mataric, 2005; Matarić et al., 2007). SARs for older adults can be grouped into service-type or companion-type robots. Service-type robots are typically designed to assist frail older people with specific Activities of Daily Living (ADL), such as bathing, dressing and eating (Beedholm et al., 2016; Durães et al., 2018; Ghafurian et al., 2021) and Instrumental Activities of Daily Living (IADL) such as housekeeping and shopping, tasks that are not mandatory for fundamental functioning, but are essential for independent living and interaction with the environment (Boyle et al., 2010; Gomes et al., 2021). Companion-type robots are used primarily to improve the user's wellbeing and provide social activities (Baisch et al., 2017; Baisch et al., 2018; Pilotto et al., 2018; Schüssler et al., 2020). Both robot types have great potential to support autonomy, aging in place, and wellbeing in later life (Fields et al., 2021). However, in order to realize these benefits and achieve successful assimilation, it is necessary to understand factors affecting older adults' evaluations of SARs (Andriella et al., 2021; Frennert, 2019) and the robots' ability to benefit them and enhance their well-being.

### 2.2. Human-Robot Interaction (HRI)

Human-robot interaction (HRI) is defined as "a field of study dedicated to understanding, designing, and evaluating robotic systems for use by or with humans" (Goodrich & Schultz, 2007, p. 204). HRI is a multi-disciplinary field that combines both human and robot factors and hence it is studied in various scientific disciplines including psychology, cognitive science, social sciences, engineering, and computer sciences (Adamides et al., 2014; Dautenhahn, 2007; Roesler et al., 2021). It addresses both technological and social aspects related to artificial intelligence, robotics, and human–computer interaction (Seibt et al., 2021). Interaction, by definition, requires communication between robots and humans to accomplish a specific goal (Goodrich & Schultz 2007).

HRI studies cover various social and physical aspects of the interaction and include theoretical and empirical research applying both quantitative and qualitative methods (Hoffman & Zhao, 2020; Veling & McGinn, 2021), athough most work to date has focused on quantitative research (Zafrani & Nimrod, 2019). As highlighted by Seibt et al. (2021), future HRI research should adopt a mixed methods approach (i.e., integrating quantitative and qualitative research). Qualitative research contributes greatly to understanding the complexity of human socio-cultural reality in different contexts, human perspectives, and the nature of interactions, which are essential factors in HRI studies (Crescitelli et al., 2019; Seibt et al., 2021; Veling & McGinn, 2021). Moreover, qualitative research brings the researcher closer to the users, and draws research attention to the human, psycho-social, cultural and multidisciplinary aspects of human experience that is particularly relevant to HRI studies (Seibt et al., 2021). It is especially important to evaluate the human aspects when the users are potentially vulnerable groups such as older adults, people with disabilities, and children (Seibt et al., 2021; Veling & McGinn, 2021; Weiss et al., 2009).

### 2.2.1. Types of research in HRI

*Snapshot vs. Longitudinal studies.* A snapshot study is carried out in a very short period of time, usually under non-natural conditions such as laboratory (Fluck & Hillier, 2014). They are often held a relatively short time after the introduction of a new technology (Viscusi, 2012), aiming of providing a picture of the current situation in a specific location at a specific time (Bishop et al., 2015), and in order to learn about users' acceptance of technology, i.e., their willingness to use technology for the tasks it is designed to support (Teo, 2011). People's acceptance of technology is an essential prerequisite for the success of HRI (Huang et al., 2021).

On the other hand, a longitudinal study involves repeated observations of specific individuals over long periods of time (Wang, 2021). This kind of study can take place over a period of weeks, months, or even years. A longitudinal study helps researchers learn and interpret people's behavior, including how they acquire new knowledge about technology, and to what extent they use and retain this knowledge over time (King, 2006). Furthermore, the unique characteristics of this type of study enables more accurate insights regarding the assimilation of new technologies in the users' lives (Cullen, 2018; Nagarajan et al., 2020). The assimilation of technology is defined as the extent to which the use of technology becomes routinized in daily activities (De Mattos & Laurindo, 2017; Purvis et al., 2001). The ability to successfully assimilate new technology depends on users' absorption or purchase of information, as well as their ability to exploit this information (Kouki et al., 2010). In reviewing the literature, Zafrani and Nimrod (2019) pointed out that the majority of HRI studies in later life were snapshot studies that lasted one day, thus, they mainly focused on acceptance rather than on assimilation (Ng et al., 2012; Šabanović et al., 2013).

*Video scenario vs. Real life condition studies vs. Laboratory studies.* A video of an HRI scenario is a tool to support viewers' imagination in order to describe to them the robot and its features (Woods et al., 2006). This video aims to predict the desired user experience in terms of use, advantages and disadvantages, created and shaped by viewers as a result of the anticipated interaction with a robot (Pillan et al., 2014; Woods et al., 2006). Video based HRI studies have several potential advantages such as the ability to reach larger numbers of individuals, easily incorporate new ideas and topics into later trials through video editing, test initial assumptions, and allow greater control over experimental conditions (Woods et al., 2006). Previous studies on HRI highlighted that although this methodology is inferior to a live HRI session, it can certainly be a very appropriate method for developing and trying out exploratory studies and pilot testing (Kidd, 2003; Paiva et al., 2004; Woods et al., 2006).

Real-world studies are typically conducted with actual robots in real life conditions such as homes, hospitals, or offices (Broadbent, 2017; Syrdal et al., 2015). In these studies, a robot must be developed and/or programmed to operate (Broadbent, 2017). In contrast to video scenario studies, real-world robot studies are more difficult to perform, usually take longer to conduct, and are much more expensive (Fink et al., 2013; Marge et al., 2009). However, they also have advantages. First and foremost, humans are actually interacting with robots, as opposed to watching videos or reading descriptions of them (Broadbent, 2017; Woods et al., 2006). Moreover, real-world studies are conducted in natural environments and remove the artificiality of the laboratory (Bethel & Murphy, 2010; Broadbent, 2017; Hoffman & Zhao, 2020).

Due to the challenge of conducting research in the real life, laboratory studies are widely used in HRI research (Babel et al., 2021; Mubin et al., 2018). Laboratory studies are conducted in an artificial, controlled and context-independent setting and thus, are limited in their generalizability to other settings (Abich et al., 2015; Hoffman & Zhao, 2020). This type of study is useful for exploring participants' perceptions regarding the reliability and safety of the robotic solution (Frennert & Jæger, 2016). However, in laboratory studies, the participants' attributes, desires, and needs are assumed to be static and stable (Lazar et al., 2017). Therefore, given that participants have more heterogeneous and dynamic characteristics, the laboratory studies do not provide deep insights of the reciprocal fit between the participant and the robot (Frennert & Jæger, 2016).

### 2.2.2. Performance measurement in HRI

Measurement of HRI studies has been conducted with a multitude of subjective and objective performance measures (Habermehl, 2017; Hoffman & Zhao, 2020; Marvel et al., 2020; Marvel et al., 2021; Murphy & Schreckenghost, 2013; Schermerhorn & Scheutz, 2011; Steinfeld et al., 2020). In this section, common performance measures in HRI studies that are relevant to this thesis are noted and defined.

*User experience (UX)* is a person's perceptions, insights, and responses that result from the use and/or anticipated use of a product, system or service (ISO 9241-210:2010). In HRI, UX is seen as a holistic concept that refers to how users experience the interaction or even only the anticipated interaction with the robot on an emotional, perceptional, mental, cognitive, physical and psychological level (Hebesberger et al., 2017; Pallot et al., 2013). UX is a complex phenomenon depending on the interaction between user profile characteristics, product system or service characteristics, sociocultural factors, and the context of use (Fu, 2004; Pallot et al., 2013; Schröppel et al., 2021). It is unique, subjective, temporary, and dynamic, as each person has their own repertoire of knowledge, prior experiences, skills and expectations (Hassenzahl & Tractinsky, 2006; Ntoa et al., 2021; Vermeeren et al., 2010). The UX is measured using both qualitative (e.g., interviews; Abbas et al., 2020) and quantitative methods (e.g., questionnaires such as UEQ; Buyukgoz et al., 2021).

*Usability*, i.e., the extent to which a product or service can be used by specified users to achieve specific goals (Lewandowski et al., 2020), is a broad concept that considers the following factors:

<u>Effectiveness.</u> The percentage of accuracy of the human-robot team in completing tasks and achieving specified goals in particular environments (Ganesan, 2017; Holm et al., 2021). This normally refers to the degree to which errors are avoided and tasks are carried out successfully, measured by "success rate" or "task completion rate", for example.

<u>Efficiency.</u> The resources expended in relation to the number of correctly completed tasks and goals per unit of time (Grabowski et al., 2021; Wojtak et al., 2021), measured for example by quantity, speed, or rate.

<u>Learnability.</u> The ease with which a novice user learns to use an interactive system to achieve a goal (Xiao et al., 2021). This is measured in terms of understandability of steps required to complete tasks and memorability of the user interface.

<u>Flexibility.</u> The ability of the human–robot system to respond to changes in its initial objectives and requirements (Arnold, 2006; Brugali, 2021). This is measured in terms of the number of possible different tasks and conditions of the system.

<u>Utility.</u> The level to which users feel that using a specific product or and service helps them achieve goals and tasks (Teggar et al., 2021). The more tasks the interface performs, the more utility it has. This is measured, for example, by the frequency with which the users feel they successfully control the system.

<u>Robustness.</u> The degree to which a human–robot system can function correctly to enable a successful achievement of tasks and goals, in a state of environment uncertainty (Shah et al., 2007; Smith et al., 2021). This is measured, for example, by the level of support provided to the user in case of uncertainty or failure to perform a task.

*Functionality* is a fundamental attribute of a system or product which indicates the ability to be used to perform specific actions for which it is intended and developed (Calisir et al., 2014; Ji et al., 2020). Functionality is a crucial step in creating or destroying trust between humans and robots, as it has a positive impact on users' perceptions and attitudes in relation to the benefit and added value of the robot (Haring et al., 2018). This is measured, for example, by the robot's "reliability" or "availability".

*Ease of use* is defined as the degree to which a person believes that using a particular system would be free of physical or mental effort (Davis, 1989; Stadler et al., 2014). Ease of use addresses users' personal needs in pursuing the functionality of robot technologies (Lu et al., 2019; Song & Kim, 2021).

*Perceived convenience* is defined as the degree to which an individual believes that a product or service would provide flexibility, availability, accessibility, and efficiency in time, energy, place, and effort (Ogbanufe et al., 2018; Okazaki & Mendez, 2013; Zhang et al., 2017). Additionally, a product or service is perceived as convenient when it reduces the emotional, cognitive, and physical burdens on their users (Chang et al., 2012).

Both *ease of use* and *perceived convenience* are measured, for example, by self-report questionnaires after using the robotic system (Liang & Lee, 2016).

*Trust* and *technophobia*, the two core concepts of this research, are defined in detail below in sections 2.5 and 2.6 respectively, along with their common measures.

### 2.3. Human-Robot Interaction (HRI) in later life

A review of the literature that examined HRI in later life (Pu et al., 2019; Rogers et al., 2021; Zafrani & Nimrod, 2019) suggests that previous research explored three major topics: uses, constraints and outcomes. Below are the principal insights concerning each topic.

*Uses.* This category included explorations of a) users' acceptance of new robotics technology, b) processes of adaptation to such technologies, and c) factors affecting user experience. Many studies suggested that although older adults and their formal caregivers were interested in robots and even excited about them, their *acceptance* of robotics technologies was somewhat ambivalent (González-González et al., 2021;

Hebesberger et al., 2017). For example, whereas older adults saw robotics as a future extension of existing communications technologies such as the Internet and smartphones, and expected robots to be widely adopted, they were also concerned that such technologies would replace and even control humans sooner or later (Liu et al., 2021; Walden et al., 2015). People with Alzheimer's Disease (AD) also demonstrated such ambivalence (Salichs et al., 2016). Expressing prospects for support in daily life activities, such individuals stated that they did not want to use robots (Wang et al., 2017). Similarly, formal caregivers reported both enthusiasm about robots (Lewis et al., 2016) and dislike of sharing their workspace with them (Hebesberger et al., 2017). In fact, according to the literature, the only audience that was entirely positive about robots was that of the informal caregivers, who demonstrated openness to robotics technology, understanding of its benefits, and a desire to use it (e.g., Abbott et al., 2019; Wang et al., 2017).

Some studies revealed that people often attribute human traits to robots and expect them to exhibit human behavior and intelligence, even though they know clearly that they are dealing with machines (Frennert et al., 2017; Onnasch & Roesler, 2021). Among persons with AD, this subject-machine duality led, in certain situations, to agitation, rejection, and displeasure (Klein et al., 2013). In addition to humanizing the robots, older adults often compared them to humans. One study, for example, reported that older adults were discerning in their approval of support for different tasks, and preferred robots for tasks related to manipulating objects, chores, and information management, but sought humans for tasks related to leisure activities and personal care (Getson & Nejat, 2021; Smarr et al., 2014). In another study, participants favored the robot instructor for physical exercise training, although they displayed strong inclinations towards humans for information delivery (Shen & Wu, 2016). In addition, a number of recent studies have shown that older adults were less receptive to intimate robotic physical assistance such as bathing, while they were more open to using robots for simpler tasks such as reminder management and communication. (e.g., Huang & Huang, 2021; Robillard & Kabacińska, 2020). Users also compared robots with pets (Bates, 2019; Lazar et al., 2016), which were more valued thanks to the reciprocity inherent in caring for them and the relationships they form, as well as with other technologies such as those of tablet computers (Mann et al., 2015) and smart home technologies (Torta et al., 2014), which were typically perceived as inferior and less enjoyable than robots.

Direct experience with robots appeared to lessen ambivalence and promote acceptance. This impact was evident in snapshot studies that enabled interaction with robots (Mehrotra et al., 2016; Shen & Wu, 2016), as well as in longitudinal studies that explored processes of adaptation to robotic technologies. The latter demonstrated that giving robots a function in the older adults' daily routines may lead to greater approval and appreciation (De Graaf et al., 2015), which, in turn, leads to increased intensity of use (Šabanović et al., 2013). If users did not ascribe specific functions to the robot, they gradually lost interest, enjoyed the interaction less (Torta et al., 2014) and eventually returned to their previous routines and habits without the robot (Frennert et al., 2017).

Reports of adaptation processes among people with cognitive impairments were somewhat different. Facing more constraints to independent use of the robots and frequently relying on their caregivers to operate them (Hebesberger et al., 2016), such users demonstrated willingness to interact with the robots that increased over time (Chang et al., 2013). They tended to treat robots as children and exhibited growing emotional attachment to them that was often expressed in various physical gestures, such as petting and hugging (Chang et al., 2013; Kim et al., 2021).

The differences in adaptation among people with varying cognitive functioning suggest that user experience depends on the individual's characteristics (Lindblom et al., 2020). Indeed, many studies in the literature reported *factors affecting user experience*, which could generally be divided into user attributes and robot attributes. User attributes significantly affecting user experience included personal factors such as age, cognitive condition, level of education and computer experience (Morillo-Mendez., 2021; Wu et al., 2016), and interpersonal factors such as perceived amount of social support (Baisch et al., 2017). Some review papers (e.g., Kachouie et al., 2014; Klein et al., 2013) also mentioned the effects of older persons' cultural backgrounds on their attitudes toward robots. Cross-cultural studies of HRI in later life are scarce, however, and the few multinational studies (e.g., Akalin et al., 2021; Jenkins & Draper, 2015; Mehrotra et al. 2016; Torta et al., 2014) have mostly focused on similarities among users rather than differences.

Robot attributes that affect user experience have also been studied extensively, including the robots' appearance, behavior, and functionality. Whereas users expressed a preference for clear distinction between humans and robots in terms of physical appearance (Walden et al., 2015), they tended to favor those who looked more like humans (Khosla et al., 2012), or displayed human-like features and gestures (Caleb-Solly et al., 2014), and focused significantly more on the robot's face and paid less attention to the rest of the body (Oh & Ju, 2020). In terms of behavior, users wanted robots to be social, intelligent, and spontaneous (Frennert et al., 2017; Tulsulkar et al., 2021), although there was some incongruity regarding the robots' playfulness. Hedonic features did increase users' willingness to interact with robots, but serious demeanor added credibility and appreciation (De Graaf et al., 2015). Similarly, users tended to like "young" robots but perceived "older" ones as more competent and safer (Marin Mejia & Lee, 2013). Finally, the robots' perceived functionality seemed to play an important role. This term describes a host of valued robot attributes such as safety, reliability, control, efficiency and satisfaction (Begum et al., 2013; Jaschinski et al., 2021; Padir et al., 2015). Users expected the robots to be useful and adjustable to their needs (Kim et al., 2021; Olatunji et al., 2020; Pripfl et al., 2016; Tsardoulias et al., 2017).

Studies showed that older adults tend to appreciate communication methods that resemble human-human interaction as well as multimodality, namely, multiple interaction possibilities (Fischinger et al., 2016; Haji Gholam Saryazdi, 2021; Siciliano & Khatib, 2018). For example, a study that applied Matilda (a companion-type humanoid robot embodied with a range of multimodal attributes such as voice, music, gestures, movement, and touch panel) in a nursing home setting showed that its multimodality was highly valued by the residents (Khosla et al., 2012).

*Constraints.* The literature also indicated a variety of constraints on robot use. This category comprised explorations of a) antecedent constraints, namely, factors that reduce or limit motivation to use robots, and b) intervening constraints that come between the desire to use robots and the actualization thereof. Among the salient antecedent constraints were uneasiness with the new technology (Erel et al., 2021; Gasteiger et al., 2021; Wu et al., 2014), and perceiving it as no more useful and/or having no added value than existing technologies (Caleb-Solly et al., 2014; Wu et al., 2016; Tonkin, 2021). It appears, however, that the most dominant antecedent constraint is the stigma associated with using a robot in old age. Trying to dissociate themselves from negative stereotypes of old age as a period of frailty and dependency, healthy older adults tended to perceive the prospective robot user as someone older, lonelier and more in need of care (Bradwell et al., 2021; Neven, 2010; Pripfl et al., 2016). Interestingly, however, even people with dementia did not think that they could benefit from using an assistive robot. At most, they could imagine themselves using one down the road if their cognitive condition worsened (Begum et al., 2013; Bradwell et al., 2021; Wu et al, 2016).

Prominent *intervening constraints* described in the literature were affordability and usability. Concern over robot costs was expressed often (Abbott et al., 2019; Koh et al., 2021; Ng et al., 2012; Padir et al., 2015). Community-dwelling adults were doubtful about buying a robot but could imagine renting one for a short period if needed (Fischinger et al., 2016), while senior home residents, who considered robots vis-à-vis human caregivers, thought it would be more reasonable for both financial and functional reasons to hire more staff than to acquire a robot (Compagna & Kohlbacher, 2015).

As much of the reviewed literature tested new devices and applications, usability, i.e., a quality attribute that assesses how easy it is to use a particular product and its user interfaces (Nielsen & Madsen, 2012), was of major interest. Accordingly, various operational difficulties were reported. Some studies, for example, described users' dissatisfaction with the robots' verbal skills, comprehension of instructions and response speed (Begum et al., 2013; Fischinger et al., 2016; Pripfl et al, 2016; Wang et al., 2019). Besides these issues, another usability factor affecting users' satisfaction was human–robot proximity, namely the physical distance between the robot and the human in their interaction (Wang et al., 2019). For example, studies revealed that approach distances preferred by people in human–robot interaction were comparable to those preferred in human–human interaction (e.g., Babel et al., 2021; Sumioka et al., 2021). Shortcomings in robot performance led to frustration (Lin et al., 2022; Pripfl et al, 2016), and some users were annoyed by the conversations that companion robots initiated autonomously (De Graaf et al., 2015). Usability was even more challenging among cognitively impaired individuals, as the robots were often unable to match the interaction abilities and speed of such users (Begum et al., 2013; Striegl et al., 2021).

Scholars also noted various reasons for usability problems, the first being the extensive heterogeneity characterizing the older population (Šabanović et al., 2013; D'Onofrio et al., 2022; McGlynn et al., 2017). Bedaf et al. (2014), for example, tried to identify which daily activities pose the greatest threat to independent living as they become more difficult for the older individual to perform. They concluded that it was often a combination of activities rather than a specific activity, and that the threat was largely specific to the person studied. Hence, a single perfect robotics technology for older adults is unlikely, and designers should develop flexible and customizable solutions (Broadbent et al., 2009; Caleb-Solly et al., 2014; Marchetti et al., 2022). Another problem is the gap between the technology developers and its users, rendering Participatory Design (PD) highly challenging. PD methods aim to develop a socially robust and responsible robot design by building on mutual learning between researchers and participants, and the active participation of older adults and/or their caregivers as "designers". Often, however, the participants' understanding of technology is limited and their expectations from the robots unrealistic (Compagna & Kohlbacher, 2015; Mehrotra et al., 2016; Woods et al., 2021).

*Outcomes*. The literature described a variety of outcomes resulting from HRI in later life, mostly divided between benefits and risks. Overall, the studies reported *positive effects* of HRI on older adults' psychological wellbeing and functioning (e.g., Broekens et al., 2009; D'Onofrio et al., 2019), and provided solid evidence that these effects can indeed be attributed to the HRI. Interacting with robots was experienced as a cognitively stimulating (Khosla et al., 2012; Louie & Nejat, 2020; Neven, 2010, Tsardoulias et al., 2017; Wu et al., 2016) and enjoyable activity (De Graaf et al., 2015; Fischinger et al., 2016; Lazar et al., 2016) and had beneficial effects on users' psychological wellbeing (Henschel et al., 2021), including better and more intensive communication with family and friends (Tsardoulias et al., 2017), elevated mood (Khosla et al., 2012), positive affect (McGlynn et al., 2017) and decreased frustration, stress, and relationship strain (Van Patten et al., 2020; Wang et al., 2017).

In addition, the robots proved efficient as a therapeutic means in long-term care settings. Often using the pet-like robot Paro (a robot designed to mimic movements and sounds of a baby harp seal in response to petting, complete with white fur) in recreation and/or occupational therapy sessions, studies showed that interactions with robots are a powerful projective screen as well as a site for working through personal and social concerns (Turkle et al., 2006). The interactions also had a positive impact on session participants' mood (Lane et al., 2016; Barata, 2019), communication interaction skills, and activity participation (Chiu et al., 2021; Koceska et al., 2019; Šabanović et al., 2013). Therapists felt that the robots are good social mediators in group sessions but considered them even more appropriate for one-on-one interaction (Chang et al., 2013).

Functional benefits primarily included the robots' contribution to older persons' independence and quality of life (Bedaf et al., 2014; Koh et al., 2021; Neven, 2010, Padir et al., 2015; Robinson & Kavanagh, 2021; Smarr et al., 2014; Tsardoulias et al., 2017; Wang et al., 2021; Wu et al., 2021). Furthermore, the robots were found useful in supporting physical exercise and/or rehabilitation thanks to their ability to correct the users' position and movements (Krakovsky et al., 2021; Tsardoulias et al., 2017) and enhance motivation (Avioz-Sarig et al., 2021), group coherence, and mood (Hebesberger et al., 2016). Studies showed that physical exercise sessions led by robots were significantly more effective than those with a virtual coach (Avioz-Sarig et al., 2021; Fasola & Matarić, 2013; Vasco et al., 2019) and even human instructors (Shen & Wu, 2016). Another study, however, revealed no positive influence on exercise behavior (Mann et al., 2015). The literature has also shown that robots even have a positive impact on caregivers, since they can improve their quality of life by reducing burden, decreasing fear, anxiety, and difficulty in challenging tasks, as well as increasing safety and confidence in performing activities that require physical assistance (Abbott et al., 2021; Jenkins & Draper, 2015; Pilotto, 2018; Smith, 2012).

Besides describing the benefits accruing from older persons' use of robots, the literature also addressed its risks and/or negative impacts insofar as both psychological wellbeing and functioning are concerned. Regarding psychological risks, concerns related primarily to robot applications in long-term care settings and dealt with damages such as discomfort or stress that may result from the robot's appearance, motion, embodiment, speech, gaze, and posture (Hussain & Zeadally, 2019; Salvini et al., 2021). One of the conceptual articles argued that robots lack emotions that are integral to the provision of such care; consequently, they cannot provide residents with essential recognition and respect (Sparrow, 2016). They may thus make care receivers feel like "problem carriers" (Beedholm et al., 2016). Other articles discussed ethical ramifications including invasion of privacy and feelings of a loss of control as a result of the presence of cameras and hearing sensors on board the robot (Caine et al., 2012; Kernaghan, 2014). The feeling of being spied on or under surveillance by robots and/or by other people can cause stress and anxiety and, in extreme cases, even manifestations of violence against the robot (Salvini et al., 2021). Furthermore, it was suggested that the robots may create tension between older people and their formal and informal caregivers. For example, robots used for monitoring formal caregivers may weaken the residents' trust in the care they receive, while robots programmed to report non-adherence to treatment may cause humiliation and anger (Jenkins & Draper, 2015).

Concerns regarding older adults' functioning were often associated with issues of safety and reliability (Ng et al., 2012). Some study participants, for example, worried about potential damage or physical harm to themselves or their environment (Beer et al., 2012). Several cases may lead older adults and robots to come into physical contact, for example, during assistive tasks, such as walking support robots (Cifuentes et al., 2014), mobility robots (Leaman & La, 2017), and person-following robots (Olatunji et al., 2020). These interactions can create hazards with different degrees of severity such as accidents or malfunctions (Mansfeld et al., 2018; Rosenstrauch & Krüger, 2017). Furthermore, although one major justification for the incorporation of robots in older people's lives is their potential to support autonomy, it was claimed that the robots may actually threaten autonomy by replacing users in tasks they would be better off performing themselves, rendering seniors even more dependent on robots (Beer et al., 2012; Jenkins & Draper, 2015).

### 2.4. Quality Evaluation of Socially Assistive Robots (SARs)

To facilitate acceptance and assimilation of SARs, realize their potential benefits, and reduce their potential risks, it is essential to understand the factors that promote Quality Evaluation (QE) of SARs and the factors that hinder such positive QE (Andriella et al., 2021). Technology QE deals with the set of a person's emotions, perceptions, and responses created, derived, and shaped as a result of interaction or anticipated interaction with a system, product, device or service (Hartson & Pyla, 2012; Hassenzahl, 2013; Jokela, 2010; Lindblom & Andreasson, 2016). The literature on the subject distinguishes between pragmatic and hedonic aspects of evaluation (e.g., Hassenzahl, 2003; Khalid, 2006; Mlekus et al., 2020; Väänänen-Vainio-Mattila et al., 2008). The pragmatic aspects of QE relate to the functionality, usability, usefulness, and utility of potential tasks that help users achieve their goals effectively and satisfactorily (da Silva et al., 2019; Hartson & Pyla, 2012; Hassenzahl & Tractinsky, 2006; Mlekus et al., 2020). Hedonic aspects refer to the users themselves and reflect the emotional benefits that users experience when interacting with the technology (Atkins, 2008; Bittner et al., 2016; Hartson & Pyla, 2012; Hornbæk & Hertzum, 2017; van de Sand et al., 2020). Attractiveness stems from both the pragmatic and hedonic evaluations of the product and describes the users' overall impression (Santoso et al., 2016).

Positive QEs are necessary to promote acceptance of SARs—a crucial condition for the assimilation process and the realization of the benefits inherent in using robots (e.g., Naneva et al., 2020). Previous studies revealed that negative perceptions of interactions with the robot might lead to negative consequences such as dissatisfaction, reluctance to use a particular robot and robots in general, loss of loyalty, and spreading of negative word-of-mouth, which may suppress the acceptance of future robots (e.g., Merkle, 2019). Thus, it is vital to design and develop SARs in a manner ensuring that interaction with them will be perceived and evaluated by users as not only appropriate, secure, and safe, but also as successful, positive, effective, and pleasurable (van Greunen, 2019). Consequently, it is essential to evaluate both pragmatic and hedonic aspects. Many studies conducted so far have indicated that trust in robots significantly enhances QE (e.g., Vandemeulebroucke et al., 2021), whereas robot-related technophobia was found to substantially decrease QE (e.g., Naneva et al., 2020; van Maris et al., 2020).

### 2.5. Trust in robots

Trust is defined as "the willingness of a party to be vulnerable to the actions of another party based on the expectations that the other will perform a particular action important to the trustor, irrespective of the ability to monitor or to control that other party" (Mayer et al., 1995, p. 712). Parallel to the growing presence of robotic technologies in our everyday environment, trust in robots is a critical factor and plays an important role in HRI research (Langer et al., 2019; Lewis et al., 2018). Trust is an essential factor in building and maintaining effective interaction with robots for an extended period of time (Naneva et al., 2020; Yang et al., 2018). Robotics literature suggests varying levels of trust in robots (Hancock et al., 2011). Appropriately calibrated trust can improve the collaboration between humans and robots (Lee & Liang, 2019; Muir, 1994; Schaefer et al., 2014) and is reached when the extent of trust matches a robot's capabilities (Babel et al., 2021; Lee & See, 2004; Parasuraman & Riley, 1997). Inappropriate levels of trust such as either distrust or overreliance can lead to negative consequences such as neglect and complacency, respectively (Kessler et al., 2017; Lee & See, 2004; Ososky et al., 2013). Such effects threaten the harmony and success of the SAR's assimilation (Hancock et al., 2011; Lee & See, 2004; Parasuraman & Manzey, 2010; Parasuraman & Riley, 1997). To efficiently implement robotic technologies in homes and workplaces, it is thus imperative to consider factors that influence trust in robots (Langer et al., 2019).

Researchers who studied trust in HRI noted that factors that may affect trust in robots generally fall within three identified categories: (a) robot-related, (b) human-related, and (c) environment-related (e.g., Akalin et al., 2022; Hancock et al., 2011;

Hancock et al., 2020; Langer et al., 2019; Schaefer et al., 2016; Lewis et al., 2018). In the review by Hancock et al. (2011), each of these three main categories were further divided into sub-categories. The robot-related category was divided into performance (e.g., reliability, false alarm rate, failure rate) and attributes (e.g., physical appearance, robot personality, and anthropomorphism); the human-related category was divided into abilities (e.g., the user's skills and competency) and demographic characteristics (e.g., age, race, and gender); and the environment-related category referred to team collaboration and task-based elements. Research shows that robot characteristics, especially, performance-based factors, are the main and most substantial impact on perceived trust in robots. Environmental factors moderately influence perceived trust, while human-related factors were not found to significantly influence perceived trust.

Schaefer (2013) indicated strong support for the above meta-analysis, but she also expanded this examination to new pathways of influence such as tenure, i.e., the length of time the human and robot work together, which was found to have a significant effect on developing trust in robots. Indeed, in their study, van Maris et al. (2017) found a significant increase in trust over six weeks of experimentation. In another meta-analysis, Schaefer et al. (2016) deepened the investigation of human-related factors affecting the development of trust by including in this category cognitive and emotional dimensions as well as human traits and states. They found a significant impact of this extended category on trust development. Moreover, the type of the robot was found to be important for trust (Schaefer, 2013; Lewis et al., 2018), especially where users and robots communicate with each other naturally, similarly to human-human interaction (Langer et al., 2019; Lewis et al., 2018). Robots that move naturalistically (Castro-González et al., 2016), with an anthropomorphic appearance (Kiesler et al., 2008) and empathic communication (Tapus & Mataric, 2007) were found to be more likely to stimulate the development of higher levels of trust (Lewis et al., 2018).

Several studies have highlighted the importance of previous experience with robots, which leads to more positive attitudes towards robots, in general, and trust in robots, in particular (Sanders et al., 2018). Two other factors found to be related to trust in robots are culture and personalization. Cultural factors may explain how humans develop trust in robots (Lewis et al., 2018; Li et al., 2010). The number of toy robots, games, TV shows, manga, and humanoid robots give Japanese culture the leading role in robotic culture and development (Bartneck et al., 2005; Conti, 2016). However, Eastern users may feel afraid or feel they have no control over the rapid progress of robot technology and its recent applications (Bartneck et al., 2007; Kaplan, 2004). Therefore, individuals from Eastern cultures were less likely to trust robots in comparison to people from Western cultures and Latin America (Chien et al., 2020). A robot's ability to personalize and adapt to user preferences and feedback is another key factor in developing trust in HRI (Langer et al., 2019).

An updated meta-analysis that validated the original categories of trust (robot-related, human-related, and environment-related) offered an extension of the initial results rather than contradictions (Hancock et al., 2020). The reported findings confirmed that factors relating to the robot, particularly robot attributes and performance, strongly influence perceived trust, compared to human-related factors. However, factors such as a user's personality, culture, comfort with robots and expertise, still constitute a significant influence on human trust in robots (Hancock et al., 2020). Furthermore, the authors identified two new human-related factors – performance expectancy and satisfaction with the interaction with robots, which have proven to be significant predictors of trust in HRI (Hancock et al., 2020). Finally, in a recent study, a robot that showed more positive emotion and apologized for its mistakes, had a positive effect on its users' trust, and their intentions to use it again (Cameron et al., 2021). This finding is consistent with the tendency of humans to rely on robots that have shown more positive emotions and attitudes (Mathur & Reichling, 2016; Oksanen et al., 2020).

### 2.5.1. Trust in robots among older adults

For older adults aged 65 and over, trust is a particularly essential component of any relationship they are involved in (Katz & Edelstein, 2018), including with robots (Schwaninger, 2020), and especially in their homes (e.g., Chen et al., 2013; Roy et al., 2000; Wada et al., 2004). In most cases, the use of autonomous SARs in this intimate space involves access to private and sensitive data (Schwaninger, 2020), which strengthens the importance of trust in creating a successful HRI. In fact, researchers have found that SARs must inspire trust in older adults, while being required to work securely and respect the privacy of users without reducing their effectiveness (De Graaf et al., 2015; Pedersen et al., 2018). In addition, studies that examined trust in robots indicated that safety concerns, particularly among older adults, lead to mistrust in robots (Scopelliti et al., 2005). Therefore, to earn and even increase older adults' trust, it is essential to alleviate their concerns (Langer et al., 2019) and enhance their sense of control over the robot (Vandemeulebroucke et al., 2021). Moreover, SARs have a responsibility to provide them a sense of safety and security (De Graaf et al., 2015), which is a key factor in building trust in HRI and influencing the older adults' intention to use robots as well as the whole technology adoption process (Allouch et al., 2009; Yu et al., 2005). Hence, trust in HRI in later life is crucial, and should be given paramount importance (Looijet et al., 2010), since without it, the initial decision to use robots would not emerge (Lazanyi & Maraczi, 2017).

Other prominent factors in creating and maintaining trust in robots among older adults are the consistency and accuracy of the robot's operations. Studies showed that when a robot demonstrated inconsistency in its behavior or provided wrong information, users' trust in the robot decreased (De Graaf et al., 2015). Generally, the older population prefers humans to take care of their personal-life needs (Stuck & Rogers, 2018; Vandemeulebroucke & Gastmans, 2021). Accordingly, to gain their trust, they expect the interaction with the social robot to be as similar as possible to the same behavioral characteristics of human-human interaction, meaning, speak politely, listen attentively, and conduct social conversations and dialogues (Cassell & Bickmore, 2000; Fischinger et al., 2016; Looijet et al., 2010).

Stuck and Rogers (2018) conducted a mixed methods study (questionnaires and a semi-structured interview) to explore the factors that reinforce and/or encourage older adults' sense of trust across four home-care tasks: Medication assistance, transferring, bathing and household tasks. Their findings showed that older adults noted three major dimensions that promoted trust: Professional skills, personal traits, and communication. Each of these main dimensions consists of sub-factors, some of which have been identified in previous HRI literature (De Graaf et al., 2015; Langer et al., 2019; Madhavan & Wiegmann, 2007; Vandemeulebroucke et al., 2021) and confirmed in this study as supporting trust such as reliability, safety, and precision of the robot, and some of which are unique factors arising from this study such as the companionability, benevolence, and material of the robot. Particularly, in tasks that involved human-robot touch (i.e., bathing and transfer), there are specific considerations to consider such as safety, gentleness, the texture and material of the robot and understanding the sensitivities of the older adult, while a key factor for medication assistance was ensuring that medications are provided at the appropriate time (Stuck & Rogers, 2018).

The functionality of the robot should meet the user's needs (David et al., 2022; Tsardoulias et al., 2017; Wiczorek et al., 2020). This point is likely to be more challenging to implement in robots designed to interact with older adults since they typically have more special needs that are often combined with restricted capabilities (Wiczorek et al., 2020; Zafrani & Nimrod, 2019). Hence, to strengthen the sense of trust among older adults, robots must be able to adjust to their current health, needs and unique desires (Stuck & Rogers, 2018).

Previous studies highlighted that creating a trusting relationship is necessary for older adults especially in demanding cognitive tasks such as managing finances (Pak et al., 2017), and in tasks that some older people have difficulty performing owing to age-related cognitive impairments such as transportation (Donmez et al., 2006; Pak et al., 2017). The authors hypothesized and argued that in these tasks, older adults are more willing to rely on automation than in the past because building trust allows them, in effect, to maintain their independence (Dellinger et al., 2001; Donmez et al., 2006; Pak et al., 2017). Finally, other aspects supporting trust in older adults include robot characteristics such as its feedback quality (Seong & Bisantz, 2008) and reliability (Madhavan & Wiegmann, 2007).

### 2.6. Technophobia

Technophobia is an umbrella term for describing fear and/or discomfort in using modern technology and concerns regarding technology's effects on society (Osiceanua, 2015). Technophobia is a prominent use prohibitor found in Information and Communication Technology (ICT) studies (Nimrod, 2018; 2021; Rosen & Weil, 1990). Despite all the benefits inherent in using SARs, some people are afraid and even avoid using them due to a belief that robots have the potential to cause both physical and emotional damage in everyday situations (e.g., Arnold & Scheutz, 2017; Haring et al., 2019; Malle & Scheutz, 2014; Scheutz, 2016). This negative attitude toward robots is a mental or psychological phenomenon that blocks people from interacting with robots, thus preventing them from being widely accepted by the masses (Nomura et al., 2004, 2006; Tussyadiah et al., 2020).

Recently, researchers have begun to explore this issue by attempting to identify the concerns regarding the use of robots (e.g., Calvert, 2017; Haring et al., 2019; Syrdal et al., 2007; Sharkey & Sharkey, 2012; Wu et al., 2012). Most of the studies have not specifically used the term "technophobia", however they have dealt with issues such as negative attitudes towards robots (Louie et al., 2014), concerns (Calo, 2011), fears (Cobaugh & Thompson, 2020), and anxiety (Sundar et al., 2016), which are derivatives, modes of expression, and aspects that represent technophobia (Di Giacomo et al., 2019; Park et al., 2010). Such studies focused, among other things, on work environments and indicated that technophobia includes a fear of being replaced by robots that is prevalent among employees such as teachers, hotel workers, older adults' caregivers, and pharmacists (e.g., Calvert, 2017; Cobaugh & Thompson, 2020; Frey & Osborne, 2017; Goudzwaard et al., 2019; Hu, 2019; Lin, 2018; Liu et al., 2017; Pino et al., 2015; Semuels, 2011; Serholt et al., 2017; Vlachos et al., 2020; Wu et al., 2012).

Another aspect related to technophobia is privacy-related ethical concerns that may arise from the use of SARs. The literature distinguishes between two types of privacy pertaining to technology use: physical privacy and informational privacy (Bygrave, 2002; Calo, 2011; Lutz & Tamó-Larrieux, 2020; Smith et al., 2011). Since SARs are most often located in homes, physical privacy may be violated by the ability of robots to enter private physical spaces such as bedrooms and bathrooms (Calo, 2011), where they might be exposed to sensitive, embarrassing, and complicated situations (Krupp et al., 2017). In these cases, the physical privacy of the individual is subject to surveillance and robots are constantly able to monitor and record of their users. When it comes to vulnerable populations such as older adults, people with disabilities and children, this is even worse, due to their limited knowledge and awareness of the subject (Lutz & Tamó-Larrieux, 2020).

Informational privacy concerns revolve around humans' ability to understand how the information shared with the robot is processed (Calo, 2011; Lutz & Tamó-Larrieux, 2020; Syrdal et al., 2007). To fulfill the users' desires, robots need to accumulate a large amount of personal information about their users (Kernaghan, 2014; Syrdal et al., 2007; Sharkey & Sharkey, 2012). Furthermore, humans tend to constantly anthropomorphize SARs (Fong et al., 2003) and as a result, they tend to consider the robots as a kind of friend, and subconsciously entrust them with private information (Lutz & Tamó-Larrieux, 2020). Users may feel threatened and uncomfortable with the fact that the robot might store sensitive personal information about them that could be transferred to a third party (Kernaghan, 2014; Lehmann et al., 2020; Lutz & Tamó-Larrieux, 2020; Sharkey & Sharkey, 2012; Syrdal et al., 2007). Moreover, as SARs are equipped with advanced processors, cameras, and sensors, they may be exploited by hackers, who infiltrate the robot's systems to spy on users' private data without their knowledge (Calo, 2011; Krupp et al., 2017; Lutz & Tamó-Larrieux, 2020).

Another ethical concern is that robotic systems such as medical robots may manage their decision-making process unfairly and as a result discriminate between individuals and groups with different demographic characteristics (Howard & Borenstein, 2018). Several studies even indicated that robots are able to deceive humans and mislead them as a result of unexpected behavior (Shim & Arkin, 2013; Terada & Ito, 2010). This is because humans perceive and treat robots as designed and algorithm-based objects, with predefined responses, and when their behavior is unpredictable, this is interpreted by users as an error and as deception (Terada & Ito, 2010), making them feel not in control (Gjersoe & Wortham, 2019). Anthropomorphizing robots, that is, attributing emotions, personalities, passions, and goals to robots, could lead to many social, psychological, and cognitive risks for the people who use them (Aicardi et al., 2020), since this process may create an illusion of social bonding between a human and a robot (Langman et al., 2021).

### 2.6.1. Technophobia among older adults

Technophobia, concerns, and negative sentiments towards robots constitute a particular barrier to the assimilation of robotic technologies among older people (e.g., Coeckelbergh et al., 2016; Frennert & Östlund, 2014; Fulmer et al., 2009; Keizer et al., 2019; Kernaghan, 2014; Khosla et al., 2021; Sharkey & Sharkey, 2012; Sparrow, 2016; Vandemeulebroucke & Gastmans, 2021). Specifically, emotional attachment and emotional deception are defined as ethical concerns in HRI in later life (e.g., Fulmer et al., 2009; Sharkey & Sharkey, 2012; Sullins, 2012; van Maris et al., 2020). Developing an emotional attachment to robots is possible since humans can become attached to objects (Keefer et al., 2012). Moreover, as SARs become more common, users are more likely to connect with them (van Maris et al., 2020). However, as older adults become more attached to the social robot, taking it away may cause emotional distress (Sharkey & Sharkey, 2010; Coeckelbergh et al., 2016).

Emotive behavior is a required feature of beneficial companion robots (Breazeal & Scassellati, 1999), whose goal is to improve communication with humans (Kirby et al., 2009). Yet, this can be a risk as older adults may believe that the social robot really experiences emotions (van Maris et al., 2020). Without exercising critical judgment (Fulmer et al., 2009), seniors may develop unrealistic expectations that can never be met (Compagna & Kohlbacher, 2015; Mehrotra et al., 2016; van Maris et al., 2020). This is actually emotional deception: Since the social emotional behavior of the robot is inconsistent with its actual abilities (Sharkey & Sharkey, 2011); it does not really experience emotions (van Maris et al., 2020) and in fact provides incorrect information about its internal emotional state (Fulmer et al., 2009). Moreover, as stated in conceptual articles, SARs are devoid of real emotions, which are an integral part of successful and effective caring for older adults. Therefore, they cannot provide them with the respect, recognition, reciprocity, attention, and human contact they need for their sense of well-being (Coeckelbergh, 2010; Decker, 2008; Kernaghan, 2014; Lehmann et al., 2020; Sharkey & Sharkey, 2012; Sparrow, 2016; Sparrow & Sparrow, 2006).

Just like among other age groups, the invasion of elders' privacy—both physical (e.g., robotic monitoring of dressing or bathing) and informational (e.g., data protection and security)—are major ethical concerns. In addition, robots may make older adults feel the loss of personal liberty and control, for example, when caregivers use robotic assistance insensitively to move, lift, wash or feed them, and make them feel like objects (Callén et al., 2009; Kernaghan, 2014; Khosla et al., 2021; Sharkey & Sharkey, 2012; Vandemeulebroucke & Gastmans, 2021; van Maris et al., 2020). In this context, Sharkey and Sharkey (2012) pointed out six main ethical concerns. In addition to the five concerns mentioned above (loss of control, privacy and personal liberty, deception and infantilization, and possible decrease in the amount of human contact), they added a sixth dimension dealing with responsibility, meaning that if robots are placed under the control of older adults, and something goes wrong and gets out of control, who bears the blame? Finally, what happens, from an ethical point of view, if the social robot has more than one older user, how should it prioritize between them, so as not to hurt the feelings (e.g., resentment, frustration, neglect) of any of them (Frennert & Östlund, 2014).

Apart from the ethical concerns, there are several other concerns related to the role of robots in older adults' lives, factors influencing older adults' acceptance of SARs and robots' appearance and aesthetics (e.g., Frennert & Östlund, 2014; Gassmann & Reepmeyer, 2008; Keizer et al., 2019; Pripfl et al., 2016). Seemingly, one fundamental justification for integrating SARs in older persons' homes is their potential to promote autonomy. However, they can also create dependence and threaten autonomy by replacing the older adults in household tasks they can perform independently (Beer et al., 2012; Jenkins & Draper, 2015).

In addition, a growing number of SARs developed for older adults are toy-like systems that resemble robots designed for children (Alhaddad et al., 2018; Keizer et al., 2019; Šabanović et al., 2013; Winfield, 2012). Older adults may be apprehensive about using these robots due to their unwillingness to be defined in the same way as children (Frennert & Östlund, 2014; Gassmann & Reepmeyer, 2008). Furthermore, previous literature revealed that even older adults who enjoyed spending time and even benefited from a semi-robotic toy, would not buy it for their home due to their concerns about negative attitudes of their social environment and their belief that "toys are for kids" (e.g., Frennert & Östlund, 2014; Kidd et al., 2006). Similarly, another dominant concern stems from the stigma associated with using a social robot in old age. Older adults who are physically and cognitively intact often perceive the potential robot user as someone older, fragile, and lonely who needs nursing care. The concern of falling into these stereotypes of old age restrains healthy older adults from accepting and interacting with SARs (Pripfl et al., 2016).

### 2.7. The effects of trust and fear on individuals' QEs of technology

To achieve successful HRI and to realize the benefits inherent in using SARs, positive experiences and evaluations are necessary (Alenljung et al., 2019). Indeed, previous studies revealed that if the interaction with the robot leads to a negative experience and a negative evaluation from the user, this can lead to negative consequences in terms of dissatisfaction, reluctance to use this particular robot and/or robots in general, loss of loyalty, and spreading a bad reputation and negative word-of-mouth, which in turn may suppress the acceptance of future robots (e.g., Alenljung et al., 2019; Carlotta et al., 2018; De Graaf & Allouch, 2013; Lindblom & Andreasson, 2016; Merkle, 2019). Thus, it is becoming increasingly vital to design and develop SARs that ensure that the interaction with them will be perceived and evaluated by users as not only appropriate, secure, and safe, but also successful, positive, effective, and pleasurable (Hartson & Pyla, 2012; Lindblom & Andreasson, 2016; van Greunen, 2019). To achieve this goal, it is critical to understand the factors that promote and inhibit successful HRI and positive evaluations.

Trust has been considered a critical factor in explaining why people accept modern technologies such as robots (Lee et al., 2018; Man et al., 2020). It plays a critical role in the relationship between the parties, since the principle is: "no trust, no use" (Man et al., 2020; Schaefer et al., 2016). In robotic technology, human trust is a highly important component and a necessary condition for seamless adaptation of technology and creating a successful HRI, as it positively influences the perceived quality, usefulness, and evaluation among users (e.g., Freedy et al., 2007; Kellmeyer et al., 2018; Lewis et al., 2018; Martelaro et al., 2016; Naneva et al., 2020; Salem et al., 2015; Torta et al. 2014; Xu et al., 2018). Trust also impacts the willingness of people to accept information and help from SARs (Freedy et al., 2007; Hancock et al., 2011) and their desire to use them for certain purposes (Naneva et al., 2020; Salem et al., 2015). As noted by Abbass et al. (2016), if people trust the technology and evaluate it positively, they are likely to delegate tasks to it that will help them make decisions in complex situations. Moreover, if the trust relationship is positively reinforced, task performance improves, and subsequently evaluations become more positive.

Additionally, trust has been considered an essential factor influencing users' adoption intentions (Huijts et al., 2012), their responses, and their evaluations regarding modern technologies such as robots (Karlin, 2012), especially in populations where knowledge of technology is limited (Chen et al., 2017; Poortinga et al., 2011) such as among older adults (Fischl et al., 2017; Goodwin et al., 2013; Larsson et al., 2013). Insufficient knowledge is a major barrier in the process of adopting technology (Shahrestani, 2018; Souders & Charness, 2016) and may lead older adults to feel unconfident and anxious about use modern technologies (Ciesla, 2020; Steelman & Wallace, 2017) for fear of ruining the device/product or causing unwanted results (Cajita et al., 2018; Kumar et al., 2013). Therefore, for older adults, trust is an even more crucial factor as it has a significant effect on perceived risk and uncertainty (Choi & Ji, 2015; Huff et al., 2019), sense of control (De Graaf et al., 2015; Schofield & Joinson, 2008) and intention to use (Maree et al., 2019). Hence, trust significantly increases the likelihood that older people will evaluate the technology more positively and even adopt it (Allouch et al., 2009; Heerink et al., 2010; Lecheva, 2017; Yu et al., 2005).

The phobia surrounding technology adoption also has a substantial impact on users' behavior and subsequent QE of a given interaction with modern technology in general, and with robots in particular (e.g., Castro-González, et al., 2016; Cornelius & Leidner, 2021; MacDorman et al., 2009; Syrdal et al., 2009; Szczuka et al., 2019). Previous studies found that concerns and negative attitudes towards robotic technologies are associated with more negative QEs of robot behavior (Syrdal et al., 2009; Tussyadiah, et al., 2020) and lead to avoidance among potential users (Nomura et al., 2004; Nomura et al., 2006). Fear of using modern technologies is a psychological or mental state (Nomura, et al., 2006; 2008; 2012; Sakamoto, et al., 1998) that prevents, reduces, or restricts people from interacting with robots, leading to a stable and enduring predisposition to evaluate robots negatively, and as a result, prevents them from being accepted by the masses (Nomura et al., 2004; 2006; Tussyadiah, et al., 2020).

In the older adult population, these concerns play a more significant role because older adults tend to be more fearful of using modern technologies than younger people (Backonja, et al., 2018; Scopelliti, et al., 2005; Weiss, et al., 2014). In particular, the literature on robotics in later life indicated that older adults have underlying concerns, barriers, and reluctance to use robots (e.g., Peek, et al., 2016; Riek, 2017; Zagler, et al., 2008), which have been shown to adversely affect the initial QE and acceptance of the robot (Beer et al., 2012; Frennert & Östlund, 2014; Kernaghan, 2014; Pripfl et al., 2016). Moreover, there is a mutual fertilization between their concerns and negative attitudes towards robots (e.g., Meissner et al. 2020; Niemelä et al. 2019; Nomura et al., 2004; Syrdal et al. 2007). This mechanism causes older adults to perceive and evaluate the whole process of adopting robotic technologies as tedious, undesirable, unknown and risky. Such perceptions may prevent them from interacting with robots in the first place (Chien et al., 2020; Mehrotra et al., 2016; Pripfl et al., 2016; Wu et al., 2014; Wu et al., 2016), and even constitute a significant obstacle to the integration of robots into seniors' daily lives (e.g., Begum et al., 2013; Caleb-Solly et al., 2014; Neven, 2010; Pripfl et al., 2016; Wu et al., 2014; Wu et al., 2016).

# 3. Methodology

## 3.1. Overview

Although studies have shown that trust and aspects of technophobia affect the QE of robots including SARs, both among the general population and among the older segment, no study has simultaneously examined the impact of these two key factors. Moreover, previous research has two significant weaknesses: (1) studies focused on only one aspect of the interaction between robots and older adults, (i.e., examining uses, constraints and outcomes separately); and (2) they were limited in the study duration, and thus mainly focused on acceptance and not assimilation. Accordingly, no study thus far has explored how the QE of SARs among older adults is shaped both with regard to anticipated interaction and by actual interaction.

This study aimed to bridge the gaps in the existing literature that explores what affects SARs' quality evaluation (QE) among older adults. The research was carried out in two parts: (a) an online survey that simultaneously explored the effects of trust and technophobia on SAR's QE as shaped by anticipated interaction; and (b) an assimilation study examining how the QE is shaped following actual interaction with the SAR by a simultaneous exploration of the SAR's uses, constraints and outcomes for older adults in real-life conditions over a long period.

The survey was conducted in order to obtain an initial empirical view of the phenomenon under investigation, and to deepen the understanding regarding the characteristics of the target population. The survey, conducted by quantitative methods, emphasized the quantification of information and analysis of causal relationships between certain variables (Babbie & Mouton, 2001; Denzin, 2003). The assimilation study applied qualitative methods, providing a behind-the-scenes glimpse of the phenomenon under investigation, and made it possible to answer not only "what happened" but also "why", and "how" it happened. Moreover, study in real-life conditions over a long period highlighted processes and meanings that are not amenable to examination in terms of quantity, amount, intensity or frequency (Garcia & Quek, 1997; Mobar & Sharma, 2011). Hence, an online survey and assimilation study, i.e., a mixed-method approach, yielded more valid and reliable results, allowed for triangulation and complementarity, and portrayed a broader and more accurate picture of SARs' QE by older adults, both through anticipated interaction and actual interaction. The SAR used in this study was a personal robotic trainer (Krakovsky, 2022) developed to enable older adults to exercise independently at home according to their schedule and physical abilities.

### 3.2. System description

The personal trainer robot developed in our lab (Krakovsky, 2022, Figure 1) was named "Gymmy" to elicit the associations of the word "gym". The robot serves as a robotic physical training coach (https://www.youtube.com/watch?v=zQ4T1NhS25Q) that demonstrates a series of physical exercises (Krakovsky et al., 2021). In addition, during the physical training, cognitive training sessions such as memory and thinking exercises, are randomly presented to the users. Furthermore, the system offers users relaxation exercises to release stress and relieve pressure, according to Jacobson's relaxation technique (Jacobson, 1938). The system includes a humanoid mechanical-looking robot (Poppy Torso) and a computer system (NUC mini-PC) to demonstrate the exercises, and an RGB-D (depth) camera to monitor the user's performance (Realsense D435). The users perform the exercise with the robot, and the camera monitors their movements and, if needed, corrects the execution. Gymmy's head is a touch screen, which served the dual purpose of delivering cognitive training exercises and providing visual feedback to users. In addition, Gymmy was equipped with speakers to offer instructions and verbal feedback. The system's algorithm was developed as parallel programming of the robot, camera, audio, and screen (for additional information, see Krakovsky et al., 2021; Krakovsky, 2022).

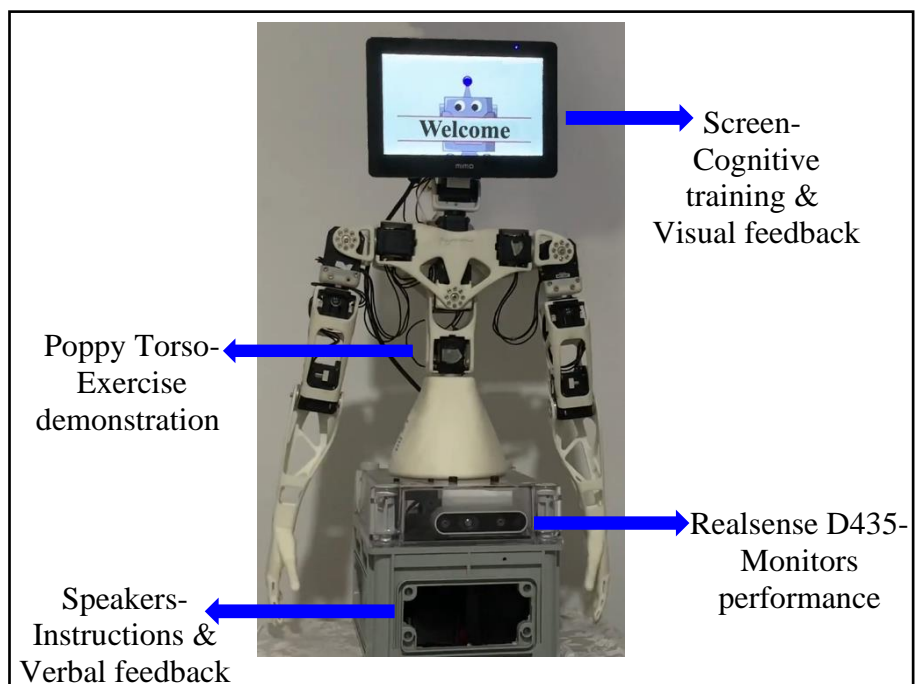

**Figure 1.** Gymmy – Personal Training Robot.

### 3.2.1. Physical Exercises

Gymmy's physical training focus was on exercises for the upper body, which matched the functionality of the Poppy robot's torso version. These exercises improve muscle strength and help older adults maintain their independence and perform daily activities such as lifting objects (Vogel et al., 2009). A total of 14 physical training exercises were developed (Avioz-Sarig, 2019; Krakovsky et al., 2021; Figure 2) according to the recommendations of the National Institute on Aging (NIH; https://go4life.nia.nih.gov/exercise-type/strength/ retrieved July 2019).

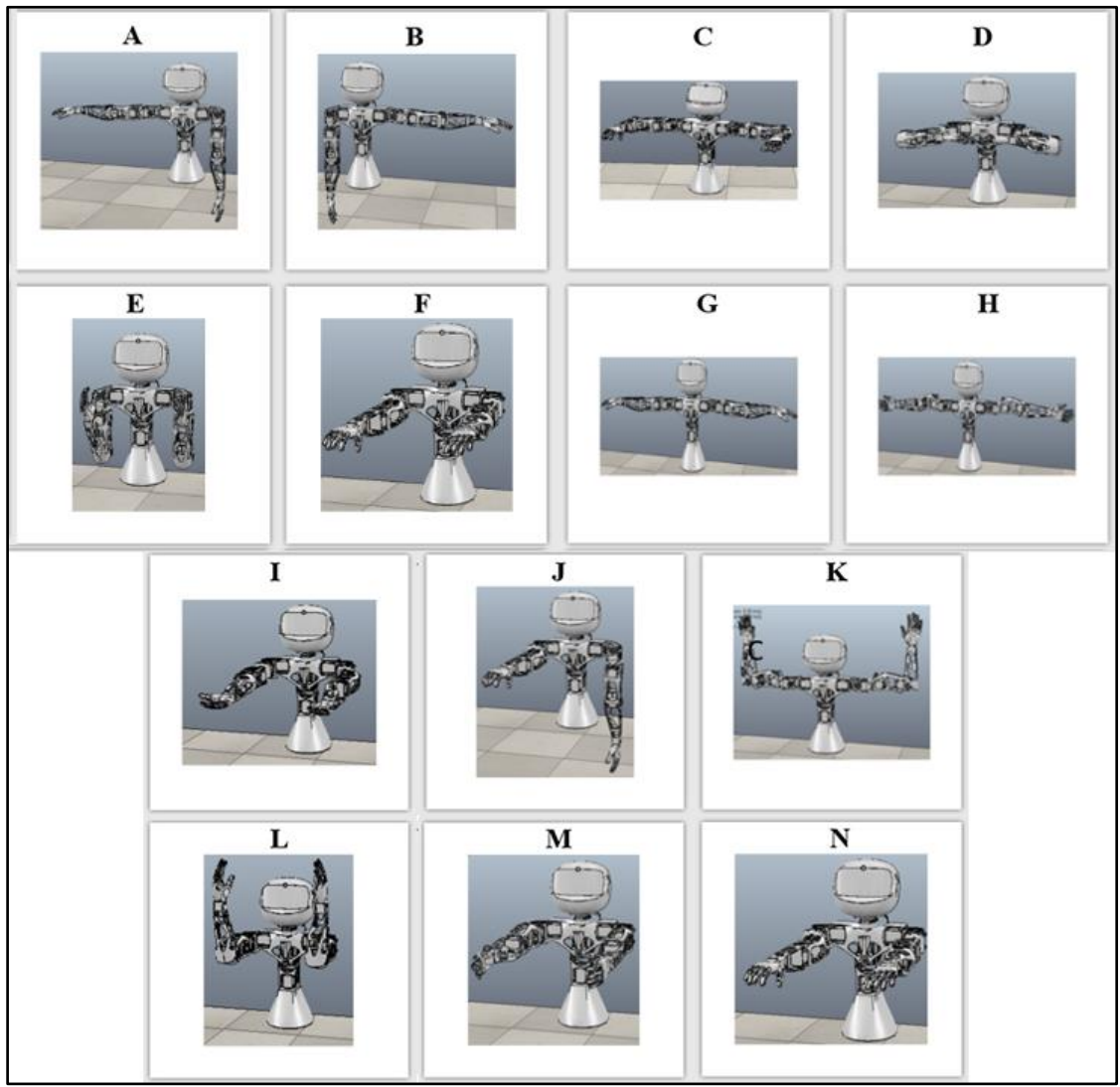

**Figure 2.** A&B- raise arms horizontally separately, C-raise arms and bend elbows forward 90, D-raise arms and bend elbows, E-bend elbows, F- raise arms forward static G-raise arms horizon-tally, H-raise arms horizontally and turn hands, I-raise arms forward and turn hands, J-raise arms forward separately, K-raise arms 90 and up, L-open and close arms 90, M- raise arms forward and to sides, , N- raise arms forward.

### 3.2.2. Cognitive Exercises

Gymmy's cognitive training was designed to address different aspects of memory, processing speed and concentration, which are crucial for older adults' ability to live independently (Arora, 2021; Eggenberger et al., 2015). Three cognitive games were randomly integrated during the physical training sessions. These games were chosen based on the literature (e.g., Ezzati et al., 2016; Nacke et al., 2009) as detailed below. Each game started with instructions and then using the touch screen, users confirmed that they were ready to start the game.

#### 3.2.2.1. Game 1: Working Memory

In this game, users were required to remember the sequence in which words were highlighted on the screen (based on Eggenberger et al., 2015). In the first part, all the words are visible, but each one was highlighted in yellow for a few seconds in a particular order. Users were asked to remember the order and then, in the next part, they were asked to mark the words in the order they were originally presented. The game screens are presented in Figure 3.

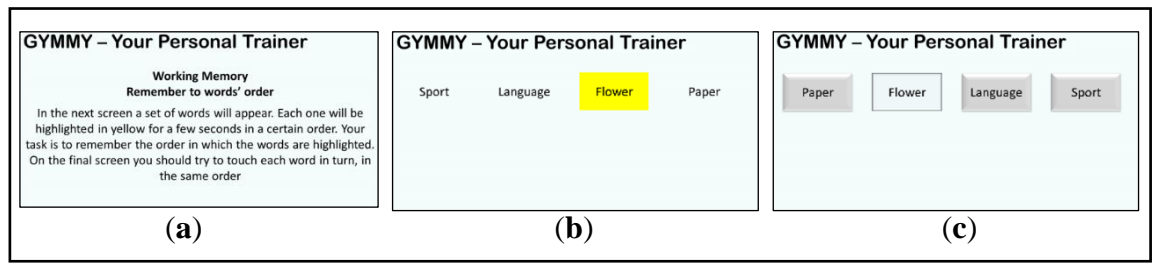

**Figure 3.** Cognitive game 1: (a) instructions screen, (b) words highlighted in a specific order and (c) the user attempts to choose the words in the same order.

#### 3.2.2.2. Game 2: Spatial Memory

The aim of this game is to remember a random spatial pattern that appears on the screen (inspired by Ezzati et al., 2016). The pattern is defined by a subset of highlighted squares within a $5 \times 5$ matrix of squares. The pattern is highlighted for a few seconds, and then, users were asked to recreate it on a blank $5 \times 5$ matrix. The game screens are presented in Figure 4.

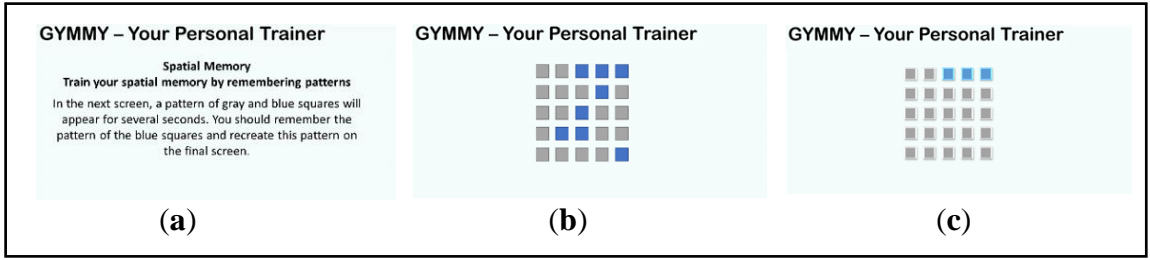

**Figure 4.** Cognitive game 2: (a) instructions screen, (b) a random pattern appears on screen and (c) the user attempts to recreate the pattern.

#### 3.2.2.3. Game 3: Mathematical Skills

The goal of this game is to solve simple mathematical equations (inspired by Nacke et al., 2009). A random mathematical equation appears on screen with four solution choices, and users are required to choose the correct solution. The game screens are presented in Figure 5.

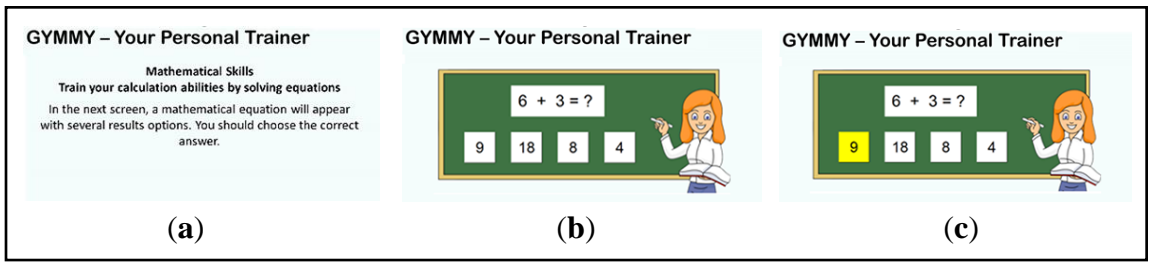

**Figure 5.** Cognitive game 3: (a) instructions screen, (b) the equation with solution choices and (c) the user attempts to select the correct solution.

### 3.2.3. Relaxation exercises

Gymmy's relaxation exercises were provided to release stress and relieve pressure, according to Jacobson's relaxation technique (Jacobson, 1938). This tool is essential for older adults' well-being (Rudnik et al., 2021), and allowed them to perform relaxation exercises for three muscle systems: arms, neck, and face (Figure 6).

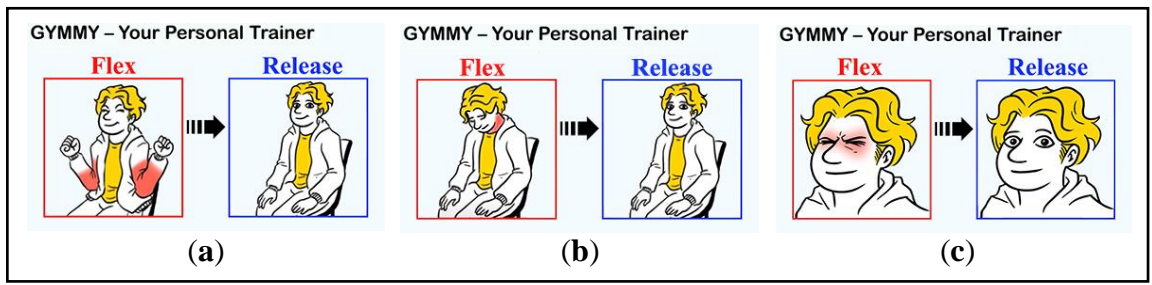

**Figure 6.** (a) Arm relaxation, (b) Neck relaxation and (c) Facial relaxation.

### 3.3. The online survey

This part of the study sought to explore the coexistence and possible relative effects of trust and technophobia on anticipated QE (pragmatic and hedonic evaluations and attractiveness) of SARs. Using Gymmy as a test case, it aimed at answering the following research questions and the following model (Figure 7):

1. How do older adults evaluate the experience of using Gymmy?
2. Do users' background variables (demographic, sociodemographic, health) associate with the QE of Gymmy among older adults, and if so, how?
3. Does trust in robots associate with QE of Gymmy among older adults, and if so, how?
4. Does robot-related technophobia associate with QE of Gymmy among older adults, and if so, how?
5. Do users' physical exercise characteristics (exercise patterns, motivation to exercise, perceived fitness, exercise limitations) associate with the QE of Gymmy among older adults, and if so, how?
6. Do trust in robots, robot-related technophobia and exercise characteristics mediate the relationship between the background variables (demographic, sociodemographic, health) and QE of Gymmy among older adults, and if so, how?
7. To what extent and how is the combination of trust in robots, robot-related technophobia and exercise characteristics associated with QE of Gymmy among older adults?
8. Do the pragmatic and hedonic quality evaluations mediate the relationship between trust in robots, robot-related technophobia, exercise variables and the attractiveness of the SAR?

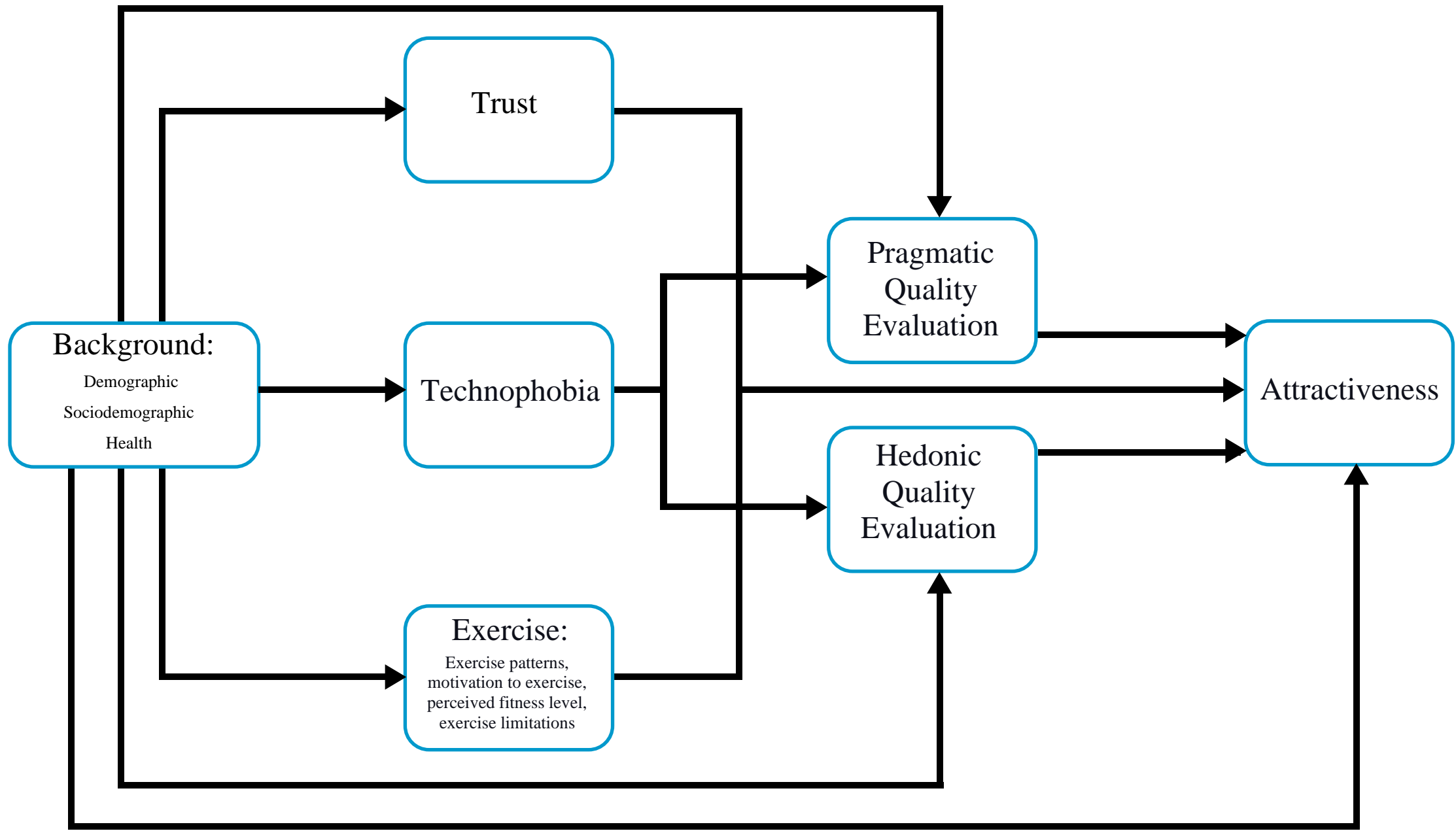

**Figure 7.** The online survey model

### 3.4. The assimilation study

This part of the study aimed to explore the process of SARs' assimilation and the factors effecting post-use QE by older adults. For that purpose, older adults were given the opportunity to use Gymmy in real-life conditions (i.e., in their homes) over a long period (six weeks). Simultaneous exploration of uses, constraints, and outcomes (including both positive and negative effects), rather than focusing on one or two of these issues, helped explain how they correlate with one another and provided a broader and more accurate picture of users' experiences. Moreover, extended simultaneous exploration explained how the HRI changed according to users' experience, to what extent the interaction was integrated in their daily lives, what factors affected frequency of use and the benefits thus accrued, and what constrained beneficial use and/or led to decreased frequency or even cessation of use.

The longitudinal study was necessary to reduce the robot's novelty and to firmly establish its functioning in more naturalistic situations where the participants were alone with the robot and communicated with it as freely as possible (Yamazaki et al., 2014). Assimilation processes helped users make sense of unfamiliar situations and provided the underlying mechanisms of dynamic change in understanding and interacting with SARs (Melson et al., 2006). Accordingly, this part was designed to answer the following research questions:

1. What are the uses, constraints and outcomes that older adults experience while assimilating a SAR into their lives?
2. Do the uses, constraints and outcomes change during the assimilation period? If so, how?
3. How do older adults' experiences with a SAR over a long period and in real life conditions affect their QE of that SAR in particular, and of SARs in general?

# 4. The online survey

## 4.1. Methods

### 4.1.1. Participants and sample description

The study applied an online survey with 384 respondents aged 65 and over. The sole criteria for participation was age, and participation was anonymous. Participants were recruited through mailing lists of various retirees' associations, aging-related websites, and academic centers for aging studies. Overall, we sent the request to participate to 1,134 people, of whom 889 entered the link to the survey, and 384 completed it. Participants' age ranged between 65 and 85 years with a mean age of 71.73 years (SD = 4.79); 59.1% were women, 62.5% were married; 96.3% had children (mean= 2.73, SD= 1.19). The mean number of years of education was 15.37 (SD = 2.65). Forty-five percent reported having a above average income and 27.6% below average; 76.3% were retirees, and 18.8% still worked at least to some extent. The majority (97.8%) were community-dwelling individuals. Seventy-two percent shared their home with at least one other person; 55% lived in cities or on the outskirts of a city, 22.9% in medium-sized or small towns, and the remainder in rural areas. Sixty-four percent were born in Israel, and 75% described themselves as secular. Forty-nine percent reported high satisfaction (i.e., eight or higher) with their physical health (Mean=7.12, SD=1.963), and 70.8% reported high satisfaction (i.e., eight or higher) with their cognitive function (Mean=8.14, SD=1.466). The sample's full sociodemographic characteristics are presented in Tables 1 and 2.

**Table 1.** *Sociodemographic characteristics of the sample (for continuous variables; N= 384).*

| Variable | Range | Mean | Std. Deviation |
|---|---|---|---|
| Age | 65-85 | 71.73 | 4.80 |
| Number of children | 0-10 | 2.73 | 1.19 |
| Number of years of education | 8-20 | 15.37 | 2.65 |
| Self-rated health | 1-10 | 7.12 | 1.96 |
| Self-rated cognitive function | 1-10 | 8.14 | 1.47 |

**Table 2.** *Sociodemographic characteristics of the sample (for ordinal and nominal variables; N= 384).*

| Variable | Frequency | Percent |
|---|---|---|
| **Gender** | | |
| Man | 157 | 40.9 |
| Woman | 227 | 59.1 |
| **Marital status** | | |
| Married | 240 | 62.5 |
| Divorced | 65 | 16.9 |
| Widowed | 53 | 13.8 |
| Single | 5 | 1.3 |
| Permanent relationship | 14 | 3.6 |
| Other | 7 | 1.8 |
| **Employment status** | | |
| Working full time | 33 | 8.6 |
| Working part-time | 39 | 10.2 |
| Retiree | 293 | 76.3 |
| Unemployed | 6 | 1.6 |
| Other | 13 | 3.4 |
| **Income level** | | |
| Much higher than average | 57 | 14.8 |
| Slightly higher than average | 118 | 30.7 |
| Similar to the average | 106 | 27.6 |
| Slightly lower than average | 50 | 13.8 |
| Much lower than average | 53 | 13.8 |
| **Residence Locality** | | |
| Big city | 185 | 48.2 |
| Outskirts of a big city | 26 | 6.8 |
| Medium or small city | 88 | 22.9 |
| Rural locality | 78 | 20.3 |
| Other | 7 | 1.8 |
| **Type of residence** | | |
| Apartment | 231 | 60.2 |
| Detached house | 144 | 37.5 |
| Assisted living | 5 | 1.3 |
| Nursing home | 1 | 0.3 |
| Other | 3 | 0.8 |
| **Living with** | | |
| Alone | 108 | 28.1 |
| Not alone (with spouse, caregiver etc.) | 276 | 71.9 |
| **Religious orientation** | | |
| Secular | 288 | 75.0 |
| Traditional | 63 | 16.4 |
| Religious | 27 | 7.0 |
| Ultra-orthodox | 6 | 1.6 |
| **Country of birth** | | |
| Israel | 249 | 64.8 |
| Western Europe, America | 43 | 11.2 |
| Asia, Africa | 40 | 10.4 |
| Eastern Europe | 46 | 12.0 |
| Other | 6 | 1.6 |

### 4.1.2. Procedure

The invitation email included a short explanation of the study and a link to an online survey site created via Qualtrics software. The survey began with a filtering question that asked respondents about their age. Participants whose age met the inclusion criterion were invited to watch a three-minute video that presented Gymmy and its functions (https://www.youtube.com/watch?v=zQ4T1NhS25Q). The video showed older persons exercising with the system while the narrator explained the system's functions and advantages. Important messages (e.g., "A camera that captures and decodes the user's body movements in real-time," "Gymmy, the robot that helps to maintain a healthy body, brain, and mind") were emphasized by using textual slides. The video was accompanied by soft background music. All images were edited using Adobe Photoshop, and videos were edited using Adobe Premiere Pro.

After watching the video, the participants were asked to answer questions related to the robot presented in the video. All measures in the questionnaire were based on common constructs used in HRI studies as detailed in the following sections. To avoid confusion among the respondents due to the transition between various scales, we applied two tactics. First, at the beginning of each part of the questionnaire, we provided information about the scale and range of possible answers. Second, when relevant, we indicated in words the meaning of each answer choice. To assess the reliability of the scales used to measure the concepts in this study and the internal consistency of the questionnaires, Cronbach Alpha was tested.

### 4.1.3. Measures
#### 4.1.3.1. Dependent variables: Quality evaluation

The QE was measured by the User Experience Questionnaire (UEQ, Laugwitz, et al., 2008; Appendix A). The main goal of the UEQ is to allow a rapid measurement of the user's evaluations of interactive products (Laugwitz et al., 2008; Santoso et al., 2016). Usually, it is used to measure post-use evaluation. In the present study, the same items were used to examine study participants' QE of *expected* use. Rather than referring to their experience with the product, they referred to their expectations of Gymmy based on what they saw in the video.

The UEQ contains six subscales with 26 bipolar items on a seven-point semantic differential scale measuring QE, i.e., the products' pragmatic and hedonic qualities and attractiveness (Hassenzahl, 2001; Laugwitz et al., 2008). Pragmatic quality describes task-related aspects (efficiency, perspicuity, dependability) and offers usability or usefulness insights (Hartson & Pyla, 2012). Hedonic quality aspects (stimulation, novelty) describe the ability of a product or system to evoke positive emotional states (Lorenz et al., 2014). Attractiveness is a pure valence dimension describing users' overall impression and stems from both the pragmatic and hedonic QEs of the product (Santoso et al., 2016).

Each UEQ item consists of two terms with opposite meanings (e.g., annoying - enjoyable), with seven answer options. The items' scale ranges from −3 to +3. Accordingly, after reverse coding several items, -3 represents the most negative answer, 0 a neutral answer, and +3 the most positive response. Scale values above +1 indicate a positive impression, while values below −1 indicate a negative one. The UEQ was translated into Hebrew by the Ph.D. candidate and independently back-translated into English by the supervisors. Their translations were compared to each other and to the original English version. The few instances in which the back-translations did not match were discussed by the team until an agreement about the exact term was achieved.

In this part of the study, the Cronbach alpha coefficients were 0.87 for both the pragmatic and hedonic evaluations, and 0.91 for attractiveness, indicating high internal consistency (Nunnally, 1994; Zinbarg et al., 2005) and implying successful interpretations of the various items (Pradana & Ferdiana, 2014).

#### 4.1.3.2. Independent variables

**Trust.** Trust in Gymmy was measured by the Human-Robot Trust Scale (Schaefer,

2013; 2016; Appendix B). This 14-items scale has two dimensions: 11 items represent the robot's performance-based functional capabilities (functions successfully, acts consistently, is reliable, predictable, dependable, follows directions, meets the needs of the mission, performs exactly as instructed, has errors, malfunctions, is unresponsive), while three items relate to social aspects (provides feedback, provides appropriate information, communicates with people) and represent the robot's "behaviors." Participants were asked to rate the robot's expected performance regarding each item on a scale ranging from 0% to 100% (Schaefer, 2016). The English version of the questionnaire was translated by applying the same procedure as that of the UEQ.

The Cronbach-Alpha coefficients in this part of the study were 0.90 and 0.84 for trust in the performance and social aspects, respectively, and 0.92 for the full scale.

**Technophobia.** Participants were asked to complete a modified version of the Technophobia Scale developed by Sinkovics et al. (2002; Appendix C), which has three dimensions: personal failure, human vs. machine ambiguity, and inconvenience. The original items were adapted to reflect participants' views of Gymmy (rather than their views of ATMs, which were the new technology in the original scale; e.g., instead of "ATMs are intimidating," we used the item "Gymmy is intimidating"). The scale was translated into Hebrew by Nimrod (2018) and included 15 items answerable on a Likert scale ranging from 1 (strongly disagree) to 5 (strongly agree).

In this part of the study, the Cronbach alpha coefficients were 0.83 for personal failure, 0.87 for human vs. machine ambiguity, 0.81 for inconvenience, and 0.91 for the full scale.

**Exercise.** The exercise patterns, motivation to exercise and perceived fitness and exercise limitations of the participants were measured using the self-assessed physical fitness questionnaire (https://www.marketest.co.uk). The questionnaire (Appendix D) contains 9 sections regarding exercise patterns (e.g., exercise frequency and duration), motivation to exercise (e.g., exercise importance and interest in exercising more often), perceived fitness and exercise limitations. The English version of the questionnaire was translated into Hebrew by the Ph.D. candidate and followed the same procedure as that of the UEQ.

**Background characteristics.** The last section of the survey included a demographic, sociodemographic, and health background questionnaire referring to participants' gender, age, marital status, number of children, residence locality, type of residence, number of people residing with the participant, religious orientation, country of birth, number of years of education, employment status, and income level. Two additional questions assessed self-rated physical and cognitive health on a 10-point Likert scale ranging from 1 ("not at all satisfied") to 10 ("completely satisfied").

### 4.1.4. Data analysis

The analysis consisted of a five-stage process, the first of which focused on the QE. The level of each of the UEQ sub-scales and the average scores for the scale as a whole were assessed according to means, standard deviations, and Pearson correlation coefficients. Stage two focused on background characteristics. The associations between users' background variables (demographic, sociodemographic, health) and all three QE dimensions (pragmatic evaluation, hedonic evaluation, and attractiveness), were evaluated using Pearson correlation coefficient, t-test and One-Way Analysis of Variance (ANOVA) tests according to the type of variable. In addition, a series of multiple regression analyses were conducted to determine the extent to which background characteristics contributed to prediction of QE variables. The independent variables were all background characteristics, and the dependent variables were pragmatic evaluation, hedonic evaluation, and attractiveness.

Stage three focused on the three independent variables: trust, technophobia and exercise. For trust, the means and standard deviations were calculated for each of the sub-scales and for the scale as a whole, and a t-test was used to explore differences between the means of the subscales. Subsequently, the associations between trust and all three QE dimensions were evaluated in a Pearson correlation matrix. Then, the associations between trust and all background variables were evaluated using Pearson correlation coefficient, t-test and ANOVA, according to the type of variable. In addition, series of linear regressions were conducted. The independent variables were trust and all background characteristics, and the dependent variables were pragmatic evaluation, hedonic evaluation, and attractiveness. Mediation analysis was also performed to determine whether the effect of background variables on QE variables was mediated by trust. This process was performed in the same manner for technophobia and exercise.

To simultaneously examine the associations of the independent variables with the QE variables, the fourth stage included performing another series of linear regressions. Again, pragmatic evaluation, hedonic evaluation, and attractiveness were the dependent variables, and all background characteristics, trust dimensions, technophobia dimensions and exercise dimensions were used as the independent variables. In the final stage, mediation analysis was performed to determine whether the effect of trust, technophobia and exercise on attractiveness was mediated by the pragmatic or hedonic quality evaluation. A Kolmogorov Smirnov (KS) and a Shapiro-Wilk normality test were performed for each variable to determine whether the data obtained were normally distributed. All variables met the criterion. In all five stages the data were analyzed using SPSS v.23 software, with a confidence interval of 95% in all tests.

### 4.2. Results
#### 4.2.1. Quality Evaluation

To answer the first research question, this analysis focused on QE. The mean scores for all scales included in the pragmatic evaluation (efficiency, perspicuity, dependability) were higher than +1, with an overall subscale mean of 1.46 (SD=1.18), indicating positive evaluation. In contrast, the means on the scales describing hedonic evaluation indicated neutral evaluation, with an overall subscale mean of 0.63 (SD=1.42; Table 3). A paired sample T-test revealed a significant difference between the pragmatic and hedonic evaluations (t=16.08, df=383, p<.001), indicating that the participants perceived Gymmy as providing more pragmatic value than hedonic experience. The mean score for the attractiveness scale was 1.00 (SD=1.43), which fits the assumption that this dimension is based on both pragmatic and hedonic evaluations of the product (Laugwitz et al., 2008). The Pearson correlations supported this notion by demonstrating strong positive correlations between the attractiveness of Gymmy and the pragmatic (r=.81) and hedonic (r=.86) evaluations. In addition, there was a strong positive correlation between the pragmatic and hedonic evaluations (r=.71). All correlations were statistically significant at a 0.01 level.

**Table 3**. *The means, standard deviations and Cronbach's alphas of the six factors of the UEQ.*

| Dimension | | Mean | Std. Deviation | Cronbach's alpha |
|---|---|---|---|---|
| Attractiveness | | 1.00 | 1.43 | 0.91 |
| Pragmatic Quality | Efficiency | 1.36 | 1.38 | 0.75 |
| | Perspicuity | 1.82 | 1.30 | 0.83 |
| | Dependability | 1.21 | 1.31 | 0.71 |
| | Dimension mean | 1.46 | 1.18 | 0.87 |
| Hedonic Quality | Stimulation | .70 | 1.51 | 0.87 |
| | Novelty | .57 | 1.51 | 0.80 |

| | | | |
|---|---|---|---|
| Dimension mean | .63 | 1.42 | 0.87 |

### 4.2.2. Background characteristics

To answer the second research question, this analysis focused on users' background variables. Relationships between continuous variables (age, number of children, number of years of education, self-rated health and cognitive function) were tested with a Pearson correlation coefficient. The analysis indicated two statistically significant correlations: Hedonic quality evaluation and attractiveness of Gymmy were negatively correlated with number of years of education (r = −.21 and r = −.19, respectively, p < 0.01 for both).

Differences in the nominal variables (gender, marital status, income, residence locality, type of residence, residing alone, religious orientation, country of birth, and employment status) were analyzed using a t-test and an ANOVA. A significant difference in the hedonic quality evaluations was found between individuals of varying religious orientations, showing that traditional, religious, and ultra-orthodox participants had higher levels of hedonic quality evaluation (mean = .90, SD = 1.54) compared to secular participants (mean = .54, SD = 1.37; t(382)=-2.137, p< .05). In addition, an independent samples t-test revealed a significant difference in the attractiveness of Gymmy according to marital status: Participants who were married or in a relationship had higher scores of attractiveness (mean = 1.10, SD = 1.34) in comparison to participants who were unmarried or not in a relationship (mean = .79, SD = 1.57; t(382)=1.874, p< .05). Furthermore, women showed higher levels of pragmatic quality evaluation (mean = 1.65, SD = 1.05) compared to men (mean = 1.19, SD = 1.30; t(382)=3.581, p< .01).

A series of multiple regression analyses were performed to determine the extent to which background characteristics contributed to prediction of the pragmatic and hedonic evaluations and the attractiveness of Gymmy. For this purpose, the ordinal variables with five or less categories and all the nominal variables in the regression were transformed into dummy codes of "0" or "1". A summary of the analyses is provided in Table 4.

**Table 4**. *Background characteristics associated with pragmatic and hedonic evaluations and the attractiveness of Gymmy: A linear regression analysis (N = 384).*

| Variable | Pragmatic quality evaluation | | | Hedonic quality evaluation | | | Attractiveness | | |
|---|---|---|---|---|---|---|---|---|---|
| | B | SE B | β | B | SE B | β | B | SE B | β |
| Gender | -0.56 | 0.13 | -0.23** | -0.39 | 0.16 | 0.13* | -0.30 | 0.16 | -0.10 |
| Age | 0.01 | 0.01 | 0.05 | 0.03 | 0.02 | 0.09 | 0.03 | 0.02 | 0.09 |
| Marital status | 0.52 | 0.25 | 0.21* | 0.25 | 0.30 | 0.08 | 0.50 | 0.30 | 0.16 |
| Education | -0.04 | 0.02 | -0.10 | -0.11 | 0.03 | -0.21** | -0.11 | 0.03 | -0.19** |
| Income | 0.07 | 0.14 | 0.03 | 0.09 | 0.16 | 0.03 | -0.03 | 0.16 | -0.01 |
| Employment status | 0.19 | 0.16 | 0.06 | 0.13 | 0.19 | 0.04 | 0.10 | 0.19 | 0.03 |
| Religious orientation | 0.12 | 0.15 | 0.04 | 0.33 | 0.19 | .100* | 0.10 | 0.18 | 0.03 |
| Self-rated health | -0.01 | 0.04 | -0.01 | -0.02 | 0.04 | -0.02 | -0.02 | 0.04 | -0.02 |
| Self-rated cognitive function | 0.06 | 0.05 | 0.07 | 0.01 | 0.06 | 0.01 | 0.04 | 0.06 | 0.04 |
| Number of children | 0.06 | 0.05 | 0.06 | -0.00 | 0.06 | -0.00 | 0.08 | 0.06 | 0.07 |
| Residence locality | -0.05 | 0.12 | -0.02 | -0.13 | 0.15 | -0.05 | -0.12 | 0.15 | -0.04 |
| Living alone | 0.30 | 0.25 | 0.11 | -0.03 | 0.30 | -0.01 | 0.16 | 0.31 | 0.05 |
| Country of birth | 0.16 | 0.13 | 0.05 | 0.02 | 0.16 | 0.00 | 0.00 | 0.20 | 0.00 |
| $R^2$ | | 0.077 | | | 0.073 | | | 0.071 | |
| F | | 2.38** | | | 2.24** | | | 2.17* | |

Note: SE: standard error. *$p< .05$; **$p< .01$; *** $p< .001$. Dummy codes: Gender, 1 = man, 0 = woman; Marital status, 1 = in a relationship, 0 = not in a relationship; Income, 1 = above average, 0 = below average; Employment status, 1 = working (part time or full time), 0 = not working (retiree or unemployed); Religious orientation, 1 = religious, 0 = secular; Residence Locality, 1 = big city, 0 = other; Living alone, 1 = yes, 0 = no; Country of birth, 1 = Israel, 0 = other.

All three regression models were found to be statistically significant (Table 4), but overall, the background variables were not strong predictors of the dependent variables: The background variables were able to explain only 7.7% and 7.3% of the variance of the pragmatic and hedonic quality evaluations (respectively), and only 7.1% of the attractiveness variance. The results of the first regression indicated that being a woman and being in a relationship were significantly associated with higher levels of pragmatic quality evaluation. The results of the second regression, however, showed that being a woman, religious, and with lower level of education were significantly associated with higher levels of hedonic quality evaluation. The third regression model demonstrated that having a lower level of education was significantly associated with high levels of attractiveness.

### 4.2.3. Trust and QE

To answer the third research question, this analysis focused on trust in robots. Participants highly trusted Gymmy's performance (Mean=0.79, SD=.15) and social aspects (Mean=0.76, SD=.19). The overall mean trust score was 0.78 (SD=.15). A paired sample t-test revealed a significant difference between the two dimensions (t=3.566, df=383, p<.001), indicating that participants trusted Gymmy's performance-based functional capabilities more than its social aspects. Pearson correlations exploring the relationships between the degree of trust in Gymmy and its QEs (Table 5) showed that all three variables were significantly positively (p<.001) associated with both dimensions of trust and with the total trust score. In other words, people with higher levels of trust in Gymmy were more likely to evaluate the expected experience of using Gymmy positively or be attracted to the robot.

**Table 5.** *Pearson correlations between the degree of trust and the quality evaluation of the expected experience of using Gymmy.*

| Variable | Pragmatic quality evaluation | Hedonic quality evaluation | Attractiveness |
|---|---|---|---|
| Trust | | | |
| Performance | 0.47** | 0.32** | 0.38** |
| Social aspects | 0.48** | 0.42** | 0.48** |
| Entire scale | 0.49** | 0.36** | 0.43** |

*p< .05; **p< .01.

The relationships between continuous background variables and degree of trust in Gymmy were tested using a Pearson correlation coefficient. Pearson correlation analysis revealed that self-rated cognitive function and number of children were significantly positively associated with both dimensions of trust and with the total trust score. In other words, participants who reported high satisfaction with their cognitive function and a greater number of children had a higher level of trust in Gymmy compared to participants who reported low levels of satisfaction with their cognitive function and had fewer children (both in the two trust-scale's dimensions as well as for the overall score, Table 6). The relationships between the nominal background variables and the degree of trust in Gymmy were analyzed using t-test and ANOVA. No significant relationships were found between these variables.

**Table 6.** *Relationships between the degree of trust in Gymmy and continuous background variables.*

| Variable | Performance | Social aspects | Trust (entire scale) |
|---|---|---|---|
| Self-rated cognitive | 0.20** | 0.11* | 0.18** |
| Number of children | 0.11* | 0.11* | 0.12* |

Note: *p< .05; **p< .01.

Three regression models (Table 7) indicated that these positive associations were maintained after controlling for background variables. In the first model, trust and background variables explained 30.1% of the pragmatic evaluation variance. The model demonstrated that being a woman and having higher levels of trust in the performance and social aspects of Gymmy were significantly associated with higher levels of pragmatic evaluation. The second model accounted for 24.3% of the variance of the hedonic evaluation. This regression showed that being a woman, less educated, and religiously observant, and having higher levels of trust in the social aspects of Gymmy were significantly associated with higher levels of hedonic evaluation. The third regression model was able to explain 29% of the variance of the attractiveness of Gymmy. This model indicated that being older, less educated and having higher levels of trust in the social aspects of Gymmy were significantly associated with high levels of attractiveness.

**Table 7.** *Trust variables and background characteristics associated with pragmatic and hedonic quality evaluations and the attractiveness of Gymmy: A linear regression analysis (N = 384)*

| Variable | Pragmatic quality evaluation | | | Hedonic quality evaluation | | | Attractiveness | | |
|---|---|---|---|---|---|---|---|---|---|
| | B | SE B | β | B | SE B | β | B | SE B | β |
| **Background** | | | | | | | | | |
| Gender | -0.50 | 0.12 | -0.21*** | -0.32 | 0.15 | -0.11* | -0.22 | 0.14 | -0.08 |
| Age | 0.01 | 0.01 | 0.06 | 0.03 | 0.01 | 0.09 | 0.03 | 0.01 | 0.10* |
| Marital status | 0.35 | 0.22 | 0.14 | 0.02 | 0.27 | 0.01 | 0.23 | 0.26 | 0.08 |
| Education | -0.04 | 0.02 | -0.09 | -0.10 | 0.03 | -0.19*** | -0.09 | 0.03 | -0.17*** |
| Income | 0.03 | 0.12 | 0.01 | 0.08 | 0.15 | 0.03 | -0.05 | 0.15 | -0.02 |
| Employment status | 0.12 | 0.14 | 0.04 | 0.09 | 0.17 | 0.02 | 0.04 | 0.17 | 0.01 |
| Religious orientation | 0.16 | 0.14 | 0.06 | 0.34 | 0.17 | 0.12* | 0.14 | 0.16 | 0.04 |
| Self-rated health | 0.00 | 0.03 | 0.01 | -0.00 | 0.04 | -0.00 | -0.00 | 0.04 | -0.00 |
| Self-rated cognitive function | -0.01 | 0.04 | -0.01 | -0.05 | 0.05 | -0.05 | -0.03 | 0.05 | -0.03 |
| Number of children | 0.00 | 0.05 | 0.00 | -0.06 | 0.06 | -0.05 | 0.01 | 0.06 | 0.01 |
| Residence locality | -0.03 | 0.11 | -0.01 | -0.11 | 0.13 | -0.04 | -0.10 | 0.13 | -0.04 |
| Living alone | 0.15 | 0.22 | 0.06 | -0.22 | 0.28 | -0.07 | -0.05 | 0.27 | -0.02 |
| Country of birth | 0.03 | 0.12 | 0.01 | -0.07 | 0.15 | -0.02 | -0.10 | 0.14 | -0.03 |
| **Trust** | | | | | | | | | |
| Performance | 2.20 | 0.56 | 0.27*** | 0.67 | 0.70 | 0.07 | 1.00 | 0.68 | 0.10 |
| Social aspects | 1.53 | 0.42 | 0.25*** | 2.73 | 0.53 | 0.37*** | 2.97 | 0.51 | 0.40*** |
| $R^2$ | | 0.301 | | | 0.243 | | | 0.290 | |
| F | | 10.58 | | | 7.89 | | | 10.02 | |

Note: SE: standard error. *$p < .05$; **$p < .01$; *** $p < .001$. Dummy codes: Gender, 1 = man, 0 = woman; Marital status, 1 = in a relationship, 0 = not in a relationship; Income, 1 = above average, 0 = below average; Employment status, 1 = working (part time or full time), 0 = not working (retiree or unemployed); Religious orientation, 1 = religious, 0 = secular; Residence locality, 1 = big city or outskirts of a big city, 0 = other; Living alone, 1 = yes, 0 = no; Country of birth, 1 = Israel), 0 = other.

To answer the sixth research question, a mediation analysis for the role of trust in robots in the effect of background variables (demographic, sociodemographic, health) on QE variables of Gymmy was conducted. No significant mediation was found for the effect of background variables on QE variables via trust in robots.

### 4.2.4. Technophobia and QE

To answer the fourth research question, this analysis focused on robot-related technophobia. The mean scores for technophobia were 1.80 (SD=.74) for the subscale of fear of personal failure, 2.54 (SD=.94) for human vs. machine ambiguity, and 2.52 (SD=.92) for inconvenience. The overall score was 2.24 (SD=.74). These values suggested that participants felt a low level of fear of using Gymmy. T-test analyses, however, indicated significant differences between the degree of technophobia in terms of fear of personal failure and the other two dimensions: human vs. machine ambiguity (t=-20.348, df=383, p<.001) and inconvenience (t=-17.840, df=383, p<.001). These results indicated that respondents' concerns regarding the growing dominance of robotic systems as human substitutes (e.g., robots for physical and cognitive training instead of human trainers) and the perceived inconvenience of using robots were significantly more intense than their worries about the sense of failure and frustrations that might arise from interacting with Gymmy.

The Pearson correlations between the three dimensions of technophobia, the overall technophobia score, and the three dependent variables (Table 8) showed that the pragmatic and hedonic evaluation and attractiveness of Gymmy were significantly negatively (p<0.01) associated with the three technophobia subscales as well as with the total technophobia score. People who expressed higher levels of technophobia were less likely to evaluate the expected experience of using Gymmy positively or be attracted to the robot.

**Table 8.** *Pearson correlations between the degree of trust, the degree of technophobia and the quality evaluation of the expected experience of using Gymmy.*

| Variable | Pragmatic quality evaluation | Hedonic quality evaluation | Attractiveness |
|---|---|---|---|
| Technophobia | | | |
| Personal failure | -0.44** | -0.37** | -0.47** |
| Human vs. machine-ambiguity | -0.39** | -0.35** | -0.44** |
| Convenience | -0.51** | -0.57** | -0.62** |
| Entire scale | -0.50** | -0.47** | -0.56** |

*p< .05; **p< .01.

The relationships between continuous background variables and degree of technophobia from using Gymmy were tested using a Pearson correlation coefficient. Pearson correlation analysis indicated a significant correlation between age and human vs. machine-ambiguity dimension (r =.104, p < 0.05). That is, as age increases, so does the degree of ambiguity towards Gymmy. Additionally, the number of years of education was significantly positively associated with the perceived inconvenience of using Gymmy (r = .135, p < 0.01). Meaning, more educated older adults perceive the use of Gymmy as less convenient and helpful compared to less educated older adults. The relationships between the nominal background variables and the degree of technophobia from using Gymmy were analyzed using a t-test and an ANOVA. No significant relationships were found between these variables.

Three regression models (Table 9) indicated that the negative associations were maintained after controlling for background variables. In the first model, technophobia and background variables explained 35.9% of the pragmatic evaluation variance. The model demonstrated that being a woman, being older, and having lower levels of technophobia in terms of personal failure and perceived inconvenience were significantly associated with higher levels of pragmatic evaluation. The second model accounted for 38.9% of the variance in hedonic evaluation. This regression showed that being a woman, older, less educated, and religiously observant, and having lower levels of technophobia in terms of perceived inconvenience were significantly associated with higher levels of hedonic evaluation. The third regression model explained 45.8% of the variance of the attractiveness of Gymmy. This model indicated that being a woman, older, and less educated, and having lower levels of technophobia in terms of perceived inconvenience and fear of personal failure were significantly associated with high levels of attractiveness.

**Table 9.** *Technophobia variables and background characteristics associated with pragmatic and hedonic quality evaluations and the attractiveness of Gymmy: A linear regression analysis (N = 384).*

| Variable | Pragmatic quality evaluation | | | Hedonic quality evaluation | | | Attractiveness | | |
|---|---|---|---|---|---|---|---|---|---|
| | B | SE B | β | B | SE B | β | B | SE B | β |
| Background | | | | | | | | | |
| Gender | -0.56 | 0.11 | -0.23*** | -0.42 | 0.13 | -0.14** | -0.32 | 0.13 | -0.11* |
| Age | 0.03 | 0.01 | 0.11* | 0.05 | 0.01 | 0.15** | 0.05 | 0.01 | 0.17*** |
| Marital status | 0.32 | 0.21 | 0.13 | 0.01 | 0.24 | 0.00 | 0.21 | 0.23 | 0.07 |
| Education | -0.02 | 0.02 | -0.05 | -0.08 | 0.02 | -0.14** | -0.07 | 0.02 | -0.13** |
| Income | 0.08 | 0.11 | 0.03 | 0.11 | 0.13 | 0.04 | -0.02 | 0.13 | -0.01 |
| Employment status | 0.22 | 0.13 | 0.07 | 0.22 | 0.15 | 0.06 | 0.16 | 0.15 | 0.04 |
| Religious orientation | 0.14 | 0.13 | 0.05 | 0.33 | 0.15 | 0.10* | 0.11 | 0.14 | 0.03 |
| Self-rated health | 0.01 | 0.03 | 0.02 | 0.01 | 0.04 | 0.02 | 0.01 | 0.03 | 0.02 |
| Self-rated cognitive function | 0.03 | 0.04 | 0.04 | -0.02 | 0.05 | -0.02 | 0.01 | 0.05 | 0.01 |
| Number of children | 0.03 | 0.05 | 0.03 | -0.04 | 0.05 | -0.03 | 0.04 | 0.05 | 0.03 |
| Residence locality | -0.06 | 0.10 | -0.03 | -0.17 | 0.12 | -0.06 | -0.17 | 0.11 | -0.06 |
| Living alone | 0.16 | 0.21 | 0.06 | -0.20 | 0.25 | -0.06 | -0.03 | 0.24 | -0.01 |
| Country of birth | 0.18 | 0.11 | 0.07 | 0.09 | 0.13 | 0.03 | 0.09 | 0.13 | 0.03 |
| Technophobia | | | | | | | | | |
| Personal failure | -0.25 | 0.10 | -0.16** | -0.10 | 0.11 | -0.05 | -0.25 | 0.11 | -0.13* |
| Human vs. machine- ambiguity | -0.10 | 0.08 | -0.08 | -0.08 | 0.09 | -0.05 | -0.15 | 0.09 | -0.10 |
| Inconvenience | -0.48 | 0.07 | -0.38*** | -0.79 | 0.09 | -0.51*** | -0.76 | 0.08 | -0.49*** |
| $R^2$ | | 0.359 | | | 0.389 | | | 0.458 | |
| F | | 12.84 | | | 14.60 | | | 19.40 | |

Note: SE: standard error. *$p < .05$; **$p < .01$; *** $p < .001$. Dummy codes: Gender, 1 = man, 0 = woman; Marital status, 1 = in a relationship, 0 = not in a relationship; Income, 1 = above average, 0 = below average; Employment status, 1 = working (part time or full time), 0 = not working (retiree or unemployed); Religious orientation, 1 = religious, 0 = secular; Residence locality, 1 = big city or outskirts of a big city, 0 = other; Living alone, 1 = yes, 0 = no; Country of birth, 1 = Israel, 0 = other.

To answer the sixth research question and further explore the effect of technophobia, we looked at the mediation analysis for the role of perceived inconvenience (technophobia) in the effect of number of years of education on the attractiveness of Gymmy. Multiple regression analyses were conducted to assess each component of the proposed mediation model. First, it was found that number of years of education was negatively associated with attractiveness of Gymmy (c-path; B=-.102, *t*(382)=-3.76, p<.001). It was also found that number of years of education was positively related to perceived inconvenience (a-path; B=.047, *t*(382)=2.66, p<.01). Lastly, results indicated that the mediator, perceived inconvenience, was negatively associated with attractiveness of Gymmy (b-path; B=-.095, *t*(382)=-15.03, p<.001). Because both the a-path and b-path were significant, the model meets the criteria according to Baron and Kenny (1986). Then, mediation analyses were tested using a bootstrapping method with bias-corrected confidence estimates (MacKinnon et al., 2004; Preacher & Hayes, 2004; Preacher et al., 2007). In the present study, the 95% confidence interval of the indirect effects was obtained with 5000 bootstrap resamples (Preacher & Hayes, 2008). Results of the mediation analysis confirmed the mediating role of perceived inconvenience in the relation between number of years of education and attractiveness of Gymmy (B=-.08; CI=-0.15, -0.02). In addition, results indicated that the direct effect of number of years of education on attractiveness of Gymmy became significant (albeit less pronounced, c′-path; B=-.058, *t*(382)=-2.66, *p*<.01) when controlling for perceived inconvenience, thus suggesting partial mediation. Figure 8 displays the results.

Apart from this partial mediation, no significant mediation was found for the effect of background variables on QE variables via robot-related technophobia.

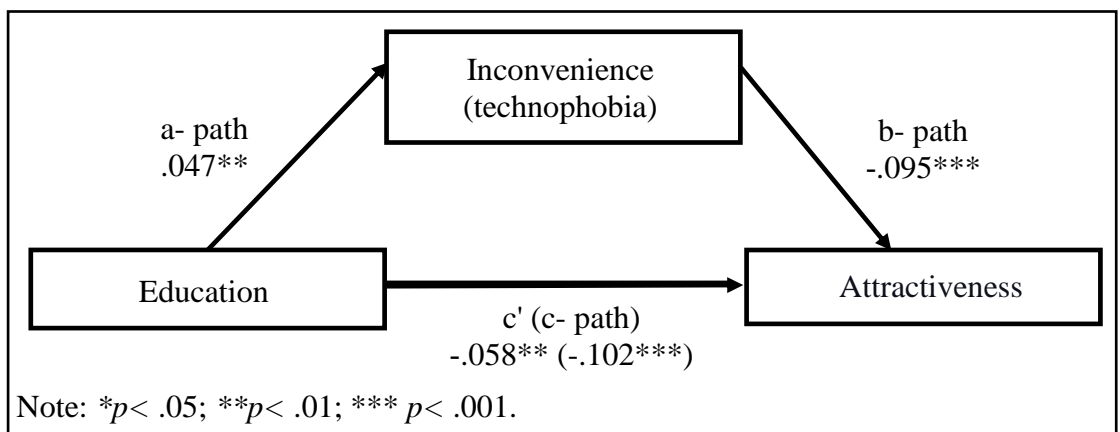

Note: *p< .05; **p< .01; *** p< .001.

**Figure 8.** Indirect effect of number of years of education on attractiveness of Gymmy through inconvenience (technophobia)

### 4.2.5. Exercise and QE

Since Gymmy functioned as a robotic physical trainer for older adults, it was important to explore the participants' existing exercise patterns that could affect the anticipated QE (pragmatic and hedonic evaluations and attractiveness) of Gymmy - the fifth research question. The sample's full exercise characteristics are presented in Tables 10-11.

**Table 10**. *Exercise characteristics of the sample (for ordinal and nominal variables).*

| Variable | Frequency | Percent |
|---|---|---|
| **Exercise importance (N=384)** | | |
| Extremely | 152 | 39.6 |
| Very | 154 | 40.1 |
| Moderately | 56 | 14.6 |
| Slightly | 22 | 5.7 |
| Not at all | 0 | 0 |
| **Perceived fitness level (N=384)** | | |
| Very good | 36 | 9.4 |
| Good | 126 | 32.8 |
| Average | 131 | 34.1 |
| Below average | 73 | 19 |
| Not fit at all | 18 | 4.7 |
| **Exercise frequency (N=384)** | | |
| Every day | 40 | 10.4 |
| 5-6 times a week | 51 | 13.3 |
| 3-4 times a week | 128 | 33.3 |
| 1-2 times a week | 91 | 23.7 |
| Less than once a week | 50 | 13 |
| Do not exercise at all | 24 | 6.3 |
| **Exercise duration (N=360)** | | |
| 2-3 hours | 6 | 1.7 |
| 1-2 hours | 67 | 18.6 |
| 46 minutes - 1 hour | 124 | 34.4 |
| 31-45 minutes | 81 | 22.5 |
| 0-30 minutes | 82 | 22.8 |
| **Interest in exercising more often (N=384)** | | |
| Yes, much more | 70 | 18.2 |
| Yes, a little more | 195 | 50.8 |
| No | 119 | 31 |

Note: Participants who reported that they did not engage in any physical activity (N=24) were not asked about their exercise patterns and motivations. Participants who reported not wanting to exercise more often (N=119) were not asked about the factors that limit them from doing so.

**Table 11.** *Exercise characteristics of the sample (for continuous variables).*

| Variable | Range | Mean | Std. Deviation |
|---|---|---|---|
| The number of types of physical activities (N=360) | 1-7 | 1.78 | 1.01 |
| The number of motives to exercise (N=360) | 1-12 | 4.62 | 2.85 |
| The number of exercise limitations (N=265) | 1-5 | 1.59 | 0.80 |

Pearson's correlation matrix between all exercise variables and the three evaluation variables was used to examine the relationships between exercise patterns, motivation to exercise and perceived fitness and exercise limitations, and the evaluation of the experience of using Gymmy. The analysis presented in Table 12 shows that the pragmatic quality evaluation was significantly positively associated with exercise importance and with the number of motives to exercise. These results indicate that the two variables that were significantly positively associated with the pragmatic quality evaluation variable are motivation variables. The more important it is for participants to exercise and the more reasons they specify as motivating them to exercise, the more they tend to evaluate Gymmy positively.

**Table 12.** *Pearson Correlations between the older adults' exercise characteristics and the quality evaluation of the experience of using Gymmy.*

| Variable | Pragmatic quality evaluation | Hedonic quality evaluation | Attractiveness |
|---|---|---|---|
| Exercise patterns | | | |
| Exercise frequency | 0.05 | -0.04 | -0.06 |
| Exercise duration | 0.01 | -0.10 | -0.09 |
| Types of exercise | -0.02 | -0.09 | -0.10 |
| Motivation to exercise | | | |
| Exercise importance | 0.14** | 0.06 | 0.03 |
| The number of motives to exercise | 0.16** | 0.09 | 0.07 |
| Perceived fitness and exercise limitations | | | |
| Perceived fitness level | 0 | -0.05 | -0.06 |
| Exercise limitations | 0.02 | 0.02 | 0.02 |

Note: *p< .05; **p< .01.

Differences in interest in exercising more often were analyzed using t-tests. A significant difference in the hedonic quality evaluations was found, as participants who expressed an interest in exercising more often had higher levels of hedonic quality evaluation (mean = .77, SD = 1.39) compared to participants who did not express such interest (mean = .34, SD = 1.46; t(382)=-2.717, p< .01). In addition, an independent samples t-test revealed a significant difference in the attractiveness of Gymmy. Participants interested in exercising more often (mean = 1.13, SD = 1.35) showed higher scores of attractiveness compared to participants without such interest (mean = .71, SD = 1.56; t(382)=-2.711, p< .01).

Since the motivation variables were the only ones that statistically associated with the evaluation variables, the rest of the analysis considered motivation only. The relationships between continuous background variables and older adults' motivation to exercise were tested using Pearson's correlation coefficient. Analysis indicated a significant negative correlation between age and the number of motives to exercise. (r =-.14, p < 0.01). That is, as age increases, the number of motives to exercise decreases. Additionally, the number of years of education was significantly positively associated with exercise importance (r = .12, p < 0.05). Engagement in physical activity is more important for more educated older adults than less educated older adults. The analysis also revealed that self-rated health was significantly positively associated with exercise importance and with the number of motives to exercise (r =.28, p < 0.01, r =.110, p < 0.05, respectively). Participants who reported high satisfaction with their physical health found it more important to engage in physical activity and had a greater number of reasons motivating them to exercise, compared to participants who reported low levels of satisfaction with their physical health. Moreover, self-rated cognitive function was significantly positively associated with exercise importance (r = .17, p < 0.01). In other words, participants who reported high satisfaction with their cognitive function found physical activity more important, compared to participants who reported low levels of satisfaction with their cognitive function.

The relationships between the nominal background variables and older adults' motivation to exercise were analyzed using t-tests and ANOVA. A significant difference in the number of motives to exercise was found between men and women. Compared to men, women specified more reasons motivating them to exercise. In addition, participants born in Israel reported higher levels of exercise importance and specified more reasons motivating them to exercise compared to participants not born in Israel. The full details of the t-tests that yielded significant results are provided in Table 13.

Chi-square tests were used to determine association between the nominal background variables and interest in exercising more often. The association between gender and interest in exercising more often was significant, $\chi^2$ (1, N = 384) = 5.39, p < .05. Women expressed more interest than men in exercising more often. Additionally, there was a significant relationship between marital status and interest in exercising more often. Participants who were not in a relationship expressed more interest in exercising more often compared to participants who were in a relationship, $\chi^2$ (1, N = 384) = 5.75, p < .05.

**Table 13.** *Nominal background variables and older adults' motivation to exercise.*

|  |  | Number of motives to exercise | | |  | Exercise importance | | |
|---|---|---|---|---|---|---|---|---|
| Variable | N | M | SD | t | N | M | SD | t |
| Country of birth | | | | | | | | |
| Israel | 233 | 4.90 | 2.83 | -2.51* | 249 | 4.22 | 0.82 | -2.51* |
| Other | 127 | 4.12 | 2.82 | | 135 | 3.99 | 0.93 | |
|  |  | Number of motives to exercise | | | | | | |
| Variable | N | M | SD | t | | | | |
| Gender | | | | | | | | |
| Men | 147 | 3.97 | 2.75 | 3.65** | | | | |
| Women | 213 | 5.07 | 2.83 | | | | | |

To explore the unique contribution of the older adults' motivation to exercise habits in explaining the dependent variables, a series of linear regressions were applied. Multiple linear regression was performed for each of the variables composing the dependent variable, the pragmatic and hedonic quality evaluation, and attractiveness of Gymmy, with the background variables and the three exercise motivation variables as the independent variables. For this purpose, the ordinal background variables with five or less categories and all the nominal background variables in the regression were transformed into dummy codes of "0" or "1".

The results show that all three regression models were found to be statistically significant. However, the older adults' motivation was not a strong predictor of the dependent variables. Only 10.6% and 12.6% of the variance of the pragmatic and hedonic quality evaluations (respectively), and only 12.0% of the attractiveness variance were explained. The results of the first regression indicated that being a woman and having less education were significantly associated with higher levels of pragmatic quality evaluation. The results of the second regression, however, showed that being a woman, having less educational, and reporting greater interest in exercising more often were significantly associated with higher levels of hedonic quality evaluation. The third regression model demonstrated that being married or in a relationship, having less education and reporting greater interest in exercising more often were significantly associated with high levels of attractiveness. A summary of the analyses is provided in Table 14.

To answer the sixth research question, a mediation analysis for the role of older adults' motivation to exercise in the effect of background variables (demographic, sociodemographic, health) on QE variables of Gymmy was conducted. No significant mediation was found for the effect of background variables on QE variables via exercise.

Table 14. *Older adults' motivation to exercise and background characteristics associated with pragmatic and hedonic evaluations and the attractiveness of Gymmy: A linear regression analysis (N = 384).*

| Variable | Pragmatic quality evaluation | | | Hedonic quality evaluation | | | Attractiveness | | |
|---|---|---|---|---|---|---|---|---|---|
| | B | SE B | β | B | SE B | β | B | SE B | β |
| Background | | | | | | | | | |
| Gender | -0.53 | 0.14 | -0.22*** | -0.37 | 0.16 | -0.13* | -0.31 | 0.17 | -0.11 |
| Age | 0.02 | 0.01 | 0.06 | 0.02 | 0.02 | 0.07 | 0.03 | 0.02 | 0.10 |
| Marital status | 0.55 | 0.26 | 0.22 | 0.44 | 0.31 | 0.15 | 0.38 | 0.17 | 0.13* |
| Education | -0.05 | 0.03 | -0.11* | -0.12 | 0.03 | -0.23*** | -0.12 | 0.03 | -0.22*** |
| Income | 0.11 | 0.14 | 0.05 | 0.14 | 0.17 | 0.05 | 0.01 | 0.17 | 0.00 |
| Employment status | 0.20 | 0.16 | 0.07 | 0.18 | 0.19 | 0.05 | 0.13 | 0.19 | 0.04 |
| Religious orientation | 0.11 | 0.16 | 0.04 | 0.34 | 0.19 | 0.11 | 0.11 | 0.19 | 0.03 |
| Self-rated health | 0.00 | 0.04 | 0.00 | -0.02 | 0.05 | -0.02 | -0.01 | 0.05 | -0.01 |
| Self-rated cognitive function | 0.05 | 0.05 | 0.06 | 0.05 | 0.06 | 0.04 | 0.07 | 0.06 | 0.06 |
| Number of children | 0.06 | 0.05 | 0.06 | -0.02 | 0.06 | -0.01 | 0.08 | 0.07 | 0.07 |
| Residence locality | -0.06 | 0.12 | -0.03 | -0.18 | 0.15 | -0.06 | -0.17 | 0.15 | -0.06 |
| Living alone | 0.31 | 0.27 | 0.12 | 0.07 | 0.32 | 0.02 | 0.13 | 0.33 | 0.04 |
| Country of birth | 0.04 | 0.14 | 0.01 | -0.13 | 0.16 | -0.04 | -0.11 | 0.17 | -0.04 |
| Exercise | | | | | | | | | |
| Exercise importance | 0.16 | 0.09 | 0.11 | 0.17 | 0.10 | 0.10 | 0.112 | 0.10 | 0.06 |
| Number of motives to exercise | 0.04 | 0.02 | 0.09 | 0.03 | 0.03 | 0.06 | 0.03 | 0.03 | 0.05 |
| Interest in exercising more often | 0.2 | 0.14 | 0.09 | 0.49 | 0.16 | 0.16** | 0.49 | 0.16 | 0.16** |
| $R^2$ | | 0.106 | | | 0.126 | | | 0.120 | |
| F | | 2.532 | | | 3.090 | | | 2.913 | |

Note: SE: standard error. *$p$< .05; **$p$< .01; *** $p$< .001. Dummy codes: Gender, 1 = man, 0 = woman; Marital status, 1 = in a relationship, 0 = not in a relationship; Income, 1 = above average, 0 = below average; Employment status, 1 = working (part time or full time), 0 = not working (retiree or unemployed); Religious orientation, 1 = religious, 0 = secular; Residence Locality, 1 = big city, 0 = other; Living alone, 1 = yes, 0 = no; Country of birth, 1 = Israel, 0 = other; Interest in exercising more often, 1 = Yes, 0 = No.

### 4.2.6. Trust, technophobia, exercise, and QE

To answer the seventh research question, this analysis focused on the simultaneous explorations of trust, technophobia, and exercise. The simultaneous explorations of these variables' correlations with the QE of Gymmy (Table 15) showed that these constructs had contradicting yet independent associations with the dependent variables. All three regression models were statistically significant, and the combination of trust variables, technophobia variables, and exercise variables contributed significantly to the ability to explain the variance in Gymmy's QEs. In the first model, trust, technophobia, exercise, and background variables explained 41.5% of the pragmatic evaluation variance. The model demonstrated that being a woman, being older, having higher levels of trust in Gymmy's performance, having lower levels of technophobia in terms of perceived inconvenience, and reporting higher level of exercise importance were significantly associated with higher pragmatic evaluation. The second model, which accounted for 43.4% of the variance, showed that being a woman, older, less educated, and religiously observant, and having lower levels of technophobia in terms of perceived inconvenience were significantly associated with hedonic evaluation. The third regression model, which explained 50.1% of the variance, indicated that being a woman, older, and less educated, and having higher levels of trust in social aspects of Gymmy and lower levels of technophobia in terms of perceived inconvenience and fear of personal failure were significantly associated with attractiveness.

**Table 15.** *Trust variables, technophobia variables, exercise variables, and background characteristics associated with pragmatic and hedonic quality evaluations and the attractiveness of Gymmy: A linear regression analysis (N = 384)*

| Variable | Pragmatic quality evaluation | | | Hedonic quality evaluation | | | Attractiveness | | |
|---|---|---|---|---|---|---|---|---|---|
| | B | SE B | β | B | SE B | β | B | SE B | β |
| **Background** | | | | | | | | | |
| Gender | -0.53 | 0.11 | -0.22*** | -0.35 | 0.13 | -0.12* | -0.25 | 0.12 | -0.09* |
| Age | 0.03 | 0.01 | 0.10* | 0.04 | 0.01 | 0.14* | 0.05 | 0.01 | 0.16** |
| Marital status | 0.18 | 0.11 | 0.07 | 0.12 | 0.13 | 0.04 | 0.19 | 0.13 | 0.06 |
| Education | -0.03 | 0.02 | -0.06 | -0.07 | 0.02 | -0.13** | -0.07 | 0.02 | -0.13*** |
| Income | 0.04 | 0.11 | 0.02 | 0.10 | 0.13 | 0.03 | -0.03 | 0.12 | -0.01 |
| Employment status | 0.17 | 0.13 | 0.05 | 0.21 | 0.152 | 0.06 | 0.15 | 0.14 | 0.04 |
| Religious orientation | 0.05 | 0.12 | 0.06 | 0.33 | 0.15 | 0.10* | 0.12 | 0.14 | 0.04 |
| Self-rated health | 0.01 | 0.03 | 0.02 | 0.02 | 0.04 | 0.02 | 0.02 | 0.03 | 0.02 |
| Self-rated cognitive function | 0.00 | 0.04 | 0.00 | -0.03 | 0.05 | -0.03 | -0.01 | 0.05 | -0.01 |
| Number of children | 0.00 | 0.04 | 0.00 | -0.06 | 0.05 | -0.05 | 0.02 | 0.05 | 0.01 |
| Residence locality | -0.06 | 0.10 | -0.03 | -0.16 | 0.12 | -0.06 | -0.15 | 0.11 | -0.05 |
| Living alone | -0.09 | 0.10 | -0.04 | -0.23 | 0.12 | -0.08 | -0.22 | 0.11 | -0.07 |
| Country of birth | 0.11 | 0.11 | 0.05 | 0.03 | 0.13 | 0.01 | 0.02 | 0.12 | 0.01 |
| **Trust** | | | | | | | | | |
| Performance | 1.60 | 0.53 | 0.20** | 0.01 | 0.63 | 0.00 | 0.11 | 0.59 | 0.01 |
| Social aspects | 0.59 | 0.40 | 0.10 | 1.21 | 0.49 | 0.16 | 1.37 | 0.45 | 0.18** |
| **Technophobia** | | | | | | | | | |
| Personal failure | -0.15 | 0.10 | -0.09 | -0.07 | 0.11 | -0.04 | -0.20 | 0.11 | -0.10* |
| Human vs. machine- ambiguity | -0.10 | 0.07 | -0.08 | -0.05 | 0.09 | -0.04 | -0.12 | 0.08 | -0.08 |
| convenience | -0.37 | 0.07 | -0.29*** | -0.70 | 0.09 | -0.45** | -0.65 | 0.08 | -0.42*** |
| **Exercise** | | | | | | | | | |
| Exercise importance | 0.15 | 0.07 | 0.11* | 0.14 | 0.08 | 0.08 | 0.09 | 0.08 | 0.05 |
| Number of motives to exercise | 0.02 | 0.02 | 0.05 | 0.02 | 0.02 | 0.03 | 0.01 | 0.02 | 0.02 |
| Interest in exercising more often | 0.28 | 0.11 | 0.11 | 0.23 | 0.13 | 0.80 | 0.21 | 0.12 | 0.70 |
| $R^2$ | | 0.415 | | | 0.434 | | | 0.501 | |
| F | | 11.41 | | | 12.32 | | | 16.18 | |

Note: SE: standard error. *$p < .05$; **$p < .01$; ***$p < .001$. Dummy codes: Gender, 1 = man, 0 = woman; Marital status, 1 = in a relationship, 0 = not in a relationship; Income, 1 = above average, 0 = below average; Employment status, 1 = working (part time or full time), 0 = not working (retiree or unemployed); Religious orientation, 1 = religious, 0 = secular; Residence locality, 1 = big city or outskirts of a big city, 0 = other; Living alone, 1 = yes, 0 = no; Country of birth, 1 = Israel, 0 = other; Interest in exercising more often, 1 = Yes, 0 = No.

To answer the eighth research question, a series of mediation analysis based on the Baron and Kenny four-step model (Hayes, 2009) were performed using Hayes' (2013) Model 4 of the SPSS PROCESS Macro.

***Trust.*** We looked at the mediation analysis for the role of the pragmatic quality evaluation in the effect of trust in the performance-based functional capabilities of Gymmy on the attractiveness of Gymmy. Multiple regression analyses were conducted to assess each component of the proposed mediation model. First, it was found that trust in the performance-based functional capabilities was positively associated with attractiveness of Gymmy (c-path; B=3.747, $t(382)$=8.089, p<.001). It was also found that trust in the performance-based functional capabilities was positively related to the pragmatic quality evaluation (a-path; B=3.770, $t(382)$=10.27, p<.001). Lastly, the results indicated that the mediator, pragmatic quality evaluation, was positively associated with attractiveness of Gymmy (b-path; B=.971, $t(382)$=23.481, p<.001). Because both the a-path and b-path were significant, the model meets the criteria according to Baron and Kenny (1986). Then, mediation analyses were tested using a bootstrapping method with bias-corrected confidence estimates (MacKinnon et al., 2004; Preacher & Hayes, 2004; Preacher et al., 2007). In the present study, the 95% confidence interval of the indirect effects was obtained with 5000 bootstrap resamples (Preacher & Hayes, 2008). Results of the mediation analysis confirmed the mediating role of the pragmatic quality evaluation in the relation between trust in the performance-based functional capabilities and attractiveness of Gymmy (B=.373; CI=0.29, 0.46). In addition, the results indicated that the direct effect of trust in the performance-based functional capabilities on attractiveness of Gymmy became insignificant (c′-path; B=.088, $t(382)$=.262, *p*=.793) when controlling for the pragmatic quality evaluation, thus suggesting full mediation.

In addition, we looked at the mediation analysis for the role of the pragmatic quality evaluation in the effect of the degree of trust in Gymmy (entire scale) on the attractiveness of Gymmy. Multiple regression analyses were conducted to assess each component of the proposed mediation model. First, it was found that trust was positively associated with attractiveness of Gymmy (c-path; B=4.144, $t(382)$=9.302, p<.001). It was also found that trust was positively related to the pragmatic quality evaluation (a-path; B=3.920, $t(382)$=11.04, p<.001). Lastly, the results indicated that the mediator, pragmatic quality evaluation, was positively associated with attractiveness of Gymmy (b-path; B=.950, $t(382)$=22.647, p<.001). Since both the a-path and b-path were significant, the model meets the criteria according to Baron and Kenny (1986). Then, mediation analyses were tested using a bootstrapping method with bias-corrected confidence estimates (MacKinnon et al., 2004; Preacher & Hayes, 2004; Preacher et al., 2007). In the present study, the 95% confidence interval of the indirect effects was obtained with 5000 bootstrap resamples (Preacher & Hayes, 2008). Results of the mediation analysis confirmed the mediating role of the pragmatic quality evaluation in the relation between the degree of trust (entire scale) and attractiveness of Gymmy (B=.386; CI=0.30, 0.47). In addition, the results indicated that the direct effect of trust in the performance-based functional capabilities on attractiveness of Gymmy became insignificant (c′-path; B=.420, *t*(382)=1.255, *p*=.210) when controlling for the pragmatic quality evaluation, thus suggesting full mediation.

*Technophobia*. No significant mediation was found for the effect of the degree of technophobia from using Gymmy on the attractiveness of Gymmy via pragmatic and/or hedonic quality evaluation.

*Exercise.* No significant mediation was found for the effect of older adults' motivation to exercise on the attractiveness of Gymmy via pragmatic and/or hedonic quality evaluation. Summary of the analyses is provided in Table 16.

**Table 16.** *Path coefficients and indirect effects for mediation analysis using Hayes' (2013) Model 4 of the SPSS PROCESS Macro.*

| | Path coefficients | | | | Bias-corrected bootstrap 95% confidence interval | |
|---|---|---|---|---|---|---|
| Mediation variables | a | b | c | c' | LLCI | ULCI |
| Trust | | | | | | |
| Performance ➡ PQE ➡ Attractiveness | 3.77*** | .97*** | 3.74*** | .09 | 0.29 | 0.46 |
| Trust ➡ PQE ➡ Attractiveness (entire scale) | 3.92*** | .95*** | 4.14*** | .40 | 0.30 | 0.47 |

Note: PQE = pragmatic quality evaluation; HQE = hedonic quality evaluation; LLCI = lower limit confidence interval; ULCI= upper limit confidence interval.
*p< .05; **p< .01; *** p< .001

## 4.3. Discussion

The simultaneous exploration of trust, technophobia and exercise yielded new insights regarding the factors predicting QE (i.e., pragmatic and hedonic evaluations and overall attractiveness) of SARs in later life. Similar to previous research, our findings indicated positive associations between trust and QE and negative associations between technophobia and QE. Regarding exercise, no significant associations were found, other than positive associations between the motivation variables and QE. The simultaneous examination of trust and technophobia—which contributed significantly to explaining the variance in the QE variables—highlighted that the relative impact of technophobia is significantly stronger than that of trust, and that the pragmatic qualities of the robot and its convenience of use, in particular, are more crucial to its QE than the emotional aspects of use.

The regression analysis conducted to explore the simultaneous impact of trust, technophobia and exercise on QE (Table 15) indicated that robot-related technophobia is a stronger predictor than trust in robots. In particular, technophobia in terms of perceived inconvenience is the variable that contributed most significantly to the ability to explain the variance in Gymmy's QE. This finding supports previous studies suggesting that ease of use, i.e., the potential user's belief that the use of technology will be convenient and effortless (Heerink et al., 2010), is a solid and strong predictor of the intention to use the technology (Venkatesh, 2000), and has a direct positive impact on perceived usefulness, attitudes, and satisfaction among consumers (De Graaf & Allouch, 2013).

To enhance older adults' QE of SARs, a special emphasis should be placed on both the actual and the perceived ease of use. Facilitating actual use is challenging, since older adults tend to have various special needs often combined with restricted capabilities (Wiczorek et al., 2020; Zafrani & Nimrod, 2019). Designing SARs for this population should consider the older users' needs to make their use of the technology as easy and convenient as possible. The convenience factor should also be included in all studies examining the variables that affect the QEs of SARs among older adults. If we could reduce their concerns and convince them that SARs are convenient to use, they will evaluate them more positively and may be more inclined to accept, use, and adopt them.

The participants perceived Gymmy as providing more pragmatic value than hedonic experience and trusted its performance-based functional capabilities more than its social aspects. This finding is consistent with previous research (Frennert et al., 2017), suggesting that older adults expect robots to be functional and adjustable to their needs, which, in turn, leads to increased intensity of use and promotes successful application (De Graaf et al., 2015; Wiczorek et al., 2020). This pragmatic approach was also evident in respondents' technophobia, since their concerns regarding the possible inconvenience of using Gymmy were significantly more intense than their worries about the potential sense of failure that might arise from interacting with it. These findings correspond with previous studies (e.g., Chen et al., 2021) which indicated a positive association between ease of use and robots' functionality. Hence, when there is a discrepancy between older adults' expectation of functionality and perceived inconvenience, QEs are likely to be more moderate.

The participants' pragmatic perception of Gymmy was also implied by the regression analyses performed. The trust factor contributing the most to evaluating Gymmy was the trust in its social aspects (Tables 7 and 15). Nevertheless, the pragmatic element of use in terms of perceived inconvenience is of greater importance to the QE than the emotional aspects of use, such as the fear of personal failure and the trust in the robot's social aspects (Tables 9 and 15). These results support previous notions of older adults' preference for functional, utilitarian, stable, and easy-to-use robotic systems (Zafrani & Nimrod, 2019).

In the current part of the study, being a woman, older, and less educated was consistently associated positively with participants' QE of the expected experience of using a SAR. These findings may have resulted from participants' digital literacy and openness to new technologies. Studies have consistently found that compared to women, men have higher levels of technological literacy and skills (Saripudin et al., 2020), and that the younger among the older adults are more technology-oriented and more willing to integrate robots into their daily lives than their older counterparts (Heerink, 2011). In addition, higher levels of education are positively related to abilities to learn, use, accept and adopt modern technologies (Nimrod, 2017; Seifert, 2020).

The literature also suggests that as age increases, older adults are typically more physically and sometimes cognitively challenged (Hammer et al., 2010), and that in comparison to older men, older women have lower levels of physical functioning (Krok et al., 2013). Additionally, engagement in physical activity is more important for older adults who are more educated (Borbón-Castro et al., 2020). The video's impression on study participants who were men, younger, and highly educated may have been that Gymmy is not physically and cognitively challenging enough. Accordingly, they may have concluded that it does not meet their needs and evaluated it less than other participants. As noted in previous studies, when robots do not meet older adults' needs, the sense of trust is not achieved (Stuck & Rogers, 2018; Wiczorek et al., 2020), and the robots are perceived as less functional, leading to a negative effect on their QE (Krakovsky et al., 2021). It is worth noting that it is possible that because of these lower levels of physical functioning, older women (vs. men) specified more reasons motivating them to exercise, a variable found to be positively associated to QE.

This part of the study has methodological, theoretical, and practical implications. Methodologically, to gain better understanding regarding older adults' QE of SARs, a more holistic and comprehensive analysis of all influencing factors should be conducted. In other words, both the pragmatic and hedonic aspects of evaluation should be considered. Furthermore, the study sheds light on the importance of simultaneous exploration of the effects of both facilitators and inhibitors, such as trust and technophobia, on older adults' QE and acceptance of SARs. From the theoretical standpoint, this part suggests that inhibiting factors have a more substantial effect than facilitating factors on QE of SARs among older adults. Specifically, it advises that the impact of technophobia is greater than the impact of trust.

Therefore, from the practical standpoint, developers and designers are advised to take steps to reduce concerns about robots to promote more positive QE and eventual acceptance of SARs in later life. In a reality of accelerated population aging, where social robots are expected to play a central role (Beer et al., 2012), it is important to invest resources and efforts in developing SARs that are easy-to-use and have unique features such as a simple structure and multi-modal communication that meet the needs of older adults. These qualities should be used in any attempt to convince older adults to use SARs.

# 5. The assimilation study

## 5.1. Methods

### 5.1.1. Participants

Participants were 19 community-dwelling older adults who resided in cities (N=12) or 'kibbutzim' (N=7) in the southern part of Israel. Participants were recruited through mailing lists of retirees, public announcements, and snowball sampling. Criteria for participation were age 75 years and over, namely, the "old-old" category (Boot et al., 2020; Kubota et al., 2012) and independent living. In this part of the study, it was decided to raise the minimum participation age from 65 to 75, as the online survey revealed that age is a variable with a consistent positive association with participants' QE, and that the younger survey respondents felt that Gymmy was not challenging enough for them. Participants' age ranged between 75 and 97 (mean = 81.05, SD = 6.19). Nine participants were married, eight were widows, one was in a permanent relationship, and one was divorced. All of the participants had children (range = 2–5, mean = 3.31, SD = 1.05). The majority had secondary education, and eight had post-secondary education. Most participants were not born in Israel (N = 17), and 14 were secular. All participants were retired except for one who still worked full-time.

### 5.1.2. Procedure

All study participants were provided with a unit of Gymmy for six weeks. In-depth semi-structured interviews were conducted with each study participant at their homes before and after the study. The first session with the participants opened with oral and written explanations about the study. After signing a consent form, the participants filled out a demographic, sociodemographic, and health background questionnaire (Appendix E) referring to their gender, age, marital status, number of children, residence locality, type of residence, number of people residing with the participant, religious orientation, country of birth, number of years of education, employment status, and income level. Two additional questions assessed self-rated physical and cognitive health on a 10-point Likert scale ranging from 1 ("not at all satisfied") to 10 ("completely satisfied").

Then, each participant was given detailed explanations about Gymmy, watched a video that presented its functions and was interviewed. In these in-depth interviews, participants shared their biographical and occupational backgrounds, daily routine, and use of Information and Communication Technology (ICT). Example questions include:

"What are the main activities you do at home? Do you feel any difficulties performing daily tasks? Which? How do you deal with them?" Specifically related to robotics, participants were asked if they had any early familiarity with this field, including direct experience with SARs. The main goal of these interviews was to explore participants' expectations of SARs in general and from Gymmy in particular. Therefore, they were also asked questions about the advantages, disadvantages, risks, and benefits that they believed existed in SARs and questions that specifically focused on their expectations from the Gymmy. For example, they were asked: "What do you expect from your interaction with Gymmy?", "Are there factors that can prevent you from using Gymmy, or influence how often you use it?"; For the full interview guideline see Appendix F.

To prevent participants' exhaustion, Gymmy's installation was done several days after the preliminarily interview. This session included installation, guidance, and demonstration of a full training with Gymmy (i.e., physical and cognitive training and relaxation exercises). Throughout the study period, participants were offered unlimited technical support. In any event of malfunctions or technical problems, they could contact the support team, which would guide them on the phone and, if needed, visit them at home to fix the problem. Only two participants contacted the support team throughout the research period.

After six weeks, at the end of the study period, concluding interviews were conducted with the participants. These interviews examined their overall experience with Gymmy vis-à-vis their initial expectations. Thus, they were asked direct questions about the frequency of use, difficulties of use, and the advantages, disadvantages, risks, and benefits that they thought existed in the use of Gymmy. For example, participants were asked: "How was your daily routine with Gymmy?", "Did the use of Gymmy encourage you to be more mindful about a healthy lifestyle? How was this reflected? How do you explain that?"; For the full interview guideline see Appendix G. All interviews were audio-recorded and transcribed verbatim.

In addition to the in-depth interviews, a short weekly telephone survey was conducted with study participants. They were asked to rate their level of satisfaction with Gymmy and the extent to which they faced operational problems using a five-point Likert scale ranging from 1 ("not at all") to 5 ("a lot"). Lastly, usage reports, which included information about the frequency and usage dates, were automatically produced by the robot. Due to a limited number of Gymmy units, the data were collected in five cycles. The first four cycles included five participants, and the fifth cycle included four participants. Data were collected between December 2020 and August 2021.

### 5.1.3. Data analysis

The data analysis followed Miles and Huberman's (1994) strategies of noting patterns, making contrasts and comparisons, and clustering as detailed below.

*Noting patterns.* Miles and Huberman (1994) explained that from the data collected, it is possible to "expect patterns of variables involving similarities and differences among categories, and patterns of processes involving connections in time and space within a context" (p. 246). Thus, to identify patterns of similarities and differences between the study participants, the analysis began with within-case analysis and proceeded to cross-case analysis. In the first stage, each participant's pre-use and concluding interviews were independently coded. Then, they were compared with other participants' interviews to obtain similarities, contrasts, and overlaps in relation to the codes found. The coding process followed an inductive coding method by using open coding and axial coding techniques to make connections and group the codes into categories according to content (e.g., risks, benefits). Applying this method allowed the findings to emerge from the text without any preexisting set of concepts, codes, or ideas.

The coding process was initially performed by the Ph.D. candidate, and then it was meticulously reviewed by the supervisors. Unclear codes and discrepancies were discussed and re-analyzed, and if necessary new categories and codes were added. In this context, it is important to note that after the interviews were carefully read and initial descriptive codes and categories were generated using open coding, focused coding filtered the data until the most meaningful and frequent codes were identified.

In this study, using a meta-matrix and visualization, advocated by Miles and Huberman (1994), helped in noting patterns and themes during the qualitative data analysis process. This process contributed to the monitoring and management of data that supports the study's objectives and provided a clearer picture of the complex phenomenon under study. The comprehensive map of the categories, codes and semantic contexts is shown in Figure 9. In addition, a search was done not only for detailed descriptions but also for explanations. For example, when participants described and summarized the period of their use of Gymmy, they used many descriptions. To correctly identify patterns and themes, it was essential to clearly understand why they felt the feelings, i.e., what underlined the feelings.

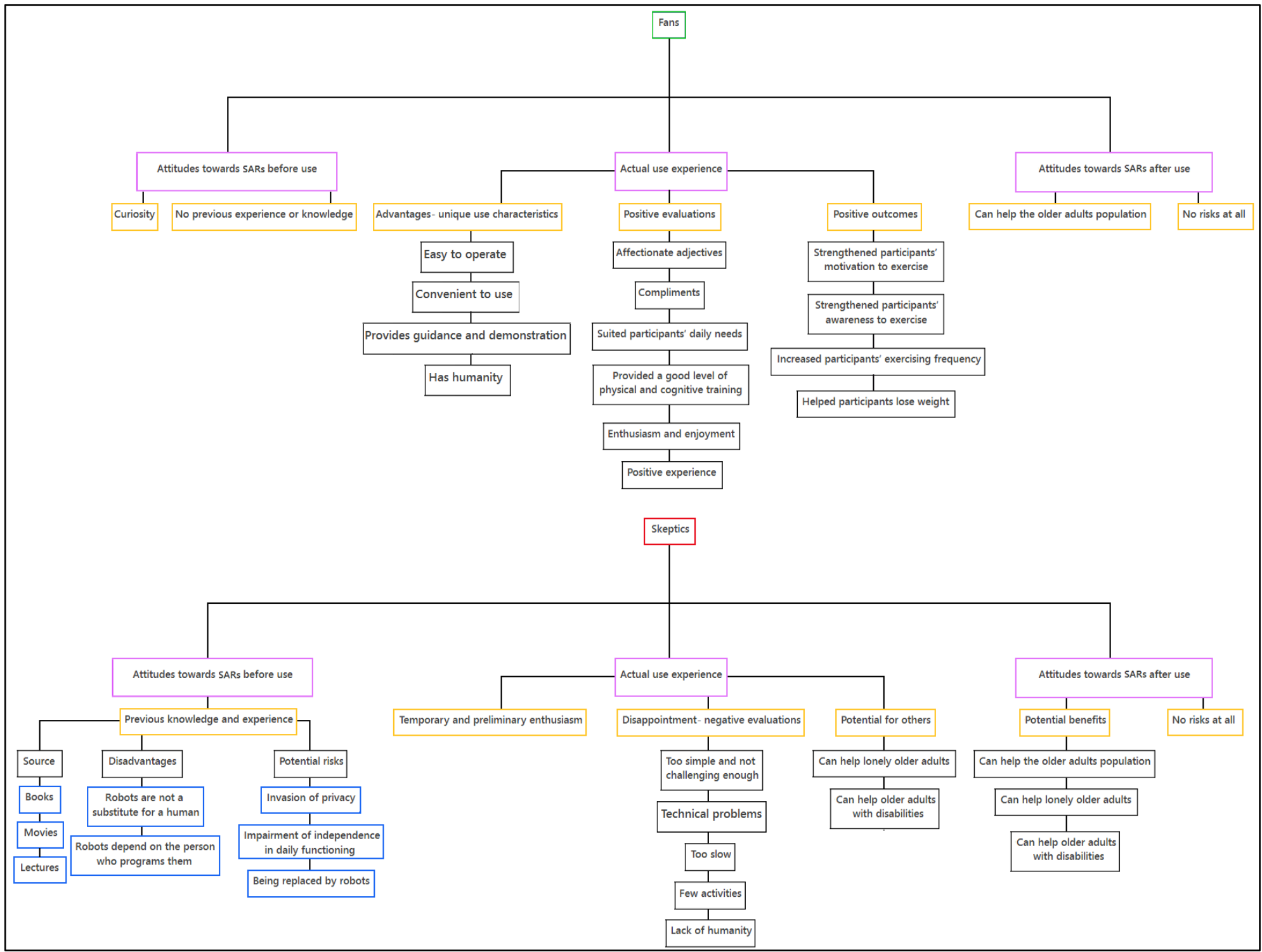

**Figure 9.** A map of semantic relations.

*Making contrasts and comparisons* is a classic way to draw and verify conclusions around emergent themes (Miles & Huberman, 1994; Williamson & Long, 2005). Drawing contrasts and making comparisons between sets of information, things, persons, activities, roles, and cases as a whole are known to differ in certain important aspects. For example, after identifying patterns and themes associated with attitudes towards SARs before using Gymmy, making contrasts and comparisons contributed greatly to concluding that coming to the study with or without early familiarity with SARs was a critical factor that influenced the feelings, mindset, and behaviors of the participants throughout the study.

*Clustering*. Miles and Huberman (1994) indicated that clustering is a tactic that can be applied at many levels to qualitative data: at the level of events or acts, of individuals or groups of individuals, of processes or cases as a whole. This tool relies on aggregation, helps in cross-case comparison, and enables to "understand a phenomenon better by grouping and then conceptualizing objects that have similar patterns or characteristics" (Miles & Huberman, p. 246). This process helps to see how the data is naturally collected into categories inductively.

In the present study, this process helped identify groups with different assimilation patterns in terms of uses, constraints, and outcomes. This was accomplished in the following manner:

- The data obtained from interviews were transcribed, and detailed line-by-line reading of each interview transcript was conducted without any a priori categories imposed on the data.
- The transcribed data were read several times to derive insights and in-depth meaning, identify patterns in the data, and form a holistic picture.
- The main concepts and themes were written down.
- Semantic units were identified and then grouped (Figure 9) in *subcategories* (shown in italics in the findings section) and 'categories' (shown in single inverted commas).
- Dendrograms of the main categories were then drawn to analyze meanings and themes deeper.

To illustrate this for the category of 'positive outcomes,' see Table 17.

Since qualitative inquiry usually requires a significant amount of data, the data analysis was supported by ATLAS.ti 9.1.5.0. This software was used to organize and interpret the raw interview data, compare data, enable the construction of concepts, and eventually formulate the overarching themes common to groups of the interviewees.

**Table 17.** *Example of Grouping of Semantic Units for Category 'positive outcomes'.*

| Semantic Units | Subcategory |
|---|---|
| "It is really a problem to move the body, and for those who work on the computer all the time it is really important to do it... I was more aware that I had to get up from the computer."<br><br>"The robot opened for me the window of real exercise, as all the exercises I did with it are basically exercises I have never done. The robot helped me by putting me in training discipline, which is good. Just like my commander in the army... it is the one who gave me the motivation to do the exercises... over time my motivation increased." | Strengthened participants' awareness and motivation to exercise |
| "Of course it added, and not at the expense of any other activity."<br>"Certainly, definitely, now more."<br>"Sure, of course."<br>"Sure, of course."<br><br>"Sometimes I went out in the sun and did some exercises alone, the same exercises ... but without the robot. Gymmy pushed me to be more active... at that time I got a bike and started using it. I decided I needed Gymmy for the arms and the bike for the legs." | Increased the frequency with which participants exercised |
| "My potbelly is gone from the activity with the robot, I actually developed some muscle in my stomach. I think my use of the robot kept me from overeating... I was satisfied with small portions of food... during this time, I felt lighter." | Helped lose weight |

## 5.2. Results

Analysis of the data collected in this assimilation study revealed two distinct groups of study participants representing two assimilation patterns (Table 18): **(A) The 'Fans'** - participants who enjoyed using Gymmy very much, trusted it, attributed added value to it, and experienced a successful assimilation process, and **(B) The 'Skeptics'** - participants who did not like Gymmy, experienced a disappointing assimilation process, and therefore expressed no interest in using it after the research period was over. The identification of these two groups occurred following the grouping of subcategories (e.g., convenient to use, technical problems) and categories (e.g., positive evaluation, negative evaluation) as and realizing that there is a consistency among the participants in the occurrence of these thematic units. The identified groups differed in their background, attitudes towards robots before and after using Gymmy, and actual use experience. Table 18 presents a summary of the factors that helped to distinguish between the two assimilation patterns. Figure 9 presents the comprehensive map of the categories, codes, and semantic contexts associated with each group. The following sections describe in detail the process of assimilation of the two groups including their characteristics, attitudes and experiences. The different assimilation patterns of the Fans and Skeptics stems from the overall distinct assimilation processes they experienced, which are presented below in three parts, attitudes towards robots before and after using Gymmy, and actual use experience.

**Table 18.** *Summary of the two types of study participants*

|  | Fans | Skeptics |
|---|---|---|
| Familiarity | No previous experience or knowledge. | Previous knowledge and experience, and familiarity. |
| Motivation | Strong. | Weak-medium. |
| Expectations | Vague. (training, energy) | Detailed. (reliability, enjoyment, diversification, without faults and bothering) |
| Concerns | None. | Detailed. (privacy, function, loss of employment) |
| Benefits | Far beyond expectations. (increased awareness, motivation, exercising frequency, enjoyment, weight loss) | Below expectations. (helpful for other people) |
| Disappointments | None. | Many. (too slow and simple, technical problems, few activities, lack of humanity) |

### 5.2.1. The Fans
#### 5.2.1.1. Personal background

The average age of the nine participants who liked Gymmy ranged between 75 and 97 years with a mean age of 82.88 years (SD = 7.57), and the mean number of years of education was 12.1 (SD = 2.89). For more information about their background, see Table 19. Most (N=7) of these participants came to the study with no previous experience or knowledge about robotics. Before the period of use of Gymmy, six of the participants in this group performed only basic physical activity several times a week, such as walking, two participants did not exercise at all, and one participant exercised in the gym. In addition, their use of various media focused mainly on traditional uses such as making phone calls and watching television.

**Table 19**. *Study participants' demographic data*

| Name | Age | Marital status | Country of birth | Number of children | Years of education | Work status |
|---|---|---|---|---|---|---|
| **Fans (N=9)** | | | | | | |
| Miley (W) | 75 | Married | Israel | 5 | 18 | Retired |
| Tom (M) | 76 | Widow | Morocco | 4 | 12 | Retired |
| Nina (W) | 77 | Widow | United States | 3 | 15 | Retired |
| Alexandra (W) | 79 | Permanent relationship | Egypt | 2 | 12 | Retired |
| Luca (M) | 79 | Married | Iraq | 4 | 11 | Retired |
| Helen (W) | 86 | Widow | Poland | 4 | 11 | Retired |
| Sami (M) | 86 | Married | Egypt | 4 | 12 | Retired |
| Paula (W) | 91 | Divorced | Austria | 2 | 10 | Retired |
| Daphna (W) | 97 | Widow | Germany | 4 | 8 | Retired |
| **Skeptics (N=10)** | | | | | | |
| Nathan (M) | 75 | Married | Slovakia | 3 | 15 | Retired |
| Clara (W) | 75 | Married | Czechoslovakia | 2 | 12 | Retired |
| Daniel (M) | 75 | Married | Austria | 3 | 16 | Full time |
| Maggie (W) | 77 | Widow | Rumania | 4 | 11 | Retired |
| Sofie (W) | 77 | Married | Rumania | 4 | 12 | Retired |
| Gabriel (M) | 80 | Married | Hungary | 5 | 18 | Retired |
| Michael (M) | 82 | Married | Libya | 2 | 15 | Retired |
| Sarah (W) | 83 | Widow | Israel | 2 | 14 | Retired |
| Joshua (M) | 85 | Widow | Hungary | 4 | 8 | Retired |
| Arik (M) | 86 | Widow | Germany | 2 | 14 | Retired |

Note: Pseudonyms were used to guarantee anonymity

### 5.2.1.2. Attitudes towards SARs before use

Seven participants from the Fans group came to the study without actual attitudes towards SARs because, as mentioned, they had no previous experience or knowledge regarding SARs. At the same time, all participants in this group had a strong sense of curiosity and a desire to experience the use of robotics. Helen (W, 86, Widow) explained: "Even at my age I still want to learn new things... I have curiosity, it's always good to know more things, and it's just interesting, to keep evolving, not to stand still," and Daphna (W, 97, Widow) shared: "I agreed to participate in the study because I am interested in new things, I am very interested in it."

### 5.2.1.3. Actual use experience

Participants in the 'Fans' group loved Gymmy very much (Figure 10; Figure 11), trusted it, appreciated its pragmatic and hedonic aspects, its 'unique use characteristics' and its 'advantages,' used it regularly throughout the study period (Figure 12), experienced 'positive evaluations' towards it, and even reported 'positive outcomes' from its use. Moreover, this group of participants directly connected Gymmy's 'unique use characteristics' and the 'advantages' they found in it. That is, its unique use characteristics are its advantages, and its advantages lie in the characteristics of its unique use. Its unique uses characteristics included pragmatic aspects such as: *Easy to operate, convenient to use,* and *provides guidance and demonstration,* and hedonic aspect such as: *Has humanity.* A powerful influence on the participants' experience was that Gymmy was *easy to operate*. The participants did not experience any use problems or difficulties during the study period (Figure 10). "It is very simple and easy, it is easy to turn it on and off," testified Paula (W, 91, Divorced), Sami (M, 86, Married) said that "anyone can use," and Miley (W, 75, Married) indicated that using it is "super easy... there is no need to be with Einstein's intelligence to operate it."

Another factor that played an essential role in the participants' experience was Gymmy's *convenience of use*. This factor contained two interrelated characteristics, Gymmy's *accessibility* and *availability*. Gymmy was placed in the most accessible place for the participants, i.e., in their homes. Therefore, they could exercise "without leaving their home" and "without investing a lot of resources." They reported that Gymmy is "accessible," "always at home," and "does not require much from you to use it." Miley (W, 75, Married) explained: "I think it's great that the robot is so accessible... for example if there is a high temperature, or I do not feel like leaving the house and drive." In the same way, Tom (M, 76, Widow) shared: "I'm glad I have a robot like this at home... on rainy days or days when I do not feel like going out at all."

Additionally, the fact that Gymmy was always available to them has allowed them to use it "whenever they want" and "as much as they want," and was a "huge advantage" for them. "I could use it at convenient times, whenever I wanted," said Tom (M, 76, Widow), Miley (W, 75, Married) described "I can use it at midnight and at five in the morning, there is no time limit," and Paula (W, 91, Divorced) shared: "I do whenever I have time, when I feel like it, in my free time and according to my will." Furthermore, Sami (M, 86, Married), for example, pointed out that "compared to a human trainer, who will not come to your home whenever you want, Gymmy is always available for you."

Another unique characteristic of Gymmy noted by these participants was that it *provided guidance and demonstration* regarding how to perform the exercises. As Helen (W, 86, Widow) avowed: "I really like getting guidance on what to do." Similarly, Daphna (W, 97, Widow) said that "it was nice that the robot told and demonstrated me exactly what to do." Finally, Gymmy's *humanity* was frequently discussed when participants regarded Gymmy as a human presence at home. For example, Paula (W, 91, Divorced) always *anthropomorphized* Gymmy when she talked about training with it and described the routine of her encounters with Gymmy as a human routine for all intents and purposes: "Every time I met him, I said hello to him, I made the movements with him, and it answered me very good," and explained: "I did the exercises according to what he said."

The 'positive evaluations' received from the Fans were for Gymmy itself, the experience of its use, and the functions it offered. First, the participants chose to describe it in many affectionate adjectives such as "so sweet," "amazing," "very nice," "pleasant," "gentle," and "friendly." In addition, they praised it with compliments for being an "excellent idea," "interesting," and "intriguing." Moreover, they indicated that Gymmy *suited their daily needs* and provided a *good physical and cognitive training level.* They experienced *enthusiasm and enjoyment* and had a *positive experience* thanks to it.

Participants described Gymmy's *suitability for their daily needs* in a variety of superlatives and explained that it "suited them exactly," "came to them in a timely manner," and was "exactly what they needed." Helen (W, 86, Widow) detailed that "Gymmy allows me to do exactly what I can, it suits me very well... it keeps me busy in a pleasant way." This adequacy between the participants' needs and the use of Gymmy, was made possible thanks to the fact that Gymmy provided the Fans with *a good level of physical and cognitive training*. From the point of view of physical training, participants in this group noted that they experienced "diversity in the type of physical activity," that "Gymmy's movements were nice," and that "the number of repetitions was good." For example, Helen (W, 86, Widow) said: "I loved all the exercises; I befriended them." Miley (W, 75, Married) shared that "the exercises were very nice, similar to things I am used to doing, like stretching the arms to the sides." The same attitudes were also reflected in the words of Luca, who indicated that "all the exercises that moved my hands were necessary, I needed them." From a cognitive training perspective, participants reported that the assignments were "good," "clear," and "without problems."

In the concluding interview, when these participants were asked to describe the period of their use of Gymmy, they pointed out that they experienced *enthusiasm and enjoyment* and had *a positive experience* thanks to it. *Enthusiasm and enjoyment* referred to positive feelings that Gymmy and its features aroused among the participants during the study period. For example, Daphna (W, 97, Widow) shared that "moving the muscles and making an effort is the best thing I can do." The enthusiasm and positive feelings that they felt caused them, among other things, to introduce Gymmy to their loved ones. Miley (W, 75, Married) said: "I showed my grandchildren how grandma exercises and they really liked it," Tom noted: "When guests arrived, I showed them how the robot works, and how I use it" (M, 76, Widow), and Daphna (W, 97, Widow) summarized that "everyone who saw Gymmy got excited."

The feelings of pleasure experienced during the study period directly affected the participants' sense at the end of the study period, which they defined as a *positive experience*. For example, Tom (M, 76, Widow) said: "This period made me feel very comfortable, I took a lot of pleasure from Gymmy... I feel much better today. In conclusion, the period with the robot was excellent and very good for me." Alex (W, 79) remembered longingly: "Gymmy made me smile in the morning when it said good morning, my name is Gymmy... I approached it happily; it was pleasant and comfortable for me. It was a good experience. I'm a little sad because Gymmy is leaving; I really like it." And Helen (W, 86, Widow) concluded: "I'm very glad I was busy with something positive, it's both fun and good for me, and it really contributed a lot to me."

As a result of the successful use they experienced, the Fans obtained several interrelated 'positive outcomes.' They reported that Gymmy strengthened their *awareness and motivation to exercise,* increased their *exercising frequency,* and helped them *lose weight.* Gymmy's presence in the participants' homes contributed to their general *exercise awareness*. As Nina (W, 77, Widow) shared: "It is really a problem to move the body, and for those who work on the computer all the time it is really important to do it... I was more aware that I had to get up from the computer."

In addition to awareness, they noted that Gymmy *motivated participants to exercise more*. Participants explained that Gymmy was actually like a training partner, one who "moves with them," "spurred," "encouraged," and "pushed them to exercise at home." As Tom's (M, 76, Widow) remarks exemplify:

> The robot opened the window of real exercise for me, as all the exercises I did with it were exercises I had never done. The robot helped me by putting me in training discipline, which is good. Just like my commander in the army... it is the one who gave me the motivation to do the exercises... over time, my motivation increased.

As the awareness and motivation to exercise increased, so did *the frequency with which participants exercised*. That is, the awareness, motivation, and Gymmy's being accessible and available to them led to an increase in the total physical activity performed by these participants during the study period. When asked by the interviewer in the concluding interview if the robot made them perform more physical activity, most of the participants in this group answered that the frequency with which they performed physical activity did increase, and shared a variety of positive responses, such as "Of course it added, and not at the expense of any other activity." – Luca (M, 79, Married); "Certainly, definitely, now more." – Helen (W, 86, Widow); and "Sure, of course." – Daphna (W, 97, Widow). Nina (W, 77, Widow) also shared: "Sometimes I went out in the sun and did some exercises alone, the same exercises ... but without the robot. Gymmy pushed me to be more active... at that time, I got a bike and started using it. I decided I needed Gymmy for the arms and the bike for the legs."

Since the participants were more active, they perceived they gained additional benefit from using Gymmy, as they felt it helped them *lose weight*. Tom (M, 76, Widow), for example, explained: "My potbelly is gone from the activity with the robot, I actually developed some muscle in my stomach. I think my use of the robot kept me from overeating... I was satisfied with small portions of food... during this time, I felt

lighter."

### 5.2.1.4 Attitudes towards SARs after use

The use of Gymmy led to the creation of a positive overall evaluation of SARs among Gymmy's Fans and positively influenced their 'perceptions.' These participants, who started the study without any knowledge of robotics and liked Gymmy, shared in the concluding interviews that they believe that using SARs "can undoubtedly" *help the older adults population.* "For older adults? For sure! In a thousand percent, it will make a great contribution to a person," stated Tom (M, 76, Widow). The same attitude was also reflected in the response of Miley (W, 75, Married): "Of course, of course, there is no question at all. By 100 percent." Furthermore, Alex (W, 79) shared: "My experience with Gymmy built-up my tolerance for living with robots in peace," and Luca (M, 79, Married) highlighted that "robotics can benefit older people in many areas... it can save time, money... it can only be beneficial."

Further evidence that the use of Gymmy has positively affected the perception and evaluation of SARs among this group of participants stemmed from the question about the dangers and risks of robots asked during the concluding interviews. During these interviews, participants were asked if, after using Gymmy, they thought that SARs (in general) could be dangerous. All the participants, without exception, indicated that "there are no risks at all," only "positive things." Luca (M, 79, Married), for example, elaborated: "I do not see an option that robots are dangerous... If humans invented them, they also know how to control them."

### 5.2.2. The Skeptics
#### 5.2.2.1. Personal background

The average age of the ten participants in the Skeptics group ranged from 75 and 86 years, with a mean age of 79.4 years (SD = 4.41). This group's mean number of years of education was 13.5 (SD = 2.28). Hence, they were somewhat younger and more educated than the Fans. For more information about their background, see Table 19. Compared to the first group, most (N=8) of the Skeptics also came to the study with knowledge, previous experience, and familiarity with the world of robotics. Maybe, as a result, they were characterized by skepticism expressed in the fact that they mentioned many disadvantages that they believed existed in the use of SARs and a variety of potential risks that may result from this use during the preliminary in-depth interviews.

The exercise habits of most of the Skeptics before the study were extensive and diverse and included activities such as swimming, biking, and practicing Feldenkrais. Finally, the media use of all participants in this group included both basic uses (e.g., making phone calls and watching television) and more advanced uses of ICTs such as computer software (e.g., Microsoft Office and games), tablet, and internet (e.g., social networking sites, online shopping, and YouTube).

### 5.2.2.2. Attitudes towards SARs before use

Eight participants from this group joined the study with previous knowledge and experience with SARs. The accumulated knowledge came from "books," "movies," and "lectures on the subject," and as a result of experience shared with them by their acquaintances. For example, Gabriel (M, 80, Married) noted: "I heard about robots, I read about robots." Similarly, Sofie (W, 77, Married) shared: "My son has an iRobot that has been cleaning their house for five or six years." Furthermore, Daniel (M, 75, Married) testified: "I am a technophile at heart, I have worked with robots... I saw all kinds of robots, I saw robots in a hospital, I saw robots on Facebook."

Three of the participants in this group indicated that SARs may "be human-friendly" and have the potential to help older adults "clean the house" and "remind them things." However, they rarely talked about the potential benefits, and it seemed that their early familiarity and being 'knowledgeable' about robotics made them come to the study with a sense of skepticism. Sofie (W, 77, Married), for example, noted: "At some point, they will be able to control the human race, I'm sure... we need to be well-guarded so that we do not reach the moment when they will be able to control us." Furthermore, Michael (M, 82, Married) shared: "We must find a way to balance the wisdom of the robots so that they cannot do everything... otherwise we will close the hospitals, kill the people, and use only robots," and summarized: "I hope I can get along with Gymmy." Following this feeling, they described several disadvantages that characterize the use of SARs and risks that they believed may be caused by this use. The 'disadvantages' included: *robots are not a substitute for a human,* and *robots depend on the person who programs them.*

*Robots are not a substitute for a human* was frequently discussed when comparing the interaction with a robot to that with a human. Participants repeatedly emphasized that "robots lack human contact." Michael (M, 82, Married), for example, explained that "a robot cannot replace humans in cases where contact is needed." Furthermore,

Maggie (W, 77, Widow) said: "There is no substitute for a look in the eyes and a hug, for laughing together, for all the things that a human being gives." In addition, the Skeptics explained that SARs cannot be a worthy substitute for humans as they are unable to experience and express emotions while "human and emotion cannot be separated," as highlighted by Clara (W, 75, Married).

The participants noted that, like other modern technologies, *robots depend on the person who programs them* for better or worse. Therefore, participants said they hope the programmer has lofty goals and aims to help users. Alongside the hope, however, there were also doubts, as Clara (W, 75, Married) explained: "They can definitely be misused... people can build robots for the purpose of harm." Similarly, Nathan (M, 75, Married) detailed: "It all depends on the programmer... he can take the robot and use it to harm the citizens of populations."

Along with the disadvantages, participants also noted three potential 'risks' that may result from the use of SARs: *Invasion of privacy, impairment of independence in daily functioning,* and *risk of being replaced by robots. Invasion of privacy* was discussed in terms of informational privacy. That is, participants shared their concerns about their ability to understand how the information shared with the robot is processed and used (or misused). For example, Nathan (M, 75, Married) said: "Actually, I have no idea what the robot is doing, and I am afraid there will be an invasion of my private life."

The second risk discussed by the participants was *impairment of independence in daily functioning,* both physically and cognitively. From a physical perspective, this risk is derived from the participants' concern of being replaced by robots in daily activities at home. Namely, despite the potential of robots to support older adults' autonomy, the participants claimed that robots may actually threaten their autonomy by replacing them in home tasks they would be better off performing themselves. Sarah (W, 83, Widow), for example, explained: "I would not want a robot... At this point in my life, I love to serve myself." Sofie (W, 77, Married) clarified: "I get along great without it, I leave it for the future." From a cognitive perspective, the fact that robots are an extremely 'intelligent' technology that can efficiently perform cognitive actions was a potential risk according to participants' perception. They claimed that the robots may threaten users' cognitive abilities in knowledge tasks they can perform on their own, as Gabriel (M, 80, Married) shared: "We will think less, and the robots will think for us."

The *risk of being replaced by robots* was discussed in the employment context. "This is a pretty difficult problem for the world... if robots come into our lives... people will lose their jobs etc.," explained Sarah (W, 83, Widow). Michael (M, 82, Married) wondered: "If robots will do all the jobs, what will people do?" He even portrayed the future as an apocalyptic scenario and explained: "Robots pose a danger of replacing people... you will have no job, and you will have nothing." Similarly, Clara (W, 75, Married) mentioned that "ever since I was young, it was said that a day will come and robots will replace humans," and Sofie (W, 77, Married) confessed, "this is my biggest fear."

### 5.2.2.3. Actual use experience

The first days of using Gymmy can be described as a success among the Skeptics. The participants learned how to use it, were satisfied with it overall, and experienced enthusiasm. However, the excitement was only temporary and preliminary, a finding reflected in their words and the usage data produced by Gymmy (Figure 12; For the full usage data, see Appendix H). For example, Daniel (M, 75, Married) explained: "At first it was nice... it was a new experience to play with a robot." Similarly, Michael (M, 82, Married) reported: "I was enthusiastic about Gymmy only at the beginning." In fact, the initial enthusiasm cooled down and turned into disappointment, which was reflected in the participants 'negative evaluations' (Figure 10). These evaluations were due to several pragmatic factors such as: *too simple and not challenging enough, technical problems, too slow,* and *few activities,* and as a result of a non-hedonic factor*: lack of humanity.*

The biggest disappointment from Gymmy among the Skeptics was that the participants, who were accustomed to performing a wide range of physical activities, perceived its use as *too simple and not challenging*, and even boring. As Clara (W, 75, Married) said: "I was expecting something more challenging, but it was not... it is too simple for my abilities." Daniel (M, 75, Married) mentioned that: "It was not a challenge... it was very easy," and Gabriel (M, 80, Married) summarized: "It's extremely boring." This was true for the physical training: "The exercises were too simple... the movements were very easy" (Sofie, W, 77, Married), for the cognitive training, which was also "at a very low level, like the level of first grade" (Nathan, M, 75, Married), and for the relaxation exercises: "they are very anemic, they are not serious" (Gabriel, M, 80, Married). This simplicity and lack of challenge made the participants feel that the use of Gymmy was "somewhat frustrating," "nagging," and "not fun."

It should be noted that throughout the study period, participants in this group expressed a somewhat lower level of satisfaction in the weekly surveys, although not by a significant margin from the 'Fans.' This small gap could result from social desirability and the wish not to disappoint the researchers. This explanation is supported in the words of three participants, who indicated that they felt a commitment to using Gymmy. Therefore, they continued to use it throughout the study period. Maggie (W, 77, Widow) explained: "I felt a commitment... I did not want to disappoint whoever is responsible for it." Similarly, Gabriel (M, 80, Married) mentioned that he kept training with Gymmy, as he "wanted to complete the task to the end," and Sofie (W, 77, Married) confessed, "I did because I promised." Nevertheless, in the concluding interviews, the participants' sense of disappointment and dissatisfaction with Gymmy was clearly evident (Figure 10). Due to the small sample size, the differences suggested by the weekly surveys are not significant. However, they offer some quantitative support for these differences between the two groups (Figure 11).

Another major issue noted by the participants was *technical problems*. As a new technological product, Gymmy had some faults. The Skeptics showed less patience for these faults, a tendency that was clearly reflected in their weekly reports (Figure 11) and directly affected their sense of disappointment. "The difficulty is that there were some exercises that the robot did not catch my movements," noted Nathan (M, 75, Married). Gabriel (M, 80, Married) shared: "It happened that the robot froze during the exercise, I waited and saw that nothing improved... then I turned it off." The same attitudes were also reflected in the words of Arik (M, 86, Widow): "There were some technical issues… It was hard to find the right place to sit."

*Too slow* referred to the time it took Gymmy to turn on and perform the exercises and the waiting time between them. Maggie (W, 77, Widow), for example, noted that "it takes a long time for Gymmy to wake up... I noticed that the operating time until it stabilizes takes more than a minute." In addition, Daniel (M, 75, Married) said: "The exercises should be much faster... more vigorous." Nathan (M, 75, Married) commented that the waiting time between exercises "is very long, so I did a run-in-place activity meanwhile," and Clara (W, 75, Married) argued: "The breaks between exercises take too long for my liking."

*Few activities* were discussed in the context of two aspects: First, regarding a too-small pool of physical exercises, and second, regarding the fact that the physical training was for the upper limbs only. In terms of the first aspect, several participants expressed their disappointment that the physical training time was too short. "I just started, and immediately it was over," indicated Arik (M, 86, Widow). Michael (M, 82, Married) summarized: "I was disappointed that the robot could not give more than four-five exercises." In terms of the second aspect, participants perceived the fact that the physical training is for upper limbs only as a disadvantage, were disappointed by this, and hoped for additional activities that would allow them to train their lower body. For example, Sarah (W, 83, Widow) said: "The exercises are minimal... the truth is that it is quite limited, there is no activity with the legs," Clara (W, 75, Married) mentioned that "the lower part of the body did not participate... it was completely missing," and Maggie (W, 77, Widow) shared: "There is a need for more exercises... to move the legs."

*Lack of humanity* as a factor that caused a negative evaluation among the Skeptics stemmed from the fact that they expected the communication with Gymmy to be as similar as possible to human-human interaction. Therefore, when expectations did not match reality, most experienced disappointment. Daniel (M, 75, Married), for example, described that "it would have been perfect if it had said, 'Come on, Daniel, time to practice.'" In almost the same way, Maggie (W, 77, Widow) explained: "The only thing I wanted was that the robot would talk to me... but it did not talk." Sofie (W, 77, Married) clarified: "Robots cannot be used in cases where human contact is required." Additionally, Nathan (M, 75, Married) explained that "compared to a human trainer, a robot does not provide you with real feedback." In this context of training with a human trainer, Joshua (M, 85, Widow) pointed out that "it is much nicer... not just practicing... there is also a conversation; we also talk during training."

Following all the factors listed above, most Skeptics felt that Gymmy "does not provide any added value" and that "there is no novelty in it." Accordingly, Gymmy "did not live up to the expectations," and they felt "no desire for permanent adoption." For instance, Sarah (W, 83, Widow) said she "had exhausted the use... it was enough for me," and Clara (W, 75, Married) concluded sarcastically: "There was nothing special here, I did not shed a tear because Gymmy was leaving."

Notwithstanding the disappointment these users experienced from Gymmy, and despite their unwillingness to adopt it permanently, their attitude towards it after use was that it had a 'potential for others.' Although Gymmy was less relevant for them, they believed that it could certainly help lonely older adults and *older adults with disabilities. Help for lonely older adults* referred to the perception that Gymmy can act as a kind of companion or friend. As Joshua (M, 85, Widow) noted: "It is very important for lonely older adults, for example for people who have been widowed and are alone... it can help alleviate loneliness." Gabriel (M, 80, Married) explained similarly: "This robot can be an advantage for lonely persons, who do not exercise and have no one to train them... Gymmy can definitely guide them and make them exercise."

Another perception that participants shared about Gymmy was that it could *help older adults with disabilities.* The fact that Gymmy is placed at home and can be used at any time allows "people who cannot leave their homes" or "people who are confined to a wheelchair" to exercise, as explained by Clara (W, 75, Married): "Gymmy is suitable for people who have some kind of disability or had stroke... people with less physical or cognitive abilities." Daniel (M, 75, Married) added: "It can strengthen the mobility of people with disabilities or in a wheelchair... people who have a hard time, who hardly move," and Maggie (W, 77, Widow) summarized: "I think this robot can make a very big contribution to people with disabilities who cannot leave their house."

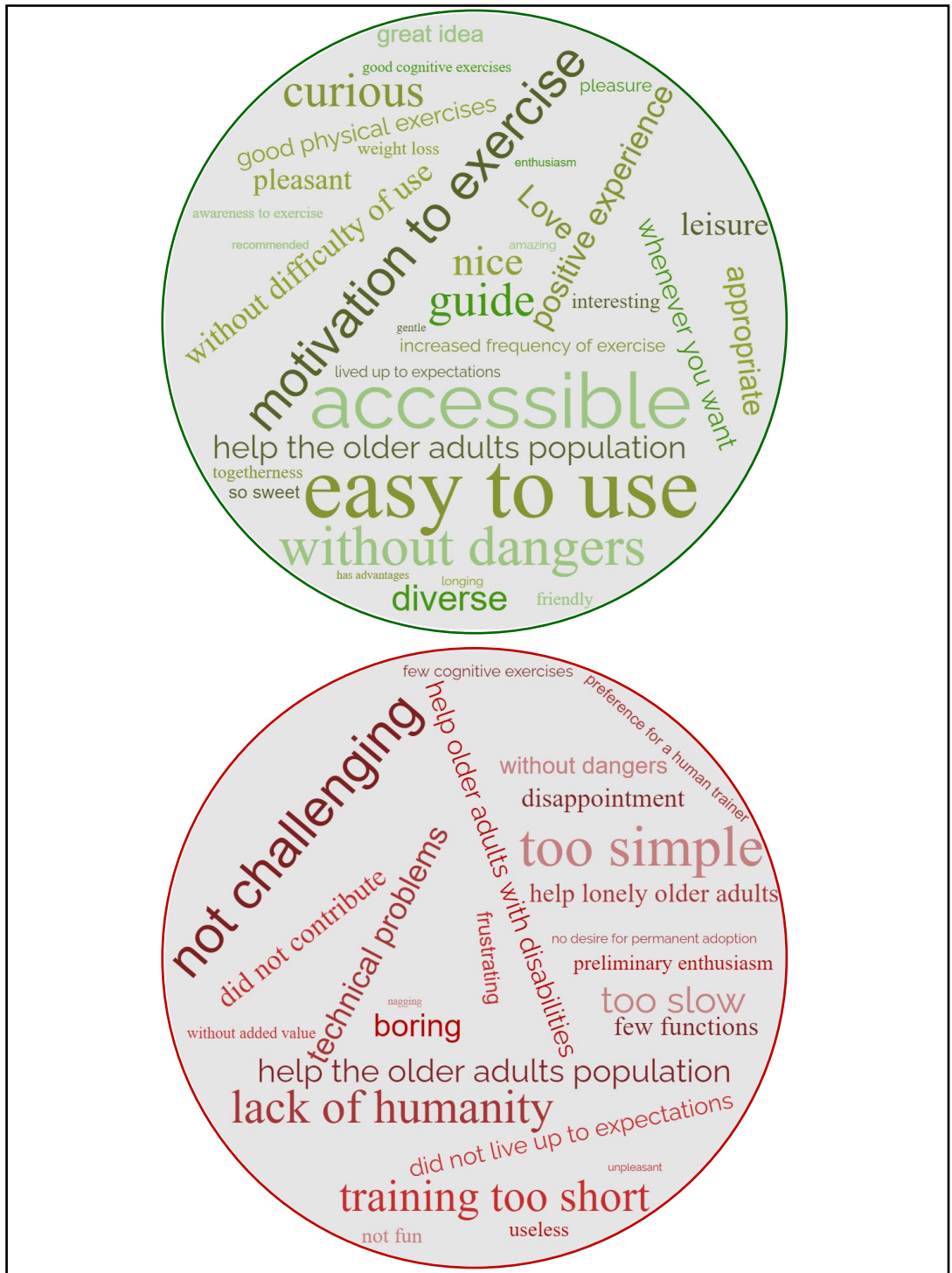

**Figure 10.** Word clouds that describe the actual use experience of the Fans (green) and the Skeptics (red).
Note: The bigger the word in the word cloud, the higher frequency it appears in the interviews.

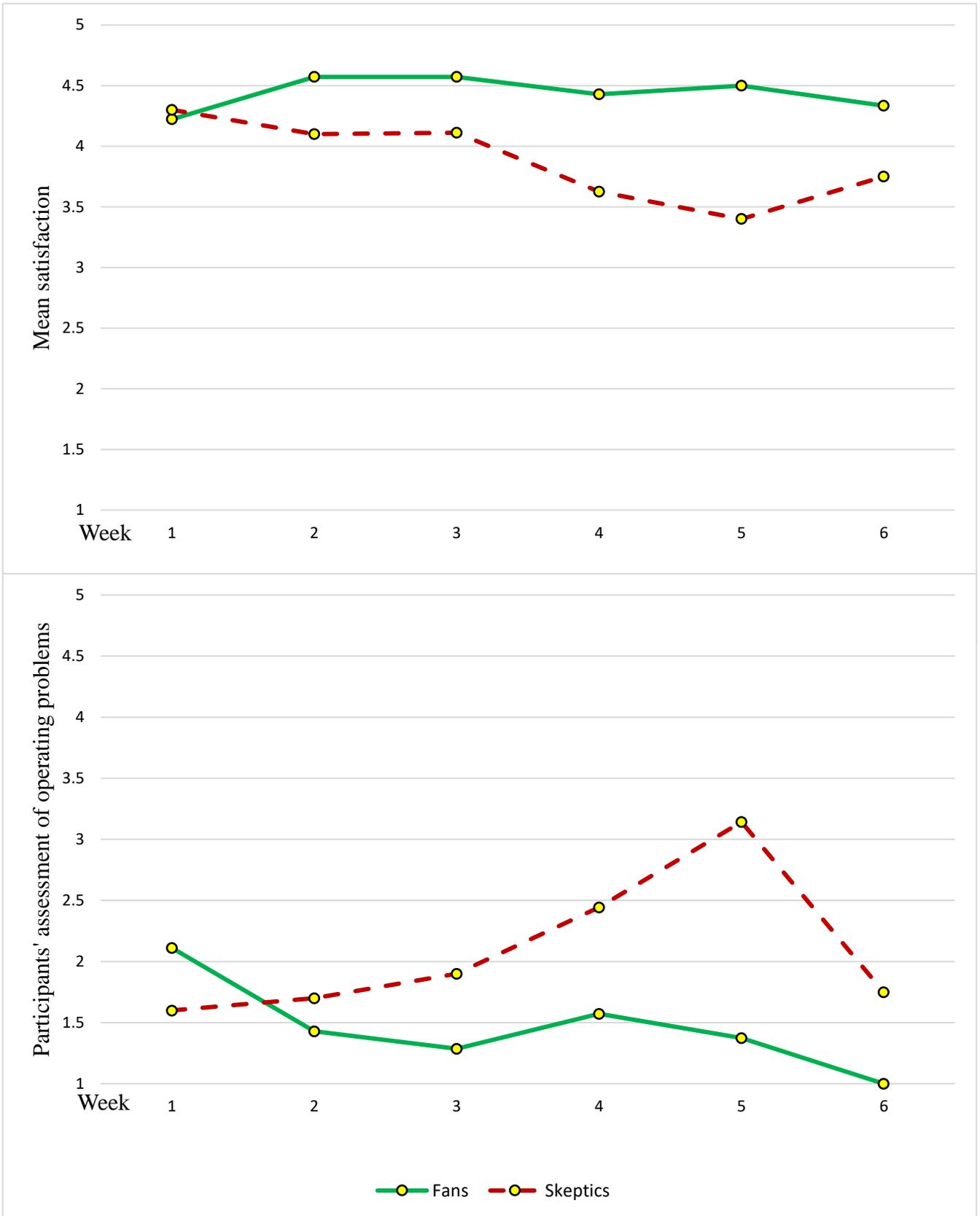

**Figure 11.** Six-weeks use trends according to the two types of study participants.

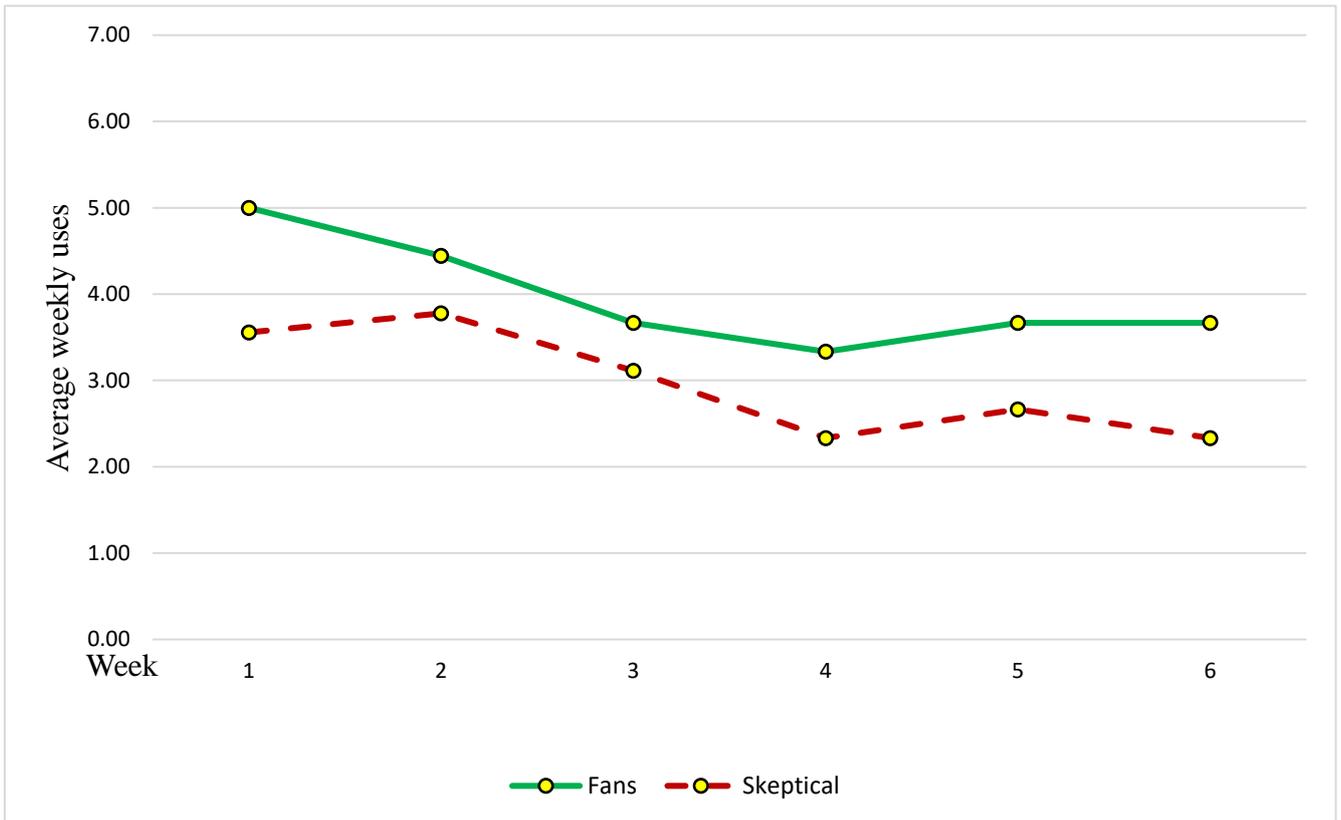

**Figure 12.** Average weekly uses according to the two types of study participants.

### 5.2.2.4. Attitudes towards SARs after use

The sense of skepticism reflected by this group in the opening interviews seemed to impact their evaluation of robots after the study period. Nevertheless, the disadvantages and risks that this group of users associated with SARs appeared to be somewhat moderated by their experiences with Gymmy. Accordingly, their evaluations of SARs at the end of the study were, in most cases, a little more positive than they were before the study. For example, Maggie (W, 77, Widow) explained at the beginning of the study period that "robots have advantages and disadvantages, but you have to get used to it, it's something so new." At the end of the study, when asked if, after using Gymmy, she is more open to future experiences with robots, she replied: "I think so... I'm open to that." Similarly, Clara (W, 75, Married) answered this question in the concluding interview by noting: "It piqued my curiosity, definitely." This attitude was considerably different from the one expressed in the opening interview, where she said that robots could be dangerous and even counted their disadvantages.

Not all Skeptics, however, demonstrated such greater openness and better evaluations of robots after the study. For example, in the opening interview, Sofie (W, 77, Married) said: "A robot might help, but I also get along without it... I get along great without it." In the concluding interview, she maintained the same mindset: "I never had any issues that required the use of robots."

Notwithstanding the differences mentioned above, all Skeptics noted 'potential benefits' they believed may result from using SARs. The benefits included *assistance to the older population in general* and *to lonely older adults and older adults with disabilities in particular.* Similar to the Fans, but less decisively, the Skeptics also believed that using SARs could help the older population. For example, Daniel (M, 75, Married) said: "They can help people who are no longer like they used to be." At the same time, their focus was on sub-populations within the older population. Like their attitude towards the specific potential of Gymmy, they believe that SARs can help mostly *lonely older adults* and *older adults with disabilities.*

*Assistance to lonely older adults* refers to the potential of SARs to help those older adults by "encouraging them" and giving them the feeling that "someone is talking to them" and "being with them at home." As Gabriel (M, 80, Married) explained: "It's wonderful for assistance needs... to a lot of lonely people. A robot that is connected to some kind of network, if a person fell in the shower, the robot can warn about it." Another potential benefit related to the ability of SARs *to help older adults with disabilities*. Some participants emphasized that especially for older adults "with disabilities," "with physical impairment," and "those who cannot leave their homes," SARs "certainly have the potential to help." Finally, when the Skeptics were asked if, after using Gymmy, they think SARs can be dangerous, apart from two persons, all the other participants stated that they were not afraid of using them or thought they might be hazardous. For example, Daniel (M, 75, Married) replied: "Dangerous? I do not believe they can be dangerous... I'm not afraid of robots."

Since their overall evaluation of SARs has not diminished, it seems that the inherent potential that the Skeptic participants identified in Gymmy neutralized the adverse effects that might have resulted from the sense of disappointment they experienced. Evidence of this stems from the fact that there was not even a single participant who evaluated SARs after the study period less than he/she did initially. The disappointment the Skeptics felt from the specific robot, Gymmy, simply reduced its relevance for them. It also led them to the insight that it is necessary to tailor the robot's ability to the person for whom it is intended, all the more so in a situation where there are adults that robots can help. As Clara (W, 75, Married) pointed out: "I emphasize, if the robot is tailored to the needs of the user... this is the formula, you cannot give someone a robot that does not suit his needs."

### 5.3. Discussion

Following the users' experiences of the Fans and the Skeptics in real-life conditions and over time allowed to identify two distinct assimilation patterns in terms of uses, constraints (both antecedent and intervening), and outcomes (including benefits and disappointments). The lengthwise assimilation patterns described in the above findings section were derived from the differences between the Fans and the Skeptics in their attitudes towards robots before and after using Gymmy and their actual use experience. The two patterns suggested that the process of SARs' assimilation is not homogeneous and provided a more profound understanding regarding the factors affecting older adults' QE of SARs following actual use. Below is a discussion of the two assimilation patterns vis-à-vis the three main topics explored with regard to HRI in later life, namely: uses, constraints, and outcomes (Zafrani & Nimrod, 2019).

*Uses*. The literature suggests that the uses category includes acceptance, adaptation, and factors affecting user experience. In the present study, the two groups of users reported a completely different experience. The Fans have benefited positively from Gymmy's use mainly due to its pragmatic use characteristics, which they have experienced as easy and convenient to operate and use. The Skeptics experienced the same robot with the same use characteristics as too simple, unchallenging, and boring.

A possible explanation for this gap in the user experience lies in the participants' exercise habits before the study. Compared to the Skeptics, whose exercise habits were extensive and diverse, the exercise habits of the Fans were only basic and mainly included walking. Gymmy provided the Fans with a new value-added function to their daily routines, i.e., physical activity, which they did not have or had very limited until the study period. In contrast, this function was adequately implemented in the Skeptics' daily routines. Accordingly, they did not attribute added value to the use of Gymmy, which, in turn, led to decreased intensity of use.

This explanation is consistent with previous research suggesting that user attributes significantly affect the user experience (e.g., Morillo-Mendez., 2021; Wu et al., 2016). It also echoes studies demonstrating that older adults expect robots to be tailored to their needs (Tsardoulias et al., 2017; Karkovsky et al., 2021). If they cannot ascribe new valuable functions to the robot, they will evaluate the interaction with it less favorably and eventually abandon its use (Frennert et al., 2017; Torta et al., 2014). Moreover, similar to the findings reported by Šabanović et al. (2013), giving Gymmy a function in the Fans' daily routines encouraged its acceptance and increased use intensity. However, if Gymmy could adapt itself to the users' fitness level, abilities, and needs, it is probable to assume that even the more physically active users (i.e., the Skeptics), would also be challenged when using it and even have a more positive experience. Consequently, it could even somewhat reduce the negative influence of technophobia among the Skeptics. Attributing human traits (Frennert et al., 2017; Onnasch & Roesler, 2021) to Gymmy was another factor that positively affected its acceptance by the Fans. Among the Skeptics, this factor caused a negative evaluation since their expectation that the interaction with Gymmy to be as similar as possible to human-human interaction did not materialize.

The present study's findings supported previous research, according to which direct experience with SARs promotes acceptance (Shen & Wu, 2016). Most of the participants in the Fans group came to the study without any explicit attitudes towards SARs because, as mentioned, they had no prior familiarity with the field of robotics. However, at the end of the study period, they developed positive attitudes towards SARs in general. They argued that SARs are not dangerous and can undoubtedly improve older adults' quality of life. Despite the disappointment from Gymmy, the direct interaction with it even reduced the ambivalence towards SARs among the Skeptics. These participants came to the study with firm attitudes towards SARs and mentioned a variety of disadvantages and risks that they believed associated with the use of SARs. Nevertheless, their overall evaluations of SARs at the end of the study period have not diminished, and, in most cases, they were even slightly more positive than they were beforehand. Furthermore, not even a single skeptic participant evaluated SARs after the study period less than they did initially.

*Constraints.* The discussion of constraints refers solely to the Skeptics, as the Fans did not report any constraints. The gap between the Fans and the Skeptics resulted not only from the differences between them in terms of exercise habits. It also stemmed from a most significant antecedent constraint found among the Skeptics. As stated above, unlike the Fans, most of the participants in the Skeptics group joined the study with previous attitudes towards SARs, which were constructed by contents to which various media exposed them. In such contents, robots are often demonized and presented as attempting to take over the world and replace humans (e.g., the Terminator; Bartneck et al., 2007; Lee et al., 2012). These negative connotations, in turn, can trigger negative attitudes and emotions towards robots (Lee et al., 2012).

In his robot novels, Isaac Asimov defines these concerns as the "Frankenstein complex" (Asimov, 1978). This phenomenon describes the anxiety that artificial robots could become competitors of humans and rebel against them (Trovato et al., 2013). Indeed, in the opening interviews, the Skeptics shared that they believed in a prophecy of wrath, which holds that robots will be able to control the human race at some point. However, this gap between the Fans and the Skeptics may also stem from the characteristics of the sample. As mentioned above, the Skeptics were somewhat more educated than the Fans. It is thus reasonable to argue that older adults with a higher level of education and prior knowledge of robotic technologies would be characterized by a more realistic perception and greater awareness of their shortcomings. At the same time, the education variable may also affect the degree of exposure to content in various media (Huffman, 2018). Higher levels of education are positively related to the ability to acquire and absorb information and content from modern technologies (Simoni et al., 2016). Therefore, the previous attitudes of the Skeptics, which may evoke negative connotations towards robots, may result from exposure to content in the various media, from being more educated, and/or as a result of the correlation between these variables.

Following their belief in the prophecies of wrath, the Skeptics pointed out several disadvantages that characterize the use of SARs and potential risks that they believed may be caused by this use. The division between disadvantages and possible risks was made to emphasize that, in the opinion of the Skeptics, the disadvantages certainly exist in the use of robots. In contrast, the potential risks, as they are called, are only potential. Three potential risks raised by the Skeptics that are consistent with the existing literature were invasion of privacy (Kernaghan, 2014; Sharkey & Sharkey, 2012; Vandemeulebroucke & Gastmans, 2021), impairment of independence in daily functioning (Beer et al., 2012; Jenkins & Draper, 2015), and risk of being replaced by robots (Calvert, 2017; Cobaugh & Thompson, 2020; Goudzwaard et al., 2019; Vlachos et al., 2020). Another potential risk arising from the Skeptics' perceptions is that robots depend on those who program them. Thus, they can take them and use them to harm people. This finding is consistent with previous literature, which holds that the older adult population has an ingrained fear of making harmful use using modern technologies such as computers and robots (Brosnan, 2002; Di Giacomo et al., 2019). These concerns about potential risks reflect emotions associated with technophobia, which refers to an "exaggerated, usually inexplicable, and illogical fear" (Sinkovics, et al., 2002, p. 478) of using modern technologies.

Technophobia consists of two chief components: Fear of using technology and concerns regarding technology's effects on society (Osiceanua, 2015). The concerns that bothered the Skeptics before the study were clearly associated with the second component. These concerns seemed to have adversely affected the Skeptics' actual use experience of Gymmy by limiting their motivation to use it and causing them to perceive it as less attractive and useful. The Fans, who did not have concerns and fears before the study, expressed positive evaluations towards Gymmy and indicated that using it suited their daily needs and provided a good level of physical and cognitive training. The Skeptics, in contrast, experienced disappointment and negatively evaluated Gymmy. This finding is consistent with previous studies that have demonstrated a negative association between technophobia and digital enthusiasm (Anderberg et al., 2019; Di Giacomo et al., 2019).

Another antecedent constraint found among the Skeptics came from the stigma associated with using a robot in old age (Bradwell et al., 2021; Neven, 2010; Pripfl et al., 2016), which is one of the most dominant constraints found in the literature on HRI in later life (Zafrani & Nimrod, 2019). In the opening interviews, these participants stated that they do not need any robotic assistance at this stage in their life and that they leave this assistance for the future, if ever. Hence, the Skeptics tended to perceive the prospective robot user as a much older person who needs substantial support with everyday tasks. This negative perception limited the Skeptics' motivation to use SARs even before using Gymmy and may have helped them dissociate themselves from ageist stereotypes.

Aside from the antecedent constraints, intervening constraints found were technical problems and slow operation. The Skeptics showed less patience for these issues than the Fans—a finding that has been reflected in the concluding interviews and directly affected their negative evaluations of Gymmy. These issues harmed the use experience even among the Skeptics who were initially quite curious about Gymmy. They constituted a buffer between the desire to use Gymmy and the realization of this desire. These findings support previous literature indicating that usability, including various operative difficulties in robot performance, constitute a significant intervening constraint (Fischinger et al., 2016; Pripfl et al., 2016; Wang et al., 2019) that leads to dissatisfaction and negative feelings such as frustration among older users (Begum et al., 2013; De Graaf et al., 2015; Pripfl et al., 2016; Wang et al., 2019).

*Outcomes.* The discussion of outcomes is divided into negative and positive outcomes (i.e., benefits). Similar to the constraints, the entire negative outcomes category refers to the Skeptics. The negative use outcomes of the Skeptics were reflected in the fact that throughout most of the study period, they experienced negative feelings of boredom, frustration, despair, lack of enjoyment, patience and interest, resentment, and dissatisfaction that resulted from the disappointment they experienced from using Gymmy. This sense of disappointment experienced by the Skeptics from Gymmy's use stemmed from its performance, the functions it offered, and from the fact that their longing for the interaction with it to be as similar as possible to human-human interaction has not been realized.

Besides describing the negative outcomes, the participants addressed the benefits gained from the period of using Gymmy, and potential benefits that they believe exist in SARs as a result of this period. Here, the division between the Fans and the Skeptics is noteworthy: Whereas the benefits mentioned by the Fans were benefits for themselves, the benefits described by the Skeptics were benefits for others. First, with Gymmy, the Fans had a positive experience and noted that they loved Gymmy and enjoyed using it. These positive feelings have also been found in previous research on HRI in later life suggesting that interacting with robots was experienced by older adults as an enjoyable activity (De Graaf et al., 2015; Fischinger et al., 2016; Lazar et al., 2016) that had positive effects on their mood (Khosla et al., 2012; McGlynn et al., 2017) and wellbeing (Henschel et al., 2021). The Fans noted additional interrelated benefits related to the central function that Gymmy provides, physical training. They reported that Gymmy strengthened their awareness and motivation to exercise, increased their exercising frequency, and even helped them lose weight. Obviously, these benefits were significant for them as their exercise habits before the study were relatively negligible.

For the Skeptics, Gymmy's specific potential benefits are the same as the potential benefits of SARs in general. They believed that both Gymmy and SARs could assist the older population, especially the lonely older adults and older adults with disabilities. It can be assumed that this perception is related to the antecedent constraint mentioned above regarding the stigma of this group associated with using a robot in old age. Like the Skeptics, but in a more decisive manner, the Fans also believed that using SARs could help the older population. However, this group did not emphasize that this assistance potential is directed primarily at "vulnerable" older.

Overall, the application of simultaneous exploration of uses, constraints, and outcomes (including positive and negative effects) over time and in real-life conditions explained how these topics correlate with one another and presented a broader and more accurate picture of factors influencing older adults' evaluation of SARs. Specifically, this exploration explained how the QE may vary according to the assimilation pattern, the factors affecting the use and the benefits gained, and the constraints to beneficial use. Moreover, the analysis of the data showed that there is an inevitable connection between these three topics, as the uses are affected and affect the constraints, and the reciprocity between these two topics affects the outcomes.

The broader and detailed analysis provided by the present study yielded three important conclusions. Briefly put, the conclusions are: (a) An actual use of SAR may increase evaluation of SARs among both skeptic and sympathetic older adults, (b) Skepticism moderates the positive impact of use on the evaluation, and (c) Disappointment from a specific robot does not necessarily detract from the general evaluation of SARs, but only reduces the relevance.

# 6. Discussion and Conclusion

Investigating the factors that promote and hinder older adults' Quality Evaluation (QE) of Socially Assistive Robots (SARs) is a critical step that may facilitate their acceptance and assimilation (Andriella et al., 2021; Frennert, 2019). To this end, this dissertation explored how the QE of SARs among older adults is shaped by both anticipated and actual interaction. First, a video-based acceptance study conducted through an online survey simultaneously examined the effect of trust and technophobia on older adults' QE of SARs following anticipated interaction using quantitative analysis. Then, an assimilation study explored how the QE is shaped following actual interaction with a SAR by a simultaneous exploration of uses, constraints, and outcomes in real-life conditions over a long period in a qualitative study. In this section, the results of the two parts of the study are discussed integratively.

## 6.1. Bridging the gaps in the existing literature
### 6.1.1. Simultaneous exploration of trust and technophobia

The simultaneous exploration of trust and technophobia yielded new insights regarding the factors predicting and influencing QE of SARs in later life. First, similar to existing literature (Kellmeyer et al., 2018; Lewis et al., 2018; Martelaro et al., 2016; Naneva et al., 2020; Syrdal et al., 2009; Tussyadiah, et al., 2020), both the online survey and the assimilation study indicated positive associations between trust and QE and negative associations between technophobia and QE.

Second, this study highlights the great importance of investigating technophobia in HRI studies, which, contrary to trust, has hardly been examined. The regression analysis conducted as part of the online survey indicated that robot-related technophobia is a stronger predictor than trust in robots, and that technophobia in terms of perceived inconvenience is the variable that contributed most significantly to the ability to explain the variance in Gymmy's QE. This finding is consistent with the results of the assimilation study, in which the Skeptics, who expressed technophobia in the opening interviews, noted their disappointment with Gymmy, which led to negative evaluations. The Skeptics' disappointment resulted from several pragmatic factors they experienced while using Gymmy such as the SAR being too simple and slow, not challenging enough, and having technical problems. All these factors represent a sense of inconvenience that the Skeptics felt during the use Gymmy. In contrast, the Fans, who had no concerns and fears before the study, trusted Gymmy and expressed positive evaluations of it after the study. They did not report any inconvenience even though they used the exact same robot.

Hence, both the acceptance and the assimilation studies suggest that the relative impact of technophobia on older adults' evaluations of SARs is stronger than that of trust. Technophobia as a starting point is a key factor in predicting older adults' QE of SARs for anticipated interaction, and in influencing their QE for actual interaction. These findings in relation to technophobia in HRI studies, constitute a significant contribution of this research, and correspond with previous literature dealing with technophobia in other modern technologies, where it has been found to have adverse effects on older adults' acceptance, performance, and competence in computers (Hogan, 2009), internet (Nimrod, 2018), and smartphones (Thalib, 2019).

Trust and technophobia inevitably affect each other. On the one hand, high levels of trust reduce aspects of technophobia (Langer et al., 2019; Pedersen et al., 2018). On the other hand, high levels of technophobia reduce trust (Baker et al., 2018). Therefore, it is important to investigate these two factors together and their mutual influence on QE of SARs among older adults. Using this simultaneous examination, we were able to understand in depth the interrelationships and the mutual influence of two factors on each other as well as their relative impact on older adults' evaluations of SARs.

### 6.1.2. Simultaneous exploration of uses, constraints, and outcomes in real-life conditions over a long period

Similar to the simultaneous examination of trust and technophobia, the simultaneous exploration of the SAR's uses, constraints, and outcomes in real-life conditions over a long period also made a significant contribution to understanding the factors influencing older adults' QE of SARs. First, the simultaneous examination enabled identification of the uses, constraints and outcomes that older adults experience while assimilating a SAR into their lives. For example, under the category of uses, which includes factors that influence the user experience, it was found that the user attributes including their exercise habits and convenience in operating Gymmy significantly influenced their perceived experience. The most influential antecedent constraint found in the assimilation study was the robot-related technophobia among the Skeptics, who believed that at some point, robots will be able to control the human race. Intervening constraints found in this part of the research were technical problems and Gymmy's slow operation. Finally, under the outcomes category, there were positive outcomes experienced by the Fans such as pleasure and a sense of satisfaction, and negative ones experienced by the Skeptics, such as feelings of boredom, frustration, despair, lack of enjoyment, and disappointment with Gymmy.

Second, this examination of these three factors, also helped identify and accurately characterize the two assimilation patterns presented in this study (Fans Vs. Skeptics). It highlighted the differences between the patterns in the various aspects of assimilation and made it possible to look at the assimilation of Gymmy as a comprehensive process with three interrelated and influencing factors. A good example of these differences between the patterns in the assimilation aspects, can be found in the factors affecting user experience (uses category). Findings show that the same robot with the same features was perceived in a fundamentally different way by the participants. That is, some participants were very satisfied with the use of Gymmy and experienced it as easy and comfortable, while others who experienced the same use found it too simple, unchallenging, and boring.

Third, as suggested by Zafrani and Nimrod (2019), the broader and detailed analysis provided by the simultaneous examination in real-life conditions over a long period made it possible to draw important conclusions based on the interrelationships between SAR's uses, constraints, and outcomes. For example, one of the study's most valuable conclusions is that disappointment with a specific robot does not necessarily detract from the general evaluation of SARs, but only reduces its relevance. This conclusion could not have been drawn without simultaneous investigation of the three factors. The disappointment of the Skeptics with Gymmy stemmed from the constraints that influenced and were influenced by the uses. However, the insight that this disappointment only reduced Gymmy's relevance to them and did not detract from their general evaluation of SARs, stemmed from an investigation of the outcomes category, in which they indicated that following the experience with Gymmy they do think SARs can assist the older population in general and lonely older adults and older people with disabilities in particular.

Fourth, conducting the study over a long period and in real-life conditions made it possible to understand the relationship between the attitudes towards SARs before and after the use of a specific model. This relationship was explained and mediated following the impact of the period of use of Gymmy on the participants, shaped by the uses, constraints and outcomes they experienced during this period. The longitudinal study has contributed greatly to capturing the behaviors that developed throughout the study period (Vad et al., 2015), and to understanding whether and how the uses, constraints, and outcomes changed during the assimilation period. For example, under the outcomes category, there were immediate benefits that Fans gained from the first day of using Gymmy, such as fun and enjoyment, and the benefits they perceived they achieved following the prolonged period of use, such as strengthening awareness and motivation to exercise and lose weight. Another example illustrating how the outcomes changed over the six weeks of the study was found in the use experience and the assimilation period of the Skeptics, who similar to the Fans, in the first days of use experienced immediate benefits of fun and enjoyment, but unlike the Fans, these benefits gradually turned into negative feelings of boredom, frustration, dissatisfaction, and lack of enjoyment and interest.

These findings are consistent with other longitudinal studies on older adults' use of modern technologies such as internet, smartphone applications, and smart speakers (Hakobyan et al., 2016; Kim & Choudhury, 2021; Nimrod & Edan, 2022; Shillair et al., 2015). These studies found that longitudinal settings are essential to give older adults time to explore new technology at their own pace (Shillair et al., 2015), fully evaluate the long-term adaptability and acceptability of the technology (Nimrod & Edan, 2022), necessary to achieve a more accurate perspective on the behavioral and perceptual changes among older adults (Hakobyan et al., 2016). Furthermore, this research demonstrated the importance of advancing longitudinal studies to evaluate parameters influencing SARs assimilation. Although only few such studies have been conducted in the field of robotics (Céspedes Gómez et al., 2021; Chang et al., 2013; Chang & Šabanović, 2015; Gasteiger et al., 2021), they indicated that longitudinal methods provide an opportunity to examine whether fear of using an unfamiliar technology such as robots is related to negative attitudes (Papadopoulos et al., 2020), and make it easier to identify how individual attitudes, perceptions and social factors affect the assimilation of HRI among older adults (Chang & Šabanović, 2015).

Research in real-life environments (the participants' homes in this case) instead of an artificial laboratory setting best suited the research objectives, as it gave participants the ability to be alone with Gymmy in familiar conditions and communicate and interact with it as naturally as possible (Yamazaki et al., 2014). Moreover, research in real-life conditions with long-term use of the new technology was required to provide a more comprehensive understanding of the effects of users' prior perceptions and attitudes on post-use evaluations and future behavioral intentions (Kujala et al., 2017). Additionally, the more "real" the conditions of the study, the more "real" the findings since research in these conditions takes into account all the constraints, limitations, irregularities and opportunities of everyday reality (Schmeeckle, 2003).

### 6.1.3. Investigation of anticipated and actual interaction utilizing quantitative and qualitative methods

Investigation of factors affecting older adults' QE of SARs regarding both anticipated and actual interaction, via mixed quantitative and qualitative methods, significantly contributed to bridging the gaps in the existing literature. First, the investigations conducted through an online survey and an assimilation study, enabled learning about older adults' expected initial and actual acceptance of SARs, as well as the whole process of SARs' assimilation and the factors effecting post-use QE by older adults. Each part of the research contributed to the understanding of the other, and, in combination with the study's other characteristics, created an innovative investigative framework that yielded important insights.

Moreover, the combination of acceptance and assimilation research suggested that an assimilation pattern can be predicted according to the level of acceptance. That is, based on online surveys/questionnaires and/or opening interviews, study participants can be classified into one of the assimilation patterns with special emphasis on whether or not they have a high level of technophobia. Following the initial classification and attribution to an assimilation pattern, efforts can be made to better adapt the SARs to the participants and, if necessary, reduce the latter's concerns before the period of use. For example, before use, guidance on the SAR and its functions should be provided. This recommendation is consistent with previous literature examining interaction between older adults and modern technologies, which suggested that providing information and guidance lead to a more positive view of ICTs among participants and help improve their acceptance and adaptation (Fields et al., 2021; Shillair et al., 2015).

The first part of this research was *quantitative* and included an online survey with a large number of participants, which yielded initial insights into the phenomenon under investigation, and contributed to the design and accuracy of the settings of the assimilation study and the in-depth interviews conducted within it. For example, after the online survey revealed that age is a variable with a consistent positive association with participants' QE, and that apparently after watching the video, the survey respondents' impression was that Gymmy is not physically and cognitively challenging enough for their age, it was decided to raise the minimum age participation for the assimilation study from 65 to 75.

The *qualitative* part—the assimilation study in general and the in-depth interviews in particular have contributed greatly to understanding what underlay the results obtained in the online survey. That is, the online survey provided answer to "**what**" questions, while the assimilation study elaborated on that understanding and made it possible to answer the "**why**" and "**how**" questions. For example, the online survey showed that the relative impact of technophobia is significantly stronger than that of trust, and that technophobia constitutes a major inhibitor on older adults' QE and acceptance of SARs. However, the answer to the question of why this was the case was not entirely clear from the online survey alone. The assimilation study made a significant contribution to answering this question, as it highlighted that technophobia is an inherent inhibitor, affecting all stages of the assimilation process, and that efforts should be made to reduce concerns about robots to promote more positive QE and eventually acceptance and successful assimilation of SARs among older adults.

Another contribution of the *mixed-methods approach* is suggested by the weekly surveys and usage data collected in the assimilation study. From a purely quantitative point of view, these data show that both the Fans and the Skeptics decreased the intensity of use after the first two weeks of the study period. Ostensibly, it could be concluded that this decrease in the average amount of weekly use was due to a decrease in participants' enjoyment, enthusiasm, and satisfaction with the SAR and its functions, and/or from feelings of disappointment and lack of interest. However, drawing this conclusion for the period of use of the Fans was fundamentally wrong, as they did not experience such feelings at all. The correct and relevant conclusion for the Fans was derived from the qualitative interviews conducted with them, in which they shared that throughout the study period they enjoyed the use of Gymmy and benefited greatly from it. The decrease in the number of uses can be explained by a novelty effect, i.e., an increased initial response of individuals to new technology following an increased initial interest, which does not necessarily indicate patterns of use over time (Cajita et al., 2020; Wright et al., 2017).

Hence, not only did the qualitative research contribute greatly to the understanding of the dynamic complexity and subtlety of human experience in HRI, it also helped to provide a reasoned interpretation of the quantitative data (Denzin, 2017; Kirkels, 2016). As highlighted by Seibt et al. (2021), qualitative research complements quantitative research by bringing the researcher closer to the participants, and through a deeper understanding of the human, psycho-social, cultural, and multidisciplinary aspects of human experience. Therefore, the method used in this research is consistent with the recommendation regarding future HRI studies, which should adopt a mixed methods approach, i.e., integrating quantitative and qualitative research in order to achieve more complete and accurate findings (Seibt et al., 2021).

### 6.2. Practical recommendations

The findings motivate two integrative practical recommendations for improving SARs' evaluation by older adults. First, to promote a more positive QE, acceptance, and successful assimilation process, developers and designers of SARs for older adults are advised to consider the needs of older adults and take steps to reduce their fears, concerns, and inconveniences about robots. In order to reduce older adults's technophobia before use, the features that make the robots pleasant to use should be stressed in all educational and marketing communication targeting older people. Moreover, a proper training session on the SAR and its functions should be provided, during which participants are given relevant information and allowed to ask questions to remove their doubts and fears. Second, as the pragmatic qualities of the SAR are more crucial to its QE than the emotional aspects, it is essential to invest efforts in developing and designing SARs that are functional, convenient, simple, provide added value, easy to use, and have unique features such as a multi-modal communication.

### 6.3. Limitations

The present research demonstrates the usefulness of the holistic approach in research on older users of technology (Zafrani & Nimrod, 2019) and the value of longitudinal methods in assimilation studies. However, despite its strengths, this research has several limitations that should be acknowledged. Although the use of an online survey in the first part of the research allowed for a large sample of participants, reliance on this method for data collection could be biased since older adults who participate in online surveys have greater digital literacy than other people of that age group. In this study, which measured technophobia, this limitation probably also meant that respondents were less afraid of technology than people who do not participate in online surveys.

In addition, in both parts of the study, the samples were not representative. In the online survey, a non-probabilistic convenience sample was performed, and in the assimilation study the main research method was qualitative, in which there is usually no attempt to get representative samples because the depth of the observations often restricts their focus to a few individuals with specific characteristics. Most of the participants in the assimilation study were healthy older adults, without physical or cognitive impairments. Therefore, we cannot generalize from its findings to frail older adults. In addition, participants in this part of the study immigrated to Israel from different countries with different cultures. However, these participants have lived in Israel for many years, and the Israeli culture has taken root and shaped them. Thus, there is certainly a possibility that different perceptions, preferences, and evaluations would have been obtained if the research had been carried out with older adults with different cultural characteristics. As previous literature has shown, different cultural values significantly affect perceptions, acceptance of and attitudes toward SARs (Bliss et al., 2021; Conti, 2016).

Another limitation related to the background characteristics of the participants in the assimilation study is related to their level of education. Most of the participants in this part of the study were educated older adults. Previous studies indicated that higher levels of education are positively associated with abilities to learn, adapt, use, accept and assimilate modern technologies (Nimrod, 2017; Seifert, 2020). It is quite possible that older adults with more varied levels of education will have other factors that will influence their QE of SARs, especially in light of the fact that the education variable had an important impact on the online survey.

The two parts of the present study were limited to one specific SAR designed for a specific purpose (physical training), and thus their findings cannot be generalized to other robotic systems. Using a SAR intended for another purpose or using a multi-purpose SAR might have yielded different results. Another limitation associated with the robot lies in the fact that only upper-body exercises were included in the system. On the one hand, it is possible that this limited version of the SAR affected participants' evaluations and the overall user experience they experienced during the study period. On the other hand, using a full-body system that includes also the lower body, could have endangered older adults due to balance problems, creating a risk of falling.

Finally, it is certainly worth noting, that the research, in both parts, was conducted during the COVID-19 pandemic, which may have affected the general mood of the participants and their assessments.

### 6.4. Directions for future research

Future research should expand the present study to explore factors affecting QE of SARs among additional older audiences including "oldest old" (85+ years; Czaja et al., 2019) participants, older individuals residing in other countries and cultural environments, and older adults with different levels of education, income, previous experience with robots, media usage and exposure, self-efficacy, and physical and cognitive functioning. Such studies should also follow up assimilation processes for longer periods to explore how uses, constraints and outcomes continue to evolve over time, compare assimilation of Gymmy by using different levels of intensity, length, and difficulty of physical training, and to a version of Gymmy with lower limbs, and compare assimilation of SARs by using additional types of robots, including multi-function vs. single-function SARs, stationary vs. mobile SARs, and proactive vs. reactive SARs. Additional research should consider additional residential contexts, compare the factors affecting QE of SARs among seniors living in nursing vs. residential homes. and involve in the research significant figures such as the older adults' caregivers and family members.

Specifically, two future assimilation studies that could be very interesting and enriching are related to cross-cultural differences among participants and differences between SARs and social robots, which will be expanded on below. First, culture provides rules, guidelines, and norms for social behavior (Kelter et al., 2004), and influences every aspect of human perception and interaction with each other (Sriramesh, 2012). Since theories of cross-cultural behavior transfer from human-human to human-robot interaction, it is assumed that culture may play a role in the assimilation process of SARs in daily lives (e.g., Evers et al., 2008; Lee et al., 2012; Li et al., 2010), and that humans' attitudes and behavior towards SARs will differ across cultures (Evers et al., 2008). Indeed, recent studies have confirmed the existence of cultural influence on the assimilation of SARs by older adults in many aspects (e.g., Bartneck, 2008; Bartneck et al., 2007; Evers et al., 2008; Lee et al., 2012; Li et al., 2010), such as acceptance and attitudes towards SARs (e.g., Bartneck et al., 2007; Bartneck, 2008; Li et al., 2010; Korn et al., 2021; Akalin et al., 2021), expectations regarding the appearance of SARs (e.g., de Graaf & Allouch, 2015; Haring et al., 2016; Kaplan, 2004), and preferred communication styles (verbal and nonverbal; e.g., Lim et al., 2021). The awareness of how cultural aspects and ethnic differences affect older adults' perspectives, attitudes, and acceptance of SARs is an essential factor that must be explored in depth in future studies to more fully understand what makes assimilation successful in different cultural contexts.

Second, although social robots and SARs, have a number of similar characteristics,

the literature points to differences between the two (e.g., Feil-Seifer & Mataric, 2005; Heerink, 2011; Henschel et al., 2021). Social robots are physically embodied autonomous artificial agents, designed to become a human-equivalent partner in social interactions, and capable of interacting naturally with people for social purposes in their everyday lives, through different ways such as communicating, cooperating, entertainment and decision making (e.g., Feil-Seifer & Mataric, 2005; Fong et al., 2003; Henkel et al., 2020; Henschel et al., 2021; Istenic et al., 2021). SARs combine features of assistive robots and social robots (Feil-Seifer & Mataric, 2005), and are designed to provide the appropriate emotional, cognitive, and social cues to encourage individuals' development, learning, or therapy (Feil-Seifer & Mataric, 2005). Due to the emphasis on social interaction, the goal of this kind of robot is to develop effective and close interactions with humans for the purpose of providing measurable assistance in convalescence, learning, rehabilitation, etc. (Feil-Seifer & Mataric, 2005). That is, the goal of future research will be to offer an understanding of the similarities and differences in processes of social robots and SARs' assimilation in later life.

Furthermore, as noted above, in order to increase the representativeness of the sample, similar studies should be advanced with other populations. As social and assistive robots are expected to penetrate the market it will be important to ensure mixed methods approaches are advanced in both acceptance and assimilation studies to ensure the success of these technologies.

# Appendices

## Appendix A: User Experience Questionnaire (UEQ)

The following items present ways in which people may describe social robots. Regarding each pair of items, please mark your overall impression with the social robot that you have used in the past weeks.

|  | -3 | -2 | -1 | 0 | +1 | +2 | +3 |  |
|---|---|---|---|---|---|---|---|---|
| annoying | ○ | ○ | ○ | ○ | ○ | ○ | ○ | enjoyable |
| not understandable | ○ | ○ | ○ | ○ | ○ | ○ | ○ | understandable |
| creative | ○ | ○ | ○ | ○ | ○ | ○ | ○ | dull |
| easy to learn | ○ | ○ | ○ | ○ | ○ | ○ | ○ | difficult to learn |
| valuable | ○ | ○ | ○ | ○ | ○ | ○ | ○ | inferior |
| boring | ○ | ○ | ○ | ○ | ○ | ○ | ○ | exciting |
| not interesting | ○ | ○ | ○ | ○ | ○ | ○ | ○ | interesting |
| unpredictable | ○ | ○ | ○ | ○ | ○ | ○ | ○ | predictable |
| fast | ○ | ○ | ○ | ○ | ○ | ○ | ○ | slow |
| inventive | ○ | ○ | ○ | ○ | ○ | ○ | ○ | conventional |
| obstructive | ○ | ○ | ○ | ○ | ○ | ○ | ○ | supportive |
| good | ○ | ○ | ○ | ○ | ○ | ○ | ○ | bad |
| complicated | ○ | ○ | ○ | ○ | ○ | ○ | ○ | easy |
| unlikeable | ○ | ○ | ○ | ○ | ○ | ○ | ○ | pleasing |
| usual | ○ | ○ | ○ | ○ | ○ | ○ | ○ | leading edge |
| unpleasant | ○ | ○ | ○ | ○ | ○ | ○ | ○ | pleasant |
| secure | ○ | ○ | ○ | ○ | ○ | ○ | ○ | not secure |
| motivating | ○ | ○ | ○ | ○ | ○ | ○ | ○ | demotivating |
| meets expectations | ○ | ○ | ○ | ○ | ○ | ○ | ○ | does not meet expectations |
| inefficient | ○ | ○ | ○ | ○ | ○ | ○ | ○ | efficient |
| clear | ○ | ○ | ○ | ○ | ○ | ○ | ○ | confusing |
| impractical | ○ | ○ | ○ | ○ | ○ | ○ | ○ | practical |
| organized | ○ | ○ | ○ | ○ | ○ | ○ | ○ | cluttered |
| attractive | ○ | ○ | ○ | ○ | ○ | ○ | ○ | unattractive |
| friendly | ○ | ○ | ○ | ○ | ○ | ○ | ○ | unfriendly |
| conservative | ○ | ○ | ○ | ○ | ○ | ○ | ○ | innovative |

**Appendix B: Human-Robot Trust Scale**

The following is a list of qualities of social robots.

**What percentage of the time, in your opinion, this social robot will…**

|  | 0% | 10% | 20% | 30% | 40% | 50% | 60% | 70% | 80% | 90% | 100% |
|---|---|---|---|---|---|---|---|---|---|---|---|
| Act consistently | | | | | | | | | | | |
| Function successfully | | | | | | | | | | | |
| Malfunction | | | | | | | | | | | |
| Have errors | | | | | | | | | | | |
| Provide feedback | | | | | | | | | | | |
| Meet the needs of the mission | | | | | | | | | | | |
| Provide appropriate information | | | | | | | | | | | |
| Communicate with people | | | | | | | | | | | |
| Perform exactly as instructed | | | | | | | | | | | |
| Follow directions | | | | | | | | | | | |
| Be dependable | | | | | | | | | | | |
| Be reliable | | | | | | | | | | | |
| Be unresponsive | | | | | | | | | | | |
| Be predictable | | | | | | | | | | | |

**Appendix C: Technophobia Scale**

The following is a list of statements describing attitude towards Gymmy. Please rate to what extent you agree with each statement.

| | Strongly disagree 1 | Disagree 2 | Neither agree nor disagree 3 | Agree 4 | Strongly agree 5 |
|---|---|---|---|---|---|
| I feel some anxiety when I approach Gymmy | | | | | |
| Gymmy will make me restless | | | | | |
| I think most people will be able to use Gymmy better than I | | | | | |
| I will feel frustrated when I use Gymmy | | | | | |
| Thinking about using Gymmy makes me nervous | | | | | |
| Gymmy is intimidating | | | | | |
| In physical training, I would rather have a human person train me, than use Gummy | | | | | |
| In cognitive training, I would rather have a human person train me, than use Gymmy | | | | | |
| I resent that social robot like Gymmy are becoming so prevalent in our daily lives | | | | | |
| I will feel more confident training with a human trainer, than with Gymmy | | | | | |
| Social robots should not handle in physical training of humans | | | | | |
| Social robots should not handle in cognitive training of humans | | | | | |
| I feel comfortable when using Gymmy | | | | | |
| Using Gymmy will make life easier | | | | | |
| I like that Gymmy is so convenient | | | | | |

**Appendix D: Exercise questionnaire**

**A) How important it is for you to engage in physical activity?**

1) Not at all important

2) Slightly important

3) Moderately important

4) Very important

5) Extremely important

**B) How do you perceive your fitness level?**

1) Not fit at all

2) Below average

3) Average

4) Good

5) Very good

**C) How often do you exercise?**

1) Do not exercise at all

2) Less than once a week

3) 1-2 times a week

4) 3-4 times a week

5) 5-6 times a week

6) Every day

**D) How long does your exercise usually last?**

1) 0-30 minutes

2) 31-45 minutes

3) 46 minutes - 1 hour

4) 1-2 hours

5) 2-3 hours

**E) What types of exercise do you usually participate in? (Please check all the appropriate options)**

1) Walking

2) Running

3) Swimming

4) Bicycle riding

5) Gym workout

6) Exercise classes such as yoga, Pilates, Feldenkrais

7) Ball games such as basketball, basketball

8) Dance

9) Other, please detail _____________

**F) What motivates you to exercise? (Please check all the appropriate options)**

1) I want to keep fit

2) I want more energy

3) I want to lose weight

4) I want to increase my muscle mass

5) I want to increase bone density

6) I want to strengthen my power

7) I want to reduce the levels of fat in my body

8) I want to improve my flexibility

9) I want to improve posture and balance

10) I want to reduce stress

11) I want to achieve a sporting goal

12) I enjoy exercising

13) Other, please detail _____________

**G) Are you interested in exercising more often?**

1) Yes, much more

2) Yes, a little more

3) No

**H) What prevents you from exercising more often? (Please check all the appropriate options)**

1) I do not have enough time

2) I lose motivation

3) I'm too tired

4) I have a health condition

5) I have no options available for exercise in my area of residence

6) I have no one to exercise with

7) I do not enjoy exercise so much

8) Other, please detail _____________

**Appendix E: Demographic, sociodemographic, and health background questionnaire**

Marital Status:

1) Married  2) Divorced  3) Widowed  4) Single  5) Permanent relationship  6) Other: ________

Number of children: ______

Residence Locality:

1) Big city  2) Outskirts of a big city  3) Medium or small city  4) Rural locality  5) Other: _______

Type of residence:

1) Apartment  2) Private House  3) Assisted living  4) Nursing Home  5) Other: _______

Living with: 1) Spouse 2) Son or daughter 3) Other family member 4) caregiver 5) Other: _______

Religious orientation:

1) Secular  2) Traditional  3) Religious  4) Ultra-orthodox

Country of birth:

1) Israel  2) Western Europe, America  3) Asia, Africa  4) Eastern Europe  5) Other: _________

Father's country of birth:

1) Israel  2) Western Europe, America  3) Asia, Africa  4) Eastern Europe  5) Other: _________

Number of years of education: ______

Employment status:

1) Working full time  2) Part-time worker  3) Retiree  4) Unemployed  5) Other: __________

Income level:

1) Much higher than average 2) Slightly higher than average 3) Similar to the average 4) Slightly lower than average 5) Much lower than average

When you think about your physical health, to what extent are you satisfied with your physical health in general?

| 1 Not at all satisfied | 2 | 3 | 4 | 5 | 6 | 7 | 8 | 9 | 10 Completely satisfied |
|---|---|---|---|---|---|---|---|---|---|
| ○ | ○ | ○ | ○ | ○ | ○ | ○ | ○ | ○ | ○ |

When you think about your cognitive functioning, to what extent are you satisfied with your cognitive functioning in general?

| 1 Not at all satisfied | 2 | 3 | 4 | 5 | 6 | 7 | 8 | 9 | 10 Completely satisfied |
|---|---|---|---|---|---|---|---|---|---|
| ○ | ○ | ○ | ○ | ○ | ○ | ○ | ○ | ○ | ○ |

**Appendix F: Opening interview guideline**

General opening question:

    1) Please tell me about yourself.

    (Family, personal history (where he was born, immigration to Israel if relevant), past and present employment, health status).

A comprehensive descriptive question:

    2) Please describe your daily routine.

Questions that invite examples:

    3) What are the main activities you do at home? Do you feel any difficulties performing daily tasks? Which? How do you deal with them?

    4) Please tell me about your current usage of communication and information technologies such as computer, Internet, and mobile phone:

    What are you using and why? (What are your main uses? Frequency of use?)

    What are you not using and why?

General questions about the research:

    5) Do you know robots? Have you ever had an experience with a certain type of robot? Which? How would you describe the experience?

    6) Have you heard of social robots? Have you had any experience with them? What? How would you describe the experience?

    7) Why did you volunteer to participate in the study on intelligent personal assistants?

    8) Do you think a robot can help you? how?

    9) Do you think that robots have advantages over the technologies we talked about? Which?

    10) Do you think that robots have drawbacks compared with these technologies? Which?

    11) Do you think robots may be dangerous? in what way?

    12) Are there certain areas where you would like to receive assistance from robots? Which?

    13) Are there certain areas where you would not want to receive assistance from robots? Which?

    14) Would you prefer a stationary or mobile robot? why?

    15) Would you prefer a proactive robot, or a robot that only respond? why?

Questions to examine expectations Gymmy:

    16) Why do you expect from your interaction with Gymmy?

    17) Are there specific uses that you would like to benefit from using Gymmy?

    18) Are there factors that can prevent you from using Gymmy, or influence how often you use it?

    19) Do you think there are risks in using Gymmy?

Summary question:

    20) Is there anything you would like to add, beyond what has already been discussed, about your expectations of experience with Gymmy?

**Appendix G: Concluding interview guideline**

General opening question:

    1) How would you describe your experience of using Gymmy.

    Did this use match your expectations? it was enjoyable? effective? dangerous? challenging?

    A comprehensive descriptive question:

    2) Please describe your daily routine with Gymmy.

Questions that invite examples:

    3) What were your main uses with Gymmy?

    4) What were the uses that disappointed you in Gymmy?

    5) Has your frequency of use of Gymmy increased/decreased over time? how? why?

    6) Do you feel that during the experience period your lifestyle has become more active?

    7) Do you feel that with the help of Gymmy you have engaged more in physical activity?

    8) Do you feel that with the help of Gymmy it was easier and more accessible for you to perform physical activity?

    9) Do you feel that with the help of Gymmy you got to engage in new exercises that you do not usually perform? how did it feel?

    10) What were the difficulties you experienced during the period of using Gymmy?

    11) Were there factors that prevented you from using Gymmy, or influenced your frequency of use? how?

    12) After experiencing Gymmy, do you think robots may be dangerous? in what way?

    13) Are there certain areas in which you expected to receive assistance from Gymmy but did not receive it? which?

    14) Do you think Gymmy has any advantages over other physical activities? which?

    15) Do you think Gymmy has any disadvantages compared to these activities? which?

    16) Do you think the period of use of Gymmy made you more open to experimenting with other robots? how?

    17) After experiencing Gymmy, do you think robots can assistance older adults? how?

Summary question:

    18) Is there anything you would like to add, beyond what has already been discussed, about your experience with Gymmy?

# Appendix H: Average weekly usage data

| Fans | Week 1 | Week 2 | Week 3 | Week 4 | Week 5 | Week 6 | Average | SD |
|---|---|---|---|---|---|---|---|---|
| Miley | 4 | 5 | 2 | 2 | 0 | 0 | 2.167 | 2.041 |
| Nina | 3 | 3 | 0 | 4 | 3 | 3 | 2.667 | 1.366 |
| Tom | 3 | 3 | 1 | 0 | 2 | 2 | 1.833 | 1.169 |
| Alexandra | 5 | 5 | 4 | 5 | 5 | 5 | 4.833 | 0.408 |
| Luca | 4 | 2 | 4 | 3 | 4 | 4 | 3.500 | 0.837 |
| Helen | 7 | 7 | 6 | 7 | 6 | 8 | 6.833 | 0.753 |
| Sami | 5 | 4 | 5 | 0 | 2 | 0 | 2.667 | 2.338 |
| Paula | 7 | 7 | 7 | 6 | 6 | 6 | 6.500 | 0.548 |
| Dafna | 7 | 4 | 4 | 3 | 5 | 5 | 4.667 | 1.366 |
| Average | 5.000 | 4.444 | 3.667 | 3.333 | 3.667 | 3.667 | | |
| SD | 1.658 | 1.740 | 2.291 | 2.449 | 2.062 | 2.693 | | |

| Skeptics | Week 1 | Week 2 | Week 3 | Week 4 | Week 5 | Week 6 | Average | SD |
|---|---|---|---|---|---|---|---|---|
| Clara | 4 | 3 | 2 | 4 | 3 | 1 | 2.833 | 1.169 |
| Daniel | 4 | 1 | 2 | 0 | 1 | 1 | 1.500 | 1.378 |
| Maggie | 3 | 6 | 5 | 3 | 4 | 3 | 4.000 | 1.265 |
| Sofie | 1 | 2 | 2 | 0 | 1 | 2 | 1.333 | 0.816 |
| Gavriel | 4 | 4 | 3 | 3 | 4 | 3 | 3.500 | 0.548 |
| Michael | 4 | 5 | 2 | 0 | 3 | 3 | 2.833 | 1.722 |
| Sarah | 4 | 5 | 4 | 7 | 4 | 4 | 4.667 | 1.211 |
| Joshua | 4 | 5 | 7 | 3 | 4 | 4 | 4.500 | 1.378 |
| Arik | 4 | 3 | 1 | 1 | 0 | 0 | 1.500 | 1.643 |
| Average | 3.556 | 3.778 | 3.111 | 2.333 | 2.667 | 2.333 | | |
| SD | 1.014 | 1.641 | 1.900 | 2.345 | 1.581 | 1.414 | | |

**תקציר**

אוכלוסיית העולם מזדקנת במהירות, ומספר המבוגרים צפוי לגדול באופן דרמטי במהלך השנים הבאות. רובוטים חברתיים מסייעים צפויים לעזור לאנושות להתמודד עם האתגרים שמציבה מגמת הזדקנות זו על ידי תמיכה בעצמאות, הזדקנות בריאה ורווחה בזקנה. כדי להשיג קבלה והטמעה מוצלחת של רובוטים חברתיים מסייעים, יש צורך להבין את הגורמים המשפיעים על הערכות האיכות שלהם בקרב מבוגרים. מחקרים קודמים שבחנו אינטראקציית אדם-רובוט בזקנה, הצביעו על כך שאמון ברובוטים משפר משמעותית את הערכות האיכות, בעוד שהיבטים של טכנופוביה מפחיתים אותן במידה ניכרת. עם זאת, הספרות הקודמת בחנה בנפרד את ההשפעות של אמון וטכנופוביה על הערכות האיכות של רובוטים חברתיים מסייעים בקרב מבוגרים, תוך התעלמות מהאפשרות שגורמי מפתח אלו יכולים להתקיים במקביל ולנטרל זה את השפעתו של זה. יתרה מכך, בניגוד לאמון, טכנופוביה כמעט ולא נחקרה בהקשר של אינטראקציית אדם-רובוט. בנוסף, הספרות הקיימת מציעה שהערכת האיכות הכוללת של רובוטים חברתיים מסייעים על ידי מבוגרים מתעצבת באמצעות שלושה היבטים: השימושים, המגבלות, ותוצאות השימוש. עם זאת, מחקרים אלו, שהיו לרוב מוגבלים בזמן, בחנו היבטי קבלה בלבד ולא הטמעה, ובדרך כלל התמקדו בהיבט אחד בלבד של האינטראקציה בין הרובוטים למבוגרים, כלומר ביצעו בחינה נפרדת של השימושים, המגבלות ותוצאות השימוש. יתרה מכך, רוב המחקרים עד כה שעסקו באינטראקציית אדם-רובוט הסתמכו על ניתוחים כמותניים או איכותניים ולא יישמו גישה של שיטות מעורבות.

עבודת הדוקטורט הזו נועדה לגשר על הפערים בספרות הקיימת תוך התבססות על שני מחקרים משלימים. ראשית, מחקר קבלה, שנערך באמצעות סקר מקוון עם 384 משתתפים, בדק בו-זמנית את ההשפעה של אמון וטכנופוביה על הערכת איכות של רובוטים חברתיים מסייעים על ידי מבוגרים. לאחר מכן, מחקר הטמעה, שנערך עם תשעה עשר מבוגרים, בחן כיצד מתעצבת הערכת האיכות בעקבות אינטראקציה ממשית עם רובוטים חברתיים מסייעים. מחקר זה בוצע באמצעות בדיקה סימולטנית של השימושים, המגבלות ותוצאות השימוש, ונערך בתנאי חיים אמיתיים ולאורך תקופה ממושכת. מחקר זה התבסס על ראיונות עומק, סקרים שבועיים ודוחות שימוש שהופקו על ידי הרובוט. בשני חלקי המחקר, נעשה שימוש ב"ג׳ימי", מערכת רובוטית לאימון גופני וקוגניטיבי של מבוגרים, שפותחה במעבדה שלנו.

התוצאות הצביעו על כך שההשפעה היחסית של טכנופוביה על הערכת איכות של מבוגרים כלפי רובוטים חברתיים מסייעים הייתה משמעותית יותר מזו של אמון, וכי טכנופוביה היוותה את המגבלה שהשפיעה בצורה המשמעותית ביותר על השימוש ברובוטים. בנוסף, במחקר ההטמעה נמצאו שני דפוסי שימוש: (א) ׳מעריצים׳ - משתתפים שנהנו מהשימוש בג׳ימי, נתנו בו אמון, ייחסו לו ערך מוסף וחוו את תהליך ההטמעה מוצלח. (ב) ה׳ספקנים׳ - משתתפים שלא אהבו את ג׳ימי, העריכו את השימוש בו באופן שלילי, וחוו תהליך הטמעה מאכזב. קבוצה זו הביעה טכנופוביה לפני ההשתתפות במחקר, בעוד שה״מעריצים״ לא דיווחו כלל על חשש או הסתייגויות בנוגע לרובוטים.

השילוב של מחקר הקבלה ומחקר ההטמעה הצביע על כך שניתן לחזות דפוס הטמעה בהתאם לרמת הקבלה. הממצאים מדגישים את החשיבות הרבה של חקר הטכנופוביה במחקרי אינטראקציית אדם-רובוט, ומציעים שאימוץ טכנולוגיות רובוטיות בקרב מבוגרים תלוי במידה רבה בהפחתת תחושת הטכנופוביה שלהם. יתר על כן, עבודת הדוקטורט הזו מדגימה את חשיבותה ותרומתה של הגישה ההוליסטית במחקר העוסק במשתמשי טכנולוגיה מבוגרים, ושופכת אור על התועלת

המופקת מחקירה סימולטנית של גורמים מקדמים ומעכבים, מחקירה בו-זמנית של השימושים, המגבלות ותוצאות השימוש, מביצוע מחקרי הטמעה בתנאי חיים אמיתיים לתקופת זמן ממושכת, ומשימוש בגישה של שיטות מעורבות במחקרי אינטראקציית אדם-רובוט.

**מילות מפתח:** הזדקנות, אינטראקציית אדם-רובוט, הערכת איכות, רובוטים חברתיים מסייעים, מאמן רובוטי, טכנופוביה, אמון, קבלה, הטמעה, מבוגרים.

## הצהרת תלמיד המחקר עם הגשת עבודת הדוקטור לשיפוט

אני החתום מטה מצהיר בזאת :

✓ חיברתי את חיבורי בעצמי, להוציא עזרת ההדרכה שקיבלתי מאת המנחים.

✓ החומר המדעי הנכלל בעבודה זו הינו פרי מחקרי <u>מתקופת היותי תלמיד/ת מחקר</u>.

תאריך : 06.03.2022    שם התלמיד : עודד זפרני    חתימה : 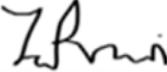

העבודה נעשתה בהדרכת

פרופ׳ גלית נמרוד
פרופ׳ יעל אידן

במחלקה להנדסת תעשייה וניהול
בפקולטה למדעי ההנדסה

אוניברסיטת בן-גוריון בנגב

# בין חשש לאמון:
# הגורמים המשפיעים על ההערכה של זקנים
# לרובוטים חברתיים מסייעים

מחקר לשם מילוי חלקי של הדרישות לקבלת תואר "דוקטור לפילוסופיה"

מאת

## עודד זפרני

הוגש לסינאט אוניברסיטת בן-גוריון בנגב

אישור המנחה: פרופ' גלית נמרוד ________
אישור המנחה: פרופ' יעל אידן ________

אישור דיקן בית הספר ללימודי מחקר מתקדמים ע"ש קרייטמן ________

ג' אדר ב', תשפ"ב                                                       06.03.2022

באר שבע

# בין חשש לאמון:
# הגורמים המשפיעים על ההערכה של זקנים
# לרובוטים חברתיים מסייעים

מחקר לשם מילוי חלקי של הדרישות לקבלת תואר "דוקטור לפילוסופיה"

מאת

עודד זפרני

הוגש לסינאט אוניברסיטת בן-גוריון בנגב

ג' אדר ב', תשפ"ב             06.03.2022

באר שבע